\documentclass{aa}

\pdfoutput=1
\usepackage{graphicx}
\usepackage{txfonts}
\usepackage[export]{adjustbox}
\usepackage{subfig}

\usepackage{longtable}
\usepackage{tabularx}
\usepackage{pdflscape}
\usepackage{multicol}
\usepackage{multirow}

\usepackage{hyperref}
 \hypersetup{
     colorlinks=true,
     linkcolor=blue,
     filecolor=magenta,
     urlcolor=cyan,
     citecolor=cyan,
 }
\usepackage{epsf}
\usepackage{rotating}

\usepackage{xcolor}

\usepackage{mathabx}

\usepackage[multiple]{footmisc}

\defcitealias{cloutier2019}{C19}
\defcitealias{kostov2019}{K19}
\defcitealias{weiss2018}{W18}
\newcommand{\cloutier}[0]{\citetalias{cloutier2019}}
\newcommand{\kostov}[0]{\citetalias{kostov2019}}
\newcommand{\weiss}[0]{\citetalias{weiss2018}}

\newcommand{\footlabel}[2]{%
    \addtocounter{footnote}{1}%
    \footnotetext[\thefootnote]{%
        \addtocounter{footnote}{-1}%
        \refstepcounter{footnote}\label{#1}%
        #2%
    }%
    $^{\ref{#1}}$%
}

\newcommand{\teff}{\mbox{$T_{\rm eff}$}}
\newcommand{\logg}{\mbox{$\log g$}}
\newcommand{\feh}{\mbox{$[\mathrm{Fe/H}]$}}
\newcommand{\afe}{\mbox{$[\mathrm{\alpha/Fe}]$}}

\newcommand{\Prot}{\mbox{$P_{\mathrm{rot}}$}}

\newcommand{\kms}{\mbox{km\,s$^{-1}$}}
\newcommand{\ms}{\mbox{m\,s$^{-1}$}}
\newcommand{\Ha}{\mbox{$H_\alpha$}}
\newcommand{\Hb}{\mbox{$H_\beta$}}
\newcommand{\Hg}{\mbox{$H_\gamma$}}
\newcommand{\NaD}{\mbox{$NaD$}}
\newcommand{\CaHK}{\mbox{$Ca\,II\,H\,\&\,K$}}

\newcommand{\MSun}{\mbox{$\mathrm{M}_\Sun$}}
\newcommand{\RSun}{\mbox{$\mathrm{R}_\Sun$}}
\newcommand{\MEarth}{\mbox{$\mathrm{M}_\Earth$}}
\newcommand{\REarth}{\mbox{$\mathrm{R}_\Earth$}}

\providecommand{\abs}[1]{\lvert#1\rvert}

\newcommand{\pdf}{PDF}
\newcommand{\prior}{prior}
\newcommand{\post}{posterior}
\newcommand{\bic}{BIC}

\newcommand{\glsp}{GLSP} 
\newcommand{\fap}{FAP}
\newcommand{\wf}{WF}

\newcommand{\tess}{TESS}
\newcommand{\jwst}{JWST}
\newcommand{\espresso}{ESPRESSO}
\newcommand{\harps}{HARPS}
\newcommand{\pfs}{PFS}
\newcommand{\vosa}{VOSA}
\newcommand{\sed}{SED}
\newcommand{\gaia}{Gaia}

\newcommand{\rv}{RV}
\newcommand{\fwhm}{FWHM}
\newcommand{\bis}{BIS}
\newcommand{\berv}{BERV}
\newcommand{\lc}{LC} 
\newcommand{\gp}{GP}
\newcommand{\ccf}{CCF}
\newcommand{\bjd}{BJD}
\newcommand{\snr}{S/N}

\renewcommand{\eqref}[1]{\ref{eq:#1}}
\newcommand{\Eq}[1]{Equation~(\eqref{#1})}

\newcommand{\eqlabel}[1]{\label{eq:#1}}

\newcommand{\sectionname}{Sect.}
\newcommand{\sectref}[1]{\ref{sect:#1}}
\newcommand{\Sect}[1]{\sectionname~\sectref{#1}}
\newcommand{\sect}[1]{\Sect{#1}}
\newcommand{\sectlabel}[1]{\label{sect:#1}}
\newcommand{\sectalt}[1]{\sectref{#1}}
\newcommand{\tabname}{Table}
\newcommand{\tabref}[1]{\ref{tab:#1}}
\newcommand{\Tab}[1]{\tabname~\tabref{#1}}
\newcommand{\tab}[1]{\Tab{#1}}
\newcommand{\tablabel}[1]{\label{tab:#1}}

\newcommand{\figname}{Fig.}
\newcommand{\figref}[1]{\ref{fig:#1}}
\newcommand{\Fig}[1]{\figname~\figref{#1}}
\newcommand{\fig}[1]{\Fig{#1}}
\newcommand{\figlabel}[1]{\label{fig:#1}}

\newcommand{\appname}{Appendix}
\newcommand{\appref}[1]{\ref{app:#1}}
\newcommand{\App}[1]{\appname~\appref{#1}}
\newcommand{\app}[1]{\App{#1}}
\newcommand{\applabel}[1]{\label{app:#1}}

\newcolumntype{P}[1]{>{\centering\arraybackslash}p{#1}}
\newcolumntype{L}[1]{>{\raggedleft\arraybackslash}p{#1}}

\newenvironment{symbolfootnotes}
  {\par\edef\savedfootnotenumber{\number\value{footnote}}
   \renewcommand{\thefootnote}{\fnsymbol{footnote}}
   \setcounter{footnote}{0}}
  {\par\setcounter{footnote}{\savedfootnotenumber}}

\let\ACMmaketitle=\maketitle
\renewcommand{\maketitle}{\begingroup\let\footnote=\thanks \ACMmaketitle\endgroup}

\usepackage{amstext}

\begin{document}

\title{Warm terrestrial planet with half the mass of Venus transiting a nearby star\thanks{Based in part on Guaranteed Time Observations collected at the European Southern Observatory under ESO programme(s) 1102.C-0744, 1102.C-0958, and 1104.C-0350 by the ESPRESSO Consortium.}\thanks{\tab{rvs} is only available in electronic form
at the CDS via anonymous ftp to cdsarc.u-strasbg.fr (130.79.128.5) or via http://cdsweb.u-strasbg.fr/cgi-bin/qcat?J/A+A/.}}

\author{Olivier D. S. Demangeon \inst{\ref{IA-Porto},\ref{UP} \thanks{\email{olivier.demangeon@astro.up.pt}}}
\and M. R. Zapatero Osorio \inst{\ref{CAB}}
\and Y. Alibert \inst{\ref{Bern}}
\and S. C. C. Barros \inst{\ref{IA-Porto},\ref{UP}}
\and V. Adibekyan \inst{\ref{IA-Porto},\ref{UP}}
\and H. M. Tabernero \inst{\ref{CAB},\ref{IA-Porto}}
\and A. Antoniadis-Karnavas \inst{\ref{IA-Porto},\ref{UP}}
\and J. D. Camacho \inst{\ref{IA-Porto},\ref{UP}}
\and A. Su\'arez Mascare\~no \inst{\ref{IAC},\ref{ULL}}
\and M. Oshagh \inst{\ref{IAC},\ref{ULL}}
\and G. Micela \inst{\ref{INAF-Palermo}}
\and S. G. Sousa \inst{\ref{IA-Porto}}
\and C. Lovis \inst{\ref{Geneve-obs}}
\and F. A. Pepe \inst{\ref{Geneve-obs}}
\and R. Rebolo \inst{\ref{IAC},\ref{ULL},\ref{CIC}}
\and S. Cristiani \inst{\ref{INAF-Trieste}}
\and N. C. Santos \inst{\ref{IA-Porto},\ref{UP}}
\and R. Allart \inst{\ref{Montreal},\ref{Geneve-obs}}
\and C. Allende Prieto \inst{\ref{IAC},\ref{ULL}}
\and D. Bossini \inst{\ref{IA-Porto}}
\and F. Bouchy \inst{\ref{Geneve-obs}}
\and A. Cabral \inst{\ref{IA-Lisboa},\ref{FCiencias}}
\and M. Damasso \inst{\ref{INAF-Torino}}
\and P. Di Marcantonio \inst{\ref{INAF-Trieste}}
\and V. D'Odorico \inst{\ref{INAF-Trieste},\ref{IFPU}}
\and D. Ehrenreich \inst{\ref{Geneve-obs}}
\and J. Faria \inst{\ref{IA-Porto},\ref{UP}}
\and P. Figueira \inst{\ref{ESO-Chile},\ref{IA-Porto}}
\and R. G\'enova Santos \inst{\ref{IAC},\ref{ULL}}
\and J. Haldemann \inst{\ref{Bern}}
\and N. Hara \inst{\ref{Geneve-obs}}
\and J. I. Gonz\'alez Hern\'andez \inst{\ref{IAC},\ref{ULL}}
\and B. Lavie \inst{\ref{Geneve-obs}}
\and J. Lillo-Box \inst{\ref{CAB}}
\and G. Lo Curto \inst{\ref{ESO-Germany}}
\and C. J. A. P. Martins \inst{\ref{IA-Porto}}
\and D. M\'egevand \inst{\ref{Geneve-obs}}
\and A. Mehner \inst{\ref{ESO-Chile}}
\and P. Molaro \inst{\ref{INAF-Trieste},\ref{IFPU}}
\and N. J. Nunes \inst{\ref{IA-Lisboa}}
\and E. Pall\'e \inst{\ref{IAC},\ref{ULL}}
\and L. Pasquini \inst{\ref{ESO-Germany}}
\and E. Poretti \inst{\ref{INAF-LaPalma},\ref{INAF-Brera}}
\and A. Sozzetti \inst{\ref{INAF-Torino}}
\and S. Udry \inst{\ref{Geneve-obs}}
}

\institute{
Instituto de Astrof\'{\i}sica e Ci\^encias do Espa\c co, CAUP, Universidade do Porto, Rua das Estrelas, 4150-762, Porto, Portugal
\label{IA-Porto}
\and Departamento de F\'{\i}sica e Astronomia, Faculdade de Ci\^encias, Universidade do Porto, Rua Campo Alegre, 4169-007, Porto, Portugal
\label{UP}
\and Instituto de Astrof\'isica e Ci\^encias do Espa\c{c}o, Faculdade de Ci\^encias da Universidade de Lisboa, Campo Grande, PT1749-016 Lisboa, Portugal
\label{IA-Lisboa}
\and Departamento de F\'isica da Faculdade de Ci\^encias da Universidade de Lisboa, Edif\'icio C8, 1749-016 Lisboa, Portugal
\label{FCiencias}
\and Observatoire de Gen\`eve,  Universit\'e de Gen\`eve, Chemin Pegasi, 51, 1290 Sauverny, Switzerland
\label{Geneve-obs}
\and Physics Institute, University of Bern, Sidlerstrasse 5, 3012 Bern, Switzerland
\label{Bern}
\and Instituto de Astrof\'{\i}sica de Canarias (IAC), Calle V\'{\i}a L\'actea s/n, E-38205 La Laguna, Tenerife, Spain
\label{IAC}
\and Departamento de Astrof\'{\i}sica, Universidad de La Laguna (ULL), E-38206 La Laguna, Tenerife, Spain
\label{ULL}
\and Consejo Superior de Investigaciones Cient\'{\i}cas, Spain
\label{CIC}
\and Centro de Astrobiolog\'\i a (CSIC-INTA), Crta. Ajalvir km 4, E-28850 Torrej\'on de Ardoz, Madrid, Spain
\label{CAB}
\and INAF - Osservatorio Astronomico di Trieste, via G. B. Tiepolo 11, I-34143 Trieste, Italy
\label{INAF-Trieste}
\and INAF - Osservatorio Astrofisico di Torino, via Osservatorio 20, 10025 Pino Torinese, Italy
\label{INAF-Torino}
\and Fundaci\'on G. Galilei -- INAF (Telescopio Nazionale Galileo), Rambla J. A. Fern\'andez P\'erez 7, E-38712 Bre\~na Baja, La Palma, Spain
\label{INAF-LaPalma}
\and INAF - Osservatorio Astronomico di Brera, Via E. Bianchi 46, I-23807 Merate, Italy
\label{INAF-Brera}
\and INAF - Osservatorio Astronomico di Palermo, Piazza del Parlamento 1, I-90134 Palermo, Italy
\label{INAF-Palermo}
\and Institute for Fundamental Physics of the Universe, Via Beirut 2, I-34151 Miramare, Trieste, Italy
\label{IFPU}
\and European Southern Observatory, Alonso de C\'ordova 3107, Vitacura, Regi\'on Metropolitana, Chile
\label{ESO-Chile}
\and European Southern Observatory, Karl-Schwarzschild-Strasse 2, 85748, Garching b. M\"unchen, Germany
\label{ESO-Germany}
\and Department of Physics, and Institute for Research on Exoplanets, Universit\'e de Montr\'eal, Montr\'eal, H3T 1J4, Canada
\label{Montreal}
}

\date{Received date / Accepted date}

\abstract
{
In recent years, the advent of a new generation of radial velocity instruments has allowed us to detect planets with increasingly lower mass and to break the one Earth-mass barrier. Here we report a new milestone in this context by announcing the detection of the lowest-mass planet measured so far using radial velocities: L\,98-59\,b, a rocky planet with half the mass of Venus. It is part of a system composed of three known transiting terrestrial planets (planets b to d). We announce the discovery of a fourth nontransiting planet with a minimum mass of $3.06_{-0.37}^{+0.33}\,\MEarth$ and an orbital period of $12.796_{-0.019}^{+0.020}$\,days and report indications for the presence of a fifth nontransiting terrestrial planet. With a minimum mass of $2.46_{-0.82}^{+0.66}\,\MEarth$ and an orbital period $23.15_{-0.17}^{+0.60}$\,days, this planet, if confirmed, would sit in the middle of the habitable zone of the L\,98-59 system.

L\,98-59 is a bright M dwarf located 10.6 pc away. Positioned at the border of the continuous viewing zone of the James Webb Space Telescope, this system is destined to become a corner stone for comparative exoplanetology of terrestrial planets. The three transiting planets have transmission spectrum metrics ranging from 49 to 255, which undoubtedly makes them prime targets for an atmospheric characterization with the James Webb Space Telescope, the Hubble Space Telescope, Ariel, or ground-based facilities such as NIRPS or ESPRESSO. With an equilibrium temperature ranging from 416 to 627 K, they offer a unique opportunity to study the diversity of warm terrestrial planets without the unknowns associated with different host stars.

L\,98-59\,b and c have densities of $3.6_{-1.5}^{+1.4}$ and $4.57_{-0.85}^{+0.77}$ $\mathrm{g.cm}^{-3}$ , respectively, and have very similar bulk compositions with a small iron core that represents only 12 to 14\,\% of the total mass, and a small amount of water. However, with a density of $2.95_{-0.51}^{+0.79}$ $\mathrm{g.cm}^{-3}$ and despite a similar core mass fraction, up to 30\,\% of the mass of \object{L 98-59 d} might be water.
}

\keywords{
Planetary systems
--
Stars: individual: L\,98-59
--
Techniques: radial velocities, high precision photometry
}

\titlerunning{L\,98-59}
\authorrunning{Demangeon et al.}

\maketitle

\section{Introduction}\sectlabel{intro}

In the past years, radial velocity (\rv) instruments such as HARPS\footnote{HARPS stands for High Accuracy Radial velocity Planet Searcher and HARPS-N for  High Accuracy Radial velocity Planet Searcher North. CARMENES stands for Calar Alto high-Resolution search for M dwarfs with Exoearths with Near-infrared and optical \'Echelle Spectrographs. ESPRESSO stands for \'Echelle SPectrograph for Rocky Exoplanets and Stable Spectroscopic Observations. \tess stands for Transiting Exoplanet Survey Satellite and TRAPPIST for TRAnsiting Planets and PlanetesImals Small Telescope.} \citep{mayor2003}, HARPS-N\footnotemark[1] \citep{cosentino2012}, and more recently, CARMENES\footnotemark[1] \citep{quirrenbach2014} and ESPRESSO\footnotemark[1] \citep{pepe2021}, have demonstrated that it is now possible to detect planets with masses similar to the mass of the Earth using \rv s \citep[e.g.,][]{astudillo-defru2017a,rice2019,zechmeister2019,suarezmascareno2020}. These results represent an important achievement in the quest for life outside the Solar System. However, it is important to keep pushing toward lower masses and longer periods to ensure that we remain capable to measure the mass of a transiting Earth analog in the habitable zone of a bright host star.

The detection of biosignatures on an exoplanet depends on our capability of studying its atmosphere; this currently relies on transit spectroscopy \citep[e.g.,][]{kaltenegger2017}. Space-based transit surveys such as Kepler/K2 \citep{borucki2010,howell2014} and \tess\footnotemark[1] \citep{ricker2015} and even ground-based surveys such as TRAPPIST\footnotemark[1] \citep{gillon2011} have revealed hundreds of transiting terrestrial planets \citep[e.g.,][]{batalha2013}. However, the community still has to detect and study the atmosphere of one of them \citep{kreidberg2019}. 
A large fraction of the known terrestrial planets are part of multiplanetary systems \citep{lissauer2011a}. Multiplanetary systems are laboratories for a variety of studies: Planet-planet interactions \citep[e.g.,][]{barros2015}, planetary formation and migration \citep[e.g.,][]{rein2012,albrecht2013,delisle2017}, and/or comparative planetology \citep[e.g.,][]{mandt2015, millholland2017a}. The discovery and accurate characterization of a system with multiple transiting terrestrial planets amenable to transit spectroscopy would thus represent a crucial milestone.

The \object{L 98-59} system, also known as the \tess\ Object of Interest 175 (TOI-175) system, is a multiplanetary system announced by \citet[][hereafter \kostov]{kostov2019} as composed of three transiting exoplanets with radii ranging from 0.8 to 1.6 Earth radii (\REarth).
The host star is a bright (magK = 7.1, \citealt{cutri2003}, magV=11.7, \citealt{zacharias2012}) nearby \citep[10.6194 pc,][]{gaiacollaboration2018b, bailer-jones2018} M-dwarf star \citep{gaidos2014}. One interesting particularity of this
system is its location, with a right ascension (\textsc{ra}) of 08:18:07.62 and declination (\textsc{dec}) of -68:18:46.80, at the border of the continuous viewing zone ($\sim 200$ days per year) of the James Webb Space Telescope \citep[\jwst,][]{gardner2006}. This system is thus a prime target for a comparative study of rocky planet atmospheres within the same system \citep{greene2016, morley2017}.

The \harps\ spectrograph \citep{mayor2003} was used to carry out an \rv\ campaign to measure the masses of these three planets \citep[][hereafter \cloutier]{cloutier2019}. The masses of the two outer planets were constrained to $2.36 \pm 0.36$ and $2.24 \pm 0.53$ Earth masses (\MEarth), leading to bulk densities of $5.3 \pm 1.2$ and $3.2 \pm 1.2\,\mathrm{g.cm}^{-3}$ for planet c and d, respectively. \citetalias{cloutier2019} were unable to constrain the mass of the inner planet b and delivered an upper limit of $1.01\, \MEarth$ (with a 95\% confidence level).
The \pfs\ spectrograph \citep{crane2006,crane2008,crane2010} was also used to attempt measuring the mass of the three planets. With only 14 \pfs\ measurements, \citet{teske2020} derived masses of $1.32 \pm 0.73$, $1.24 \pm 0.95,$ and $2.11 \pm 0.72\, \MEarth$ for planets b, c, and d, respectively. These mass estimates are less precise, but roughly compatible with those of \cloutier . Due to the low number and the lower precision of the \pfs\ data, we did not include these measurements in our analysis.

We report here the results of a follow-up \rv\ campaign with ESPRESSO \citep{pepe2021} aimed at refining the mass of the planets in the L 98-59 system.
In \sect{datasets} we present the \rv\ and photometric data sets. In \sect{star} we characterize the host star. We describe our analysis of the data sets in \sect{plchar}. Finally, in Sects. \sectref{disc} and \sectref{conc} we discuss the particularities and the importance of this system.

\section{Datasets}\sectlabel{datasets}

Our analysis of the L\,98-59 system relies on \rv\ and photometric time series. The \rv s were obtained with the HARPS (\sect{harpsdata}) and ESPRESSO (\sect{espdata}) instruments. The light curve (\lc) was acquired by \tess (\sect{tessdata}) space telescope.

\subsection{High-resolution spectroscopy}

\subsubsection{HARPS}\sectlabel{harpsdata}

\citetalias{cloutier2019} obtained 165 spectra with \harps, which is\ installed at the 3.6 telescope of the \textsc{eso} La Silla Observatory (programs 198.C-0838, 1102.C-0339, and 0102.C-0525) between October 17, 2018 (barycentric Julian date, \bjd\ = 2458408.5), and April 28, 2019 (\bjd\ = 2458601.5).
\harps\ is a fiber-fed cross-dispersed echelle spectrograph operating in a temperature- and pressure-regulated vacuum chamber. It covers wavelengths from 380 to 690 nm with an average spectral resolution of $R = 115\,000$.
We obtained the \rv s from \citetalias{cloutier2019} and refer for details of the observations and their processing to this publication. However, we caution that in order to reproduce the results presented by \cloutier , in particular the \rv\ time series and its generalized Lomb-Scargle periodogram \citep[\glsp,][]{zechmeister2009}, we had to exclude four measurements obtained at 2\,458\,503.795048, 2\,458\,509.552019, 2\,458\,511.568314, and 2\,458\,512.581045 \bjd . We identified these measurements with a $4\,\sigma$ iterative sigma clipping. These measurements were excluded from all the analyses in this paper. All measurements were obtained with fiber B pointed at the sky (no simultaneous observation of a calibration source). One hundred and forty measurements were obtained with an exposure time of 900 s, resulting in an average signal-to-noise ratio (\snr) of 41 per resolution element at 650 nm. For the remaining 21 measurements, the exposure time varied from 500 to 1800 s, resulting in a median \snr\ of 49. The \rv s were extracted from the spectra through template matching \citep{astudillo-defru2017}. Their median precision ($1\,\sigma$ uncertainty) is $2.08\,\ms$.

In addition to the \rv\ measurements, C19 provided the measurement of several stellar activity indicators: the full width at half maximum (\fwhm) of the cross-correlation function (\ccf), the bisector span of the \ccf\ (\bis), the depth of the \Ha, \Hb, \Hg\ lines, the depth of the sodium doublet \NaD,\ and the S-index based on the depth of the \CaHK\ doublet. All these indicators are sensitive to chromospheric or photospheric activity.

\subsubsection{ESPRESSO}\sectlabel{espdata}

We obtained 66 spectra with \espresso, which is\ installed at the \textsc{VLT} telescopes of the \textsc{eso} Paranal Observatory between November 14, 2018 (\bjd\ = 2458436.5), and March 4, 2020 (\bjd\ = 2458912.5), as part of the \espresso\ Guaranteed Time Observation (programs 1102.C-0744, 1102.C-0958, and 1104.C-0350).
\espresso\ \citep{pepe2021} is also a fibre-fed high-resolution echelle spectrograph operating in a temperature- and pressure-regulated vacuum chamber. It covers wavelengths from 380 to 788 nm with an average spectral resolution of $R = 140\,000$ in its single UT high-resolution mode (HR21, slow-readout mode) that was used for these observations. All measurements were obtained with the sky on fiber B. All measurements were obtained with a 900 s exposure time, resulting in an average \snr\ of 70 per resolution element at 650 nm. The \rv s were extracted from the spectra using version 2.2.1 of the \espresso\ pipeline Data-Reduction Software (\textsc{drs})\footnote{A detailed description of the \espresso\ \textsc{drs} can be found in the \espresso\ pipeline user manual available at \href{https://www.eso.org/sci/software/pipelines/espresso/espresso-pipe-recipes.html}{espresso-pipe-recipes}
}.
It computes the \ccf\ of the sky-subtracted spectra with a stellar line mask to estimate the \rv\ \citep{baranne1996}. In this case, the mask was optimized for stars of spectral type M2 V.
The \ccf\ was then fit with an inverted Gaussian model. The parameters of the profile are the continuum level; the center of the Gaussian profile, which provides the measurement of the \rv ; and its \fwhm.
Finally, the amplitude provides a measure of the contrast of the \ccf. The uncertainties on the measured \rv s are computed using the algorithms described in \citet[][]{bouchy2001} and reflect the photo-noise-limited precision. The uncertainties on the \fwhm\ are estimated as the double of \rv\ uncertainties.
In addition to the \rv, \fwhm,\ and contrast measurements, we computed several activity indicators: the \bis\ \citep{queloz2001}
, the depth of the $H_{\alpha}$ line, the sodium doublet \citep[\NaD, ][]{diaz2007}, and the S-index \citep{lovis2011, noyes1984}.

From the 66 measurements, we discarded three measurements, obtained at 2\,458\,645.496, 2\,458\,924.639, and 2\,458\,924.645 BJD$_\textrm{TDB}$, due to their high \rv\ uncertainties (identified through an iterative $4\,\sigma$ clipping). An inspection of the night reports indicates that these measurements were obtained under poor observing conditions: Strong wind, poor seeing, 
and cirri and bright moon for the first measurement. The last measurement was even interrupted by high winds.
The median precision ($1\,\sigma$ uncertainty) obtained on the \espresso\ \rv s is $0.8\,\ms$ (a factor 2.6 better than the \harps\ \rv s).
At about the middle of our \rv\ campaign, in June 2019, the fiber-link of \espresso\ was replaced. This resulted in an increased throughput, but required us to consider an \rv\ offset between the data taken before and after this intervention \citep{pepe2021}.

\subsection{High-precision photometry with \tess}\sectlabel{tessdata}

L 98-59 (TIC 307210830, TOI-175) was observed by \tess\ in short cadence (2 min) in 9 sectors (2, 5, 8, 9, 10, 11, 12, 28, and 29) with cameras 4 and 3. These observations correspond to $\sim$ 243 days of noncontinous observations taken between August 22, 2018 (\bjd\ = 2458352.5), and September 22, 2020 (\bjd\ = 2459114.5). We downloaded the \lc s from the Mikulski Archive for Space Telescopes (MAST) using the python package \texttt{astroquery}.
The \lc\ data products provided by the \tess\ pipeline \citep{jenkins2016} provide two \lc s, the simple aperture photometry \textsc{sap} \lc\ and the pre-search data-conditioned simple aperture photometry \textsc{pdcsap} \lc\ \citep{smith2012, stumpe2014}. In contrast to the \textsc{sap} \lc, the \textsc{pdcsap} \lc\ is detrended using common basis vectors computed over all stars observed on the same \textsc{ccd}. For our analyses, we exclusively used the \textsc{pdcsap} \lc.
From the \lc, we removed the data points whose quality flags where showing the bits 1, 2, 3, 4, 5, 6, 8, 10, and 12 following the example provided by the \tess\ team.
Following a procedure inspired by \citetalias{kostov2019}, we detrended the \lc\ from the residual stellar activity signal and instrument noise using a Gaussian process (\gp).
We masked all the transits of the three planets using the ephemerides and transit durations provided by \citetalias{kostov2019} and fit the resulting \lc\ with a \gp\ model using the \texttt{celerite} Python package \citep{foreman-mackey2017, foreman-mackey2018} and a mean shift between sectors. The functional form of the kernel we used was the one of a damped harmonic oscillator chosen for its flexibility and smooth variations, allowing us to model the unknown mixture of stellar activity and residual instrumental noise. Its equation is
\begin{equation}
    S(\omega) = \sqrt{\frac{2}{\pi}} \frac{S_0 \, \omega_{0}^{4}}{(\omega^{2} - \omega_{0}^{2})^2 + \omega^{2} \, \omega_{0}^{2} / Q^2},
\end{equation}
where Q, the quality factor, was fixed to $\frac{1}{\sqrt{2}}$, $S_0$ is the amplitude, and $\omega_0$ is the angular frequency corresponding to the break point in the power spectral density of this kernel.

The fit was performed using an affine-invariant ensemble sampler for \textsc{mcmc} \citep{goodman2010} implemented in the Python package \texttt{emcee} \citep{emcee}, which samples the \post\ probability density function. We used a multidimensional Gaussian distribution for the likelihood. For the \prior s, we used a uniform \prior\ between -20 and 15 on $\ln S_0$ , and we obtained a \post\ providing an estimate of $S_0 = 82.38^{+6.59}_{-5.77}$ ppm, using the median and the 68\,\% confidence interval. For $\omega_0$, we used a uniform \prior\ between -20 and 15 on $\ln \omega_0$ , and we obtained an estimate of $\ln \omega_0 = 1.17^{+0.08}_{-0.09}$ (in $\ln \textrm{day}^{-1}$). We did not attribute \prior s to the offset between sectors, and the retrieved values are compatible with the values provided in \tab{tessoffsets}. We used 32 walkers (for 11 free parameters) and performed a first run of 500 iterations as burn-in. The initial positions for this first run were drawn from the \prior\ for $S_0$ and $\omega_0$ and set to 0 for the offset between sectors. After this first run, we reset the \texttt{emcee} sampler and performed a second run of 2000 iterations, which started from the last positions of the previous run. After this second run, we examined the histogram of the acceptance fraction of the chains to identify chains that had significantly lower acceptance fractions than the others. A lower acceptance fraction implies a stronger correlation between consecutive iterations, which will increase the sampling error of the \post\ \pdf\ inferred from the histograms of the chains. We also examined the histogram of the logarithm of the \post\ probability of the chain (estimated by the average of this value computed over the last 1\,\% of the iterations of each chains). The objective was to understand whether all the chains have converged toward regions of the parameter space that have similar \post\ probability density values. In this case, both histograms are mono-modal, indicating that all chains have similar acceptance fractions and sample regions of the parameter space with similar \post\ probability density values.
We confirmed that all the chains converged and converged to the same region of the parameter space using the Geweke criterion \citep{geweke1992}. All the chains indeed converged to the same region of the parameter space after the first 750 iterations of the second run. We further confirmed that the remaining parts of the chains converged and were long enough by computing the integrated autocorrelation time using the method implemented in \texttt{emcee} and verifying that it was ten times shorter than the remaining number of iterations.

We normalized the \lc\ by dividing it by the best \gp\ model, whose parameters values are the median values of the converged \textsc{mcmc} chains. Finally, we cut the \lc\ to keep only data points within 1.5 transit durations on either side of each mid-transit time. This reduced the number of data points and the computation time.

\section{Characterization of the M dwarf L\,98-59\,A}\sectlabel{star}

According to \kostov, L 98-59 A is an M3V star. The derivation of accurate stellar properties through high-resolution spectroscopy for M stars is complicated because blended lines prevail.
We thus used several approaches to characterize L 98-59 A in order to assess and discuss the homogeneity and the accuracy of the outcomes. This analysis is presented in detail in \app{stelchar}, and we summarize the results in this section.

\subsection{Stellar atmospheric parameters}\sectlabel{spectro}

To derive the stellar parameters, effective temperature (\teff), surface gravity (\logg), and metallicity (\feh), we chose to fit the combined spectrum of L\,98-59\,A constructed using 61 \espresso\ spectra ($\snr = 1063$ at 7580 \AA) with the latest version of the spectral synthesis code \textsc{SteParSyn} \citep[][see \app{specsyn} for more details]{tab18,tab20b}. We adopted the estimates provided by \textsc{SteParSyn} except for the uncertainty on \teff\ , which we identified as underestimated (see \app{atmostelchar}). We enlarged this uncertainty to encompass the best values provided by the other methods within $1\,\sigma$. The set of adopted estimates is provided in \tab{syspar}.

\subsection{Stellar modeling: Mass, radius, and age}\sectlabel{starmod}

Thanks to the high precision and accuracy of \gaia\ parallactic distances \citep[$10.6194 \pm 0.0032$ pc inferred from the \gaia-\textsc{dr2} parallax by][]{bailer-jones2018} and the well-sampled photometric spectral energy distribution (\sed, see \app{vosa}), we can derive a reliable estimate of the absolute bolometric luminosity of L 98-59: $0.01128 \pm 0.00042\, \mathrm{L_\Sun}$. Added to our estimate of \teff\ (see \sect{spectro}, \app{atmostelchar} and \tab{syspar}), we infer the radius of L 98-59 A to be $0.303^{+0.026}_{-0.023}\,\RSun$ using the Stefan-Boltzmann law. This agrees well (better than $1\,\sigma$) with the literature value derived by \kostov\ from the mass-radius relations for M and K dwarfs of \citet{boyajian2012}.
We derived the mass of L\,98-59\,A using the Virtual Observatory \sed\ Analyzer online tools\footnote{\textsc{vosa} is publicly available online \href{http://svo2.cab.inta-csic.es/theory/vosa/}{http://svo2.cab.inta-csic.es/theory/vosa/}}  \citep[\vosa , ][see \app{starmod} for more details]{bayo2008}. \vosa\ derives the mass by comparing the measured \teff\ and bolometric luminosity to BT-Settl evolutionary tracks \citep{allard2012}.
Finally, we determined the age of L\,98-59\,A using the photometry and distance provided by Gaia (see \app{starmod} for more details). We compared the location of L\,98-59\,A in the color-magnitude diagram (see \fig{colormag}) to mean sequences of stellar members of the $\beta$ Pictoris moving group ($\sim$20 Myr, \citealt{miret-roig2020}), the Tucana-Horologium moving group ($\sim$45 Myr, \citealt{bell2015}), the Pleiades open cluster ($\sim$120 Myr, \citealt{gossage2018}), and the field (possible ages in the range 0.8--10 Gyr). This comparison allowed us to infer that L\,98–59 has an age consistent with that of the field. This age estimate is confirmed by our kinematics analysis, which indicates that L\,98–59\,A is a thin-disk star that does not belong to any known young moving group (see \app{starmod} and \tab{kinematics}).
The adopted radius, mass, and ages of L\,98-59\,A are provided in \tab{syspar}.

\begin{figure}[!htb]
    \resizebox{\hsize}{!}{\includegraphics[]{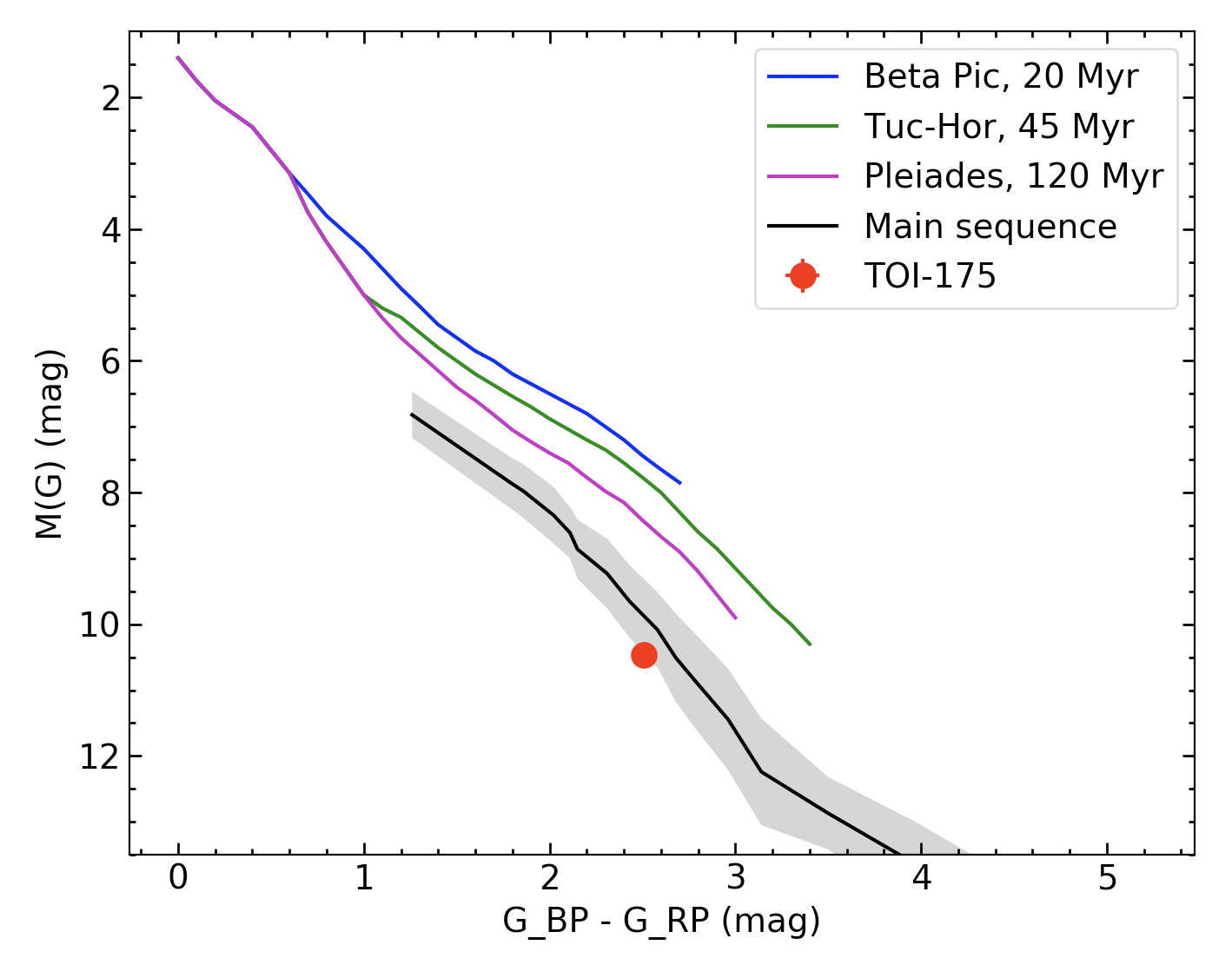}}
    \caption{\figlabel{colormag} Absolute magnitude (in the G Gaia bandpass) vs. color (magnitude difference between the Gaia bands $G_{BP}$ and $G_{RP}$): L\,98–59\,A  (TOI-175) is located in the Gaia color-magnitude diagram together with the mean sequences of young clusters and moving groups \citep{luhman2018} and the main sequence of stars \citep{cifuentes2020}. The error bars of L\,98–59\,A are smaller than the symbol size. The gray area represents the $1\,\sigma$ dispersion of field M dwarfs.
}
\end{figure}

\subsection{Stellar Mg and Si abundances}\sectlabel{abundances}

Stellar abundances of Mg and Si are valuable constraints for modeling the interior of  planets (see \sect{intcomp}). 
However, deriving individual abundances of M dwarfs from visible spectra is a very difficult task \citep[e.g.,][]{Maldonado-20}.
We estimated the abundances of Mg and Si following the procedure described in \citet{Adibekyan-17}. From the APOGEE DR16 \citep{Jonsson-20}, we selected cool stars ($T_{\mathrm{eff}} <$ 5500 K, the choice of this temperature limit does not have a significant impact) with metallicities similar to that of L\,98-59\,A within 0.05 dex. We considered only stars with the highest S/N (> 500) spectra to guarantee the high quality of the extracted parameters and abundances of these stars. Because L\,98-59\,A is a member of the Galactic thin-disk population (see \tab{kinematics}), only stars belonging to the thin-disk population were selected. The selection of the thin-disk stars was based on the [Mg/Fe] abundance of the APOGEE stars \citep[see, e.g.,][]{Adibekyan-12}. With these constraints, we obtained a sample of about 1000 thin-disk stars with properties similar to those of our target. The mean abundances of Mg and Si of these stellar analogs were adopted as proxy for the empirical abundances, and their standard deviation (star-to-star scatter) was adopted as the uncertainty (see \tab{syspar}).

\subsection{Stellar rotation and activity periods}\sectlabel{stelact}

As mentioned in Sects. \sectref{harpsdata} and \sectalt{espdata}, the \harps\ and \espresso\ instruments give access to the time series of several activity indicators. These activity indicators are sensitive to variations in the stellar chromosphere, but not to the presence of planets in the system. They are therefore ideal for identifying periodicities that arise from stellar chromospheric activity. To identify these periods, we computed the \glsp\ of all available activity indicators, see \fig{actindgls}. This figure also includes the \glsp\ of the \rv\ measurements.

The \glsp s of the \espresso\ activity indicators suggest that the rotation period ($P_{\mathrm{rot}}$) of L 98-59 A is $80.9^{+5.0}_{-5.3}$ days, measured on the highest peak of the \fwhm\ \glsp , in agreement with \cloutier. The \glsp s of the \fwhm, the contrast of the \ccf,\ and the S-index all show peaks at this period with a false-alarm probability (\fap) below $0.1\,\%$. The \fap\ levels were computed using the analytical relation described in \citet{zechmeister2009} for the Zechmeister-K\u rster (ZK) normalization. Our \glsp s of the \harps\ activity indicators are consistent with those presented by \cloutier. The \glsp s of the \bis, the S-index, and \Ha\ show peaks with an \fap\ below $0.1\,\%$, but not at the same period. However, as noted by \cloutier, the peak with the highest significance, which is found in the \glsp\ of \Ha, is close to 80 days. This period is used by \cloutier\ as an estimate of the rotation period.

\begin{figure*}[!htb]
    \centering
    \subfloat[\espresso]{\includegraphics[valign=t]{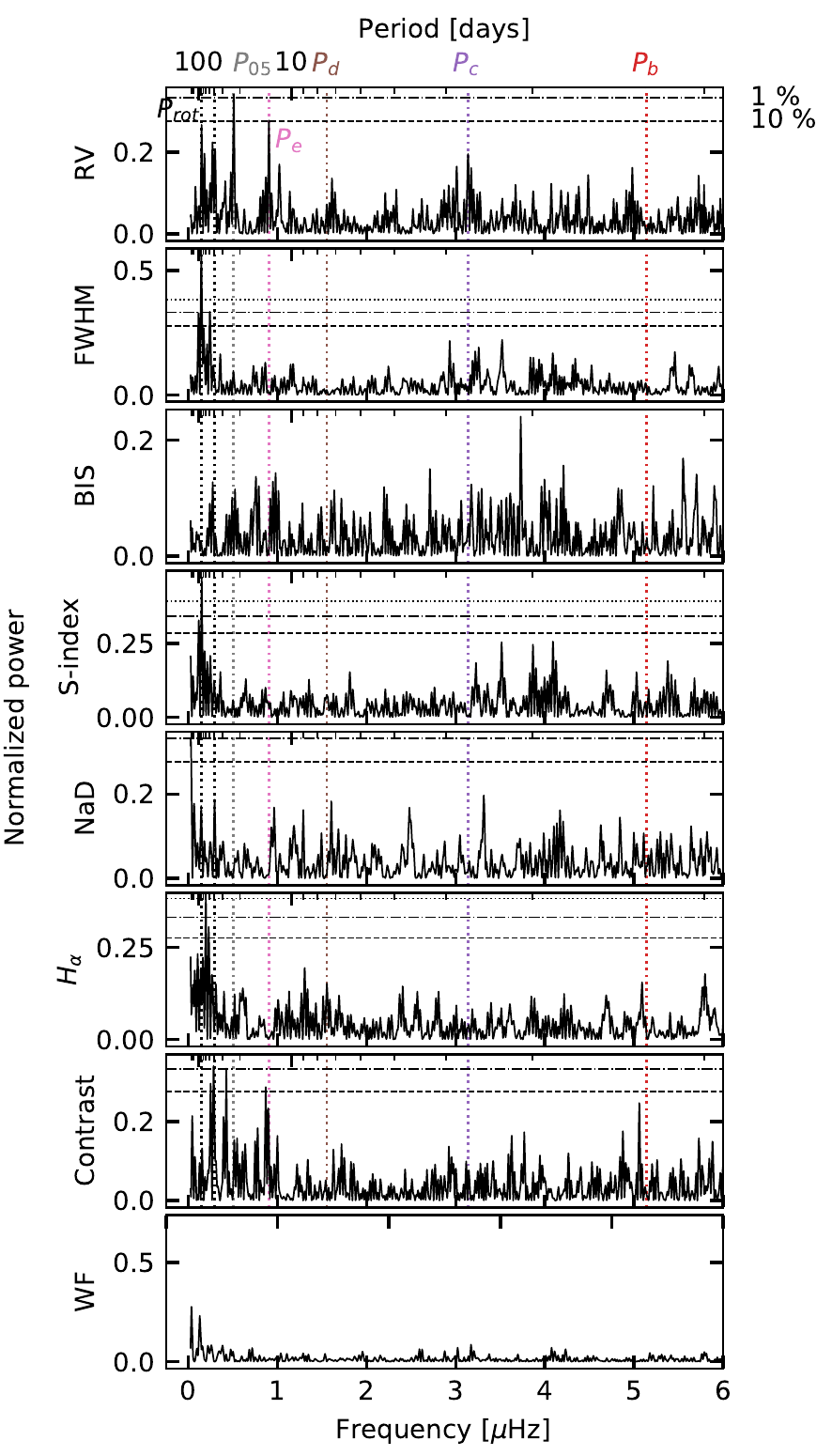}}
    \quad
    \subfloat[\harps]{\includegraphics[valign=t]{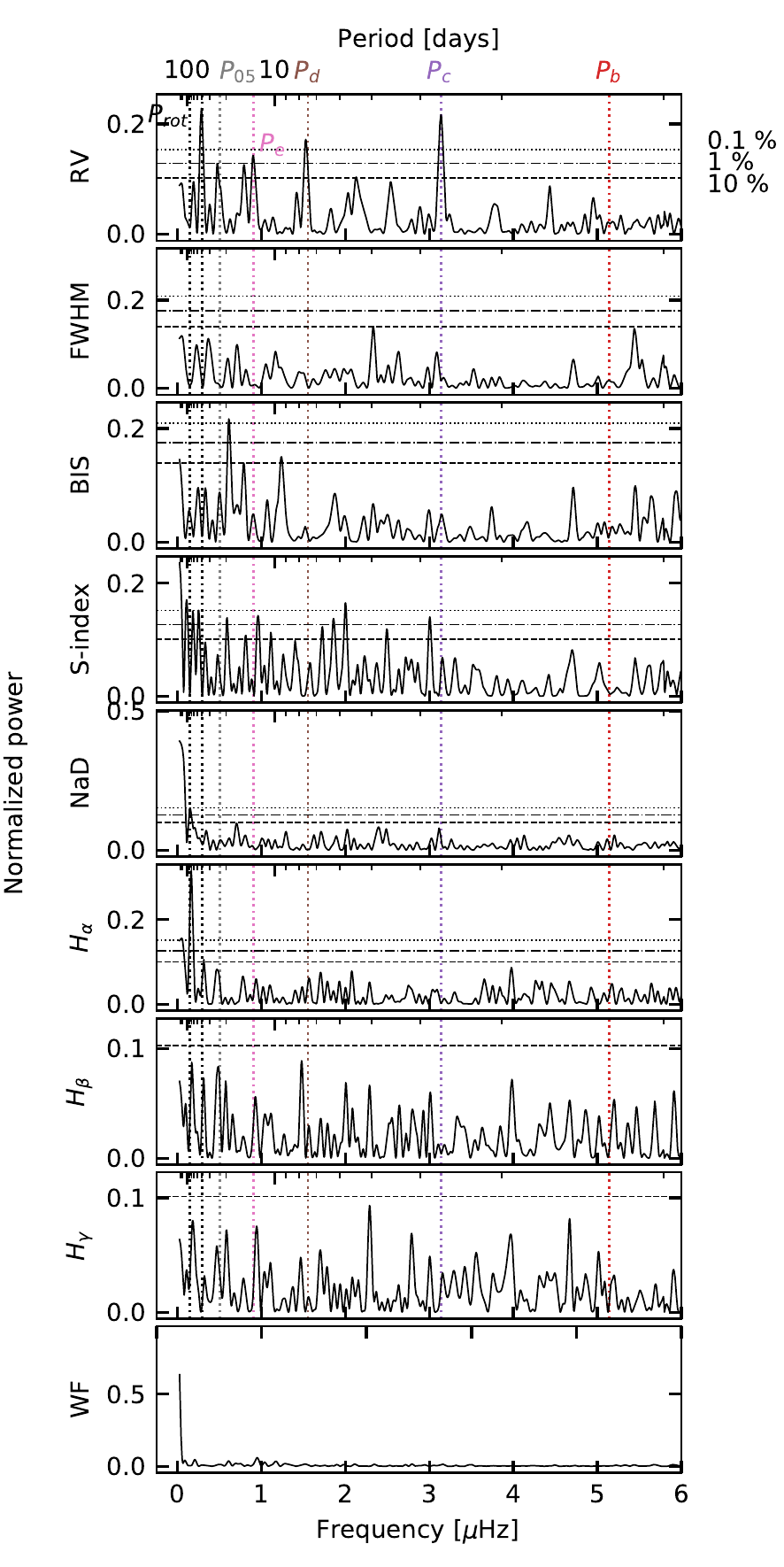}} 
    \caption{\figlabel{actindgls}. \glsp\ of the \rv\ and activity indicators from \espresso\ (a) and \harps\ (b) data. The last row for both instruments presents the window function. The vertical dotted lines indicate from right to left the orbital period of the planets b, c, d, and e, the planetary candidate 05, and half and the full stellar rotation period (assumed here to be 80 days). The horizontal lines indicate the amplitude levels corresponding to 10\%\ (dashed line), 1\%\ (dot-dashed line) and 0.1\,\% (dotted line) of the \fap. The amplitudes of the \glsp s are expressed using the ZK normalization described in \citet[][eq. 5]{zechmeister2009}. The \fap\ levels are computed using the analytical relation also described in \citet{zechmeister2009} for this normalization. 
    We display the \glsp\ of the \bis\ for completness and comparison with \cloutier, but we caution that the reliability of \bis\ measurements from \ccf s for M dwarfs is uncertain \citep{rainer2020}.
    }
\end{figure*}

Photometric time series can also provide insight into the stellar rotation periods. The appearance and disappearance of dark and bright active regions due to stellar rotation, such as spots and plages, produce a modulation in the \lc . To investigate the rotational modulation in the \tess\ \lc, we first fit the \textsc{pdcsap} \tess\ \lc\ with a \gp\ and an offset for each sector. Using the retrieved offsets between the sectors, we computed the \glsp\ of the \tess\ \lc\ presented in \fig{glsptess} (see also \app{protfromphotometry}). The three highest peaks in this periodogram in order of decreasing amplitude are at 93, 115, and 79 days. The 79-day periodicity is a confirmation of the 80-day period identified in the \glsp s of the spectroscopic time series presented in \fig{actindgls}. However, the 93- and 115-day periodicities are absent in these periodograms. 

Overall, the spectroscopic and photometric time series all exhibit power at a period of 80\,days. This is thus our best guess for the rotation period of L\,98-59. However, the power spectrum of all these stellar activity indicators depicts a complex activity pattern that does not seem to be fully described by only one periodicity and its harmonics.

\begin{figure}[!htb]
    \resizebox{\hsize}{!}{\includegraphics[]{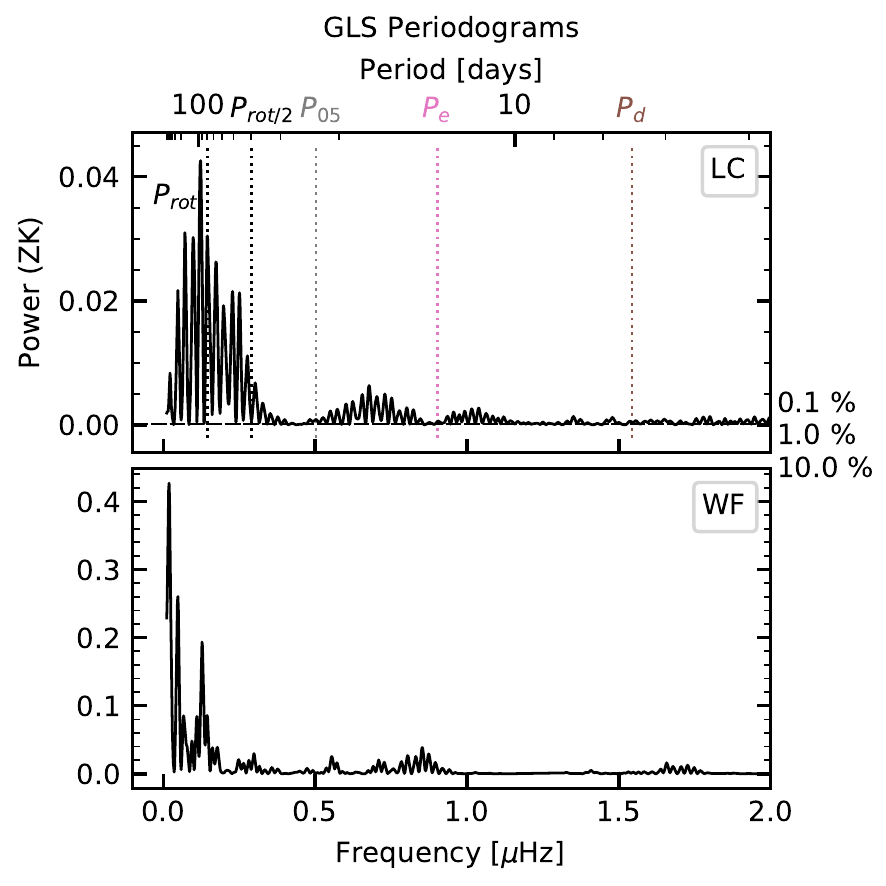}}
    \caption{\figlabel{glsptess}. \glsp\ of the \tess\ \lc . The format of the this figure is identical to that in \fig{actindgls}. In particular, the power of the \glsp\ is normalized using the ZK normalization. The highest peak in this periodogram is for a period of 93 days.}
\end{figure}

\section{Radial velocity and light-curve modeling}\sectlabel{plchar}

\subsection{Search for additional planets in the L 98-59 system}\sectlabel{firstana}

\kostov\ and \cloutier\ confirmed the presence of three transiting planets in the L 98-59 system. Using the new sectors from \tess\ and the new \rv\ data from \espresso, we wish to improve the precision of the planetary parameters and search for additional planets.

The \glsp\ of the \harps\ \rv\ data (see \fig{actindgls}-b) shows six peaks above an \fap\ of 10\% at about 3.7 (orbital period of planet c), 7.6 (orbital period of planet d), 13, 15, 23, and 40 ($\sim \Prot / 2$) days. The \glsp\ of the \espresso\ \rv\ data (see \fig{actindgls}-a) shows two narrow peaks above an \fap\ of 10\% at about 13 and 23 days. The fact that the two peaks identified in the \espresso\ data are also present in the \harps\ data and are not an obvious fraction of the stellar rotation period indicates that there might be two additional planets in the system.

Due to the high computational cost linked to the analysis of the nine \tess\ sectors, we divided our analysis into three steps. In the first step (\sect{lconly}), we analyze the \tess\ \lc\ alone in order to refine the properties of the three known transiting planets and in particular their ephemerides. In the second step (\sect{rvonly}), we use these ephemerides as \prior\ for the analysis of the high-resolution spectroscopy data. The main objective of this second step is to assess the presence of additional planets in system L\,98-59 (\sect{modselect}). Finally, in a third step (\sect{finalana}), we perform a final joint analysis of the \rv s and the \lc\ to obtain the final parameters of the system.  

\subsubsection{\lc\ analysis}\sectlabel{lconly}

To model the planetary transits, we used a modified version\footnote{The modified version of \texttt{batman} is available at \url{https://github.com/odemangeon/batman}. It prevents the code to stay trapped in an infinite loop for highly eccentric orbits.}
of the Python package \texttt{batman}\footlabel{opensourcesoftwares}{Several of the Python packages used for this work are publicly available on Github: \texttt{radvel} at \url{https://github.com/California-Planet-Search/radvel}, \texttt{george} at \url{https://github.com/dfm/george}, \texttt{batman} at \url{https://github.com/lkreidberg/batman}, \texttt{emcee} at \url{https://github.com/dfm/emcee}, \texttt{ldtk} at \url{https://github.com/hpparvi/ldtk}.}
\citep{kreidberg2015}. The parameters used for each planets are the orbital period $P$, the time of inferior conjunction ($t_{\mathrm{ic}}$), the products of the planetary eccentricity by the cosine and sine of the stellar argument of periastron ($e\cos\omega$ and $e\sin\omega$), the ratio of the planetary radius to that of the star (${R_p / R_*}$), and the cosine of the planetary orbital inclination ($\cos i_{p}$). The model also included the stellar density ($\rho_*$). For the limb-darkening law, we used the four coefficients of the nonlinear model ($u_{1,\tess}$, $u_{2,\tess}$, $u_{3,\tess}$ , and $u_{4,\tess}$).
To this set of parameters, we added one additive jitter term ($\sigma_{\textrm{\tess}}$) for the photometry in all \tess\ sectors to account for a possible underestimation of the error bars \citep[][]{baluev2009}.

To infer the values of these parameters, we maximized the \post\ probability density function (\pdf) of the model as prescribed by the Bayesian inference framework \citep[e.g.,][]{gregory2005a}. The likelihood functions we used were multidimensional Gaussians.
To obtain robust error bars, we explored the parameter space with an affine-invariant ensemble sampler for \textsc{mcmc} implemented in the Python package \texttt{emcee} \citep{emcee}.
We adapted the number of walkers to the number of free parameters in our model.
As a compromise between speed and efficiency, we used $\lceil \textrm{n}_{\textrm{free}} \times 2.5  \times 2\rceil / 2$  walkers, where $\textrm{n}_{\textrm{free}}$ is the number of free parameters and $\lceil~\rceil$ is the ceiling function.
This allowed us to have an even number of walkers that was at least twice ($\sim 2.5$ times) the number of free parameters, as suggested by the authors of \texttt{emcee}.
The initial values of each walker were obtained from the output of a maximization of the \post\ \pdf\ made with the Nelder-Mead simplex algorithm \citep{NelderMead} implemented in the Python package \texttt{scipy.optimize}. The initial values for the Nelder-Mead simplex maximization were drawn from the \prior s of the parameters. The objective of this pre-maximization was to start the \texttt{emcee} exploration closer to the best region of the parameter space and thus reduce its convergence period. Our experience is that this usually results in a reduction of the overall computational time because the Nelder-Mead simplex algorithm usually converges faster than \texttt{emcee}.

The \prior\ \pdf\ assumed for the parameters were noninformative and are given in \tab{syspar} (column \prior), along with references justifying their use when needed (column Source \prior ). Along with the \post\ \pdf\ provided in the same table, this allowed a qualitative assessment of the impact of the \prior\ on the \post\ (inferred values). A detailed description of the reasons for the choice of each \prior\ is given in \app{priors}.

To choose the initial values for the analysis, those used to start the pre-minimization, we usually use values drawn from the \prior s. However, here, we did not analyze the full \tess\ \lc, only small portions of it around the location of the transits (see \sect{tessdata}). Consequently, drawing initial values from noninformative \prior s would very likely cause the simulated transits to fall outside of the selected portions of the \lc\ and make the optimization impossible. To prevent this, we drew the initial values for $P$, $t_{\mathrm{ic}}$, ${R_p / R_*}$ , and $\cos i_{p}$ from the \post\ \pdf s obtained by \kostov.

We used 50,000 \textsc{mcmc} iterations and analyzed the chains using the same procedure as is described in \sect{tessdata}.
The \post\ distributions of the parameters of the three transiting planets were then used as \prior s for the analysis of the \rv s.

\subsubsection{\rv\ analysis}\sectlabel{rvonly}

Our model of the \rv s is composed of three main components: The planetary model, the stellar activity model, and the instrumental model. 
In their analysis of the HARPS\ data, \cloutier \  demonstrated the importance of stellar activity mitigation for this system. They inferred an amplitude of $\sim 7\,\ms$ for the stellar activity signal compared to $\lesssim 2\,\ms$ for the semi-amplitude of the three planetary Keplerians.
We thus paid particular care to the stellar activity mitigation and used two different approaches. The first approach is similar to the one used by \cloutier. We fitted the \rv\ data using Keplerians for the planetary signals and a \gp\ with a quasi-periodic kernel for the stellar activity. The mathematical expression of the kernel of this \gp\ is
\begin{equation}
    K_{\rv}(t_{i}, t_{j}) = {A_{\rv}}^2 \exp\left[-\frac{(t_{i} - t_{j})^2}{2 {\tau_{\mathrm{decay}}^2}} -\frac{\sin^2\left(\frac{\pi}{P_{\mathrm{rot}}} \abs{t_i - t_j}\right)}{2 \gamma^2} \right]
,\end{equation}
where $A_{\mathrm{rv}}$ is the amplitude of the covariance, $\tau_{\mathrm{decay}}$ is the decay timescale, $P_\mathrm{rot}$ is the period of recurrence of the covariance, and $\gamma$ is the periodic coherence scale \citep[e.g.,][]{grunblatt2015}. We used the Python package \texttt{george}\footref{opensourcesoftwares} \citep{ambikasaran2015} for the implementation. For the interpretation of the results, it is valuable to understand the impact of these hyperparameters on the stellar activity model that this kernel produces \citep[e.g.,][]{angus2018, haywood2014}. $A_{\mathrm{rv}}$ scales with the amplitude of the stellar activity signal. $P_{\mathrm{rot}}$ indicates its main periodicity and is considered as a measure of the stellar rotation period \citep{angus2018}. $\tau_{\mathrm{decay}}$ and $\gamma$ are two indicators of the coherence of the stellar activity signals. $\tau_{\mathrm{decay}}$ governs the aperiodic coherence, the coherence between one period and the next periods. It is considered a measure of the timescale of growth and decay of the active regions \citep{haywood2014}. If it is longer than $P_{\mathrm{rot}}$, the stellar activity pattern will change slowly from one rotation period to the next. $\gamma$ controls the periodic coherence, that is, the coherence of the signal within a stellar rotation period. It is considered an indicator of the number of active regions. The larger $\gamma$ , the weaker the correlation between two points within a rotation period. $\gamma$ governs the complexity of the harmonic content of the stellar activity signal \citep[][]{angus2018}.

For the second approach, we used the same model, but we jointly fit the \rv s and the \fwhm\ values that accompany each \rv\ measurement. The \fwhm\ was fit with a \gp\ with a quasi-periodic kernel. This kernel is independent of the one used for the \rv, but it uses the same hyperparameters, except for the amplitude ($A_{\fwhm}$). This approach, inspired by \citet{suarezmascareno2020} and subsequently \citet{lillo-box2020}, relies on the assumption that the variations in \fwhm\ are solely due to stellar activity and that their periodicity and coherence are the same as the stellar activity component of the \rv. Under these assumptions, the joint fit of the \rv\ and \fwhm\ data sets allows constraining the hyperparameters of the quasi-periodic kernel better. In contrast to a first fit of the \fwhm s followed by a second fit of the \rv s using the marginalized \post\ of the first fit as \prior\ for the second, this approach preserves the correlation between the hyperparameters.

For the planetary model, we used a constant systemic velocity ($v_0$) and one Keplerian function per planet in the system. The parameters of each Keplerian are 
the semi-amplitude ($K$) of the \rv\ signal, and similarly to \sect{lconly}, the orbital parameters $P$, $t_{\mathrm{ic}}$, $e\cos\omega$ , and $e\sin\omega$. The Keplerians were implemented using the Python packages \texttt{radvel}\footref{opensourcesoftwares} \citep{fulton2018}.

For the instrumental model, as mentioned in \sect{espdata}, we considered three instruments in our model because of the fiber-link
change of \espresso: \harps, \espresso\ before ($\mathrm{pre}$), and \espresso\ after the intervention ($\mathrm{post}$). We used $\textrm{ESPRESSO}_{\mathrm{pre}}$ as \rv\ reference, meaning that $v_0$ was measured with the data coming from this instrument. We modeled the \rv\ offsets with the other two instruments with two offset parameters ($\Delta \mathrm{RV}_{\harps/\mathrm{pre}}$ and $\Delta \mathrm{RV}_{\mathrm{post/pre}}$).
The \fwhm\ is also subject to offsets between instruments, and our model includes a constant level for each instrument ($C_{\mathrm{pre}}$, $C_{\mathrm{post}}$, and $C_{\harps}$).
Finally, for the \rv\ and \fwhm\ and for each instrument, we considered one additive jitter parameter to account for a potential underestimation of the measurement errors due to underestimated or even nonconsidered noise sources \citep[][]{baluev2009} ($\sigma_{\rv, \mathrm{pre}}$, $\sigma_{\rv, \mathrm{post}}$, $\sigma_{\rv, \harps}$, $\sigma_{\fwhm, \mathrm{pre}}$, $\sigma_{\fwhm, \mathrm{post}}$, and $\sigma_{\fwhm, \harps}$).

To infer the values of these parameters, we performed a pre-minimization followed by an \textsc{mcmc} exploration as described in \sect{lconly}. The only difference was that this time, the initial values were all drawn from the \prior s. The \prior\ \pdf s assumed for the parameters are given in \tab{syspar} , except for the \prior\ of $P$ and $t_{\mathrm{ic}}$ of the three transiting planets. For these, we used the \post\ \pdf s of our analysis of the \tess\ \lc\ (provided in a footnote of \tab{syspar}). A detailed description of the reasons for the choice of each \prior\ is given in \app{priors}.

\subsubsection{Evidence for additional planets in the L 98-59 system}\sectlabel{modselect}

We analyzed our \rv\ data with six different models by varying the number of planets in the system from three to five and the stellar mitigation approach with or without the \fwhm\ data (see \sect{rvonly}). After each analysis, we inspected the output of the fit using plots such as those provided in Figs. \figref{4plTS} and \figref{4plGLS}. Figure \figref{4plTS} shows the \rv\ time series including the data from both instruments, the best planetary plus activity model, and the residuals of this model. Figure \figref{4plGLS} displays the \glsp\ of the combined \rv\ data, the residuals, the planetary and stellar activity models sampled at the same times as the \rv\ data, and the window function (\wf).

\begin{figure*}[!htb]
  \resizebox{\hsize}{!}{\includegraphics[]{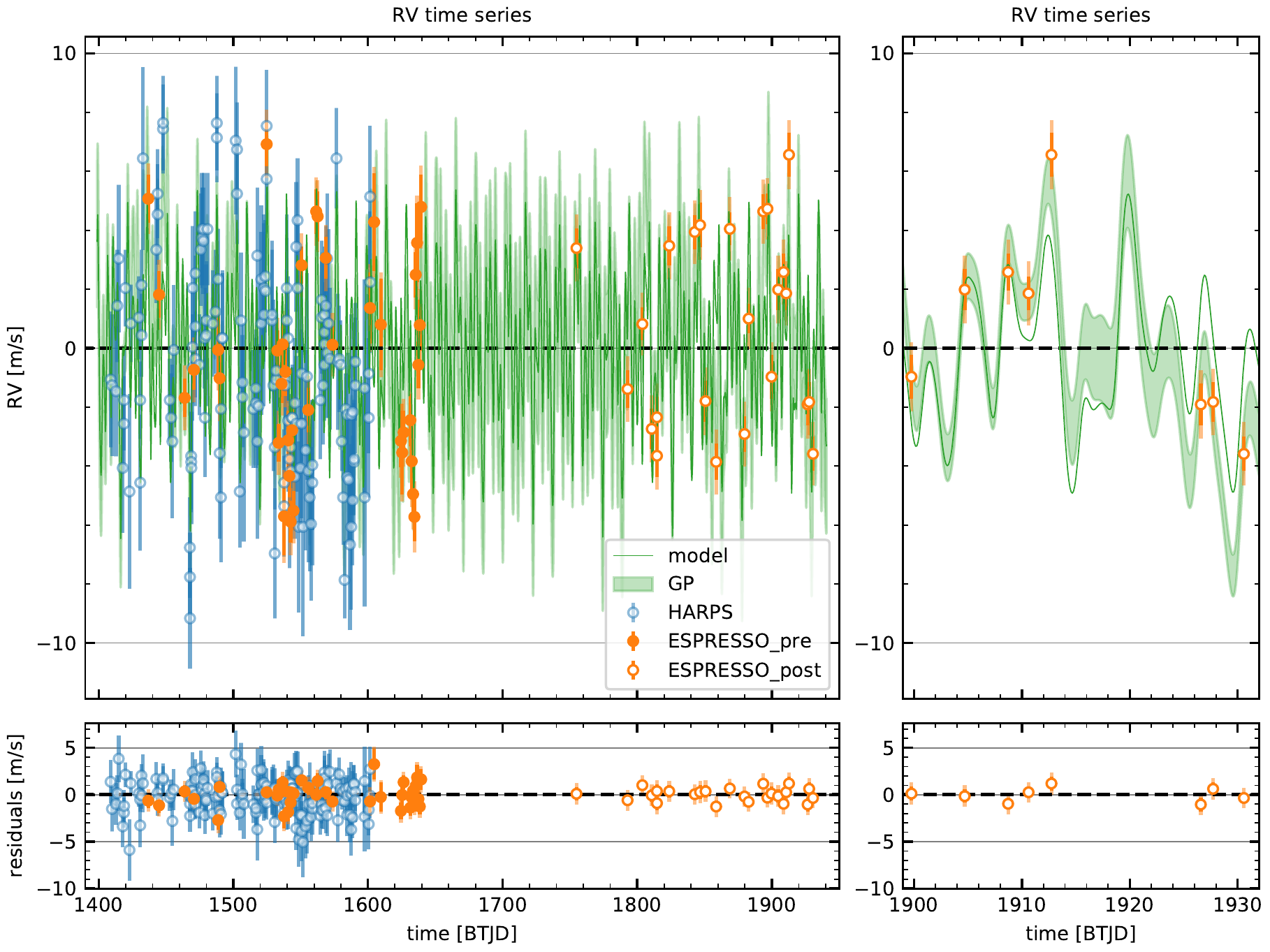}}
  \caption{\figlabel{4plTS} Outcome of the fit of the four planets model: (Top left) \rv\ time series along with the best model (solid green line), which includes the planetary signals and best prediction from the \gp\ stellar activity model. The $1\,\sigma$ uncertainties from the \gp\ prediction are also displayed (shaded green area). For this plot, we subtracted the systemic velocity and the instruments offsets from the RV data (see values in \tab{syspar}). (Bottom left) Time series of the residuals of the best model. (Right) Zoom on a small portion of the time series to better visualize the short-timescale variations.}
\end{figure*}

\begin{figure*}[!htb]
  \resizebox{\hsize}{!}{\includegraphics[]{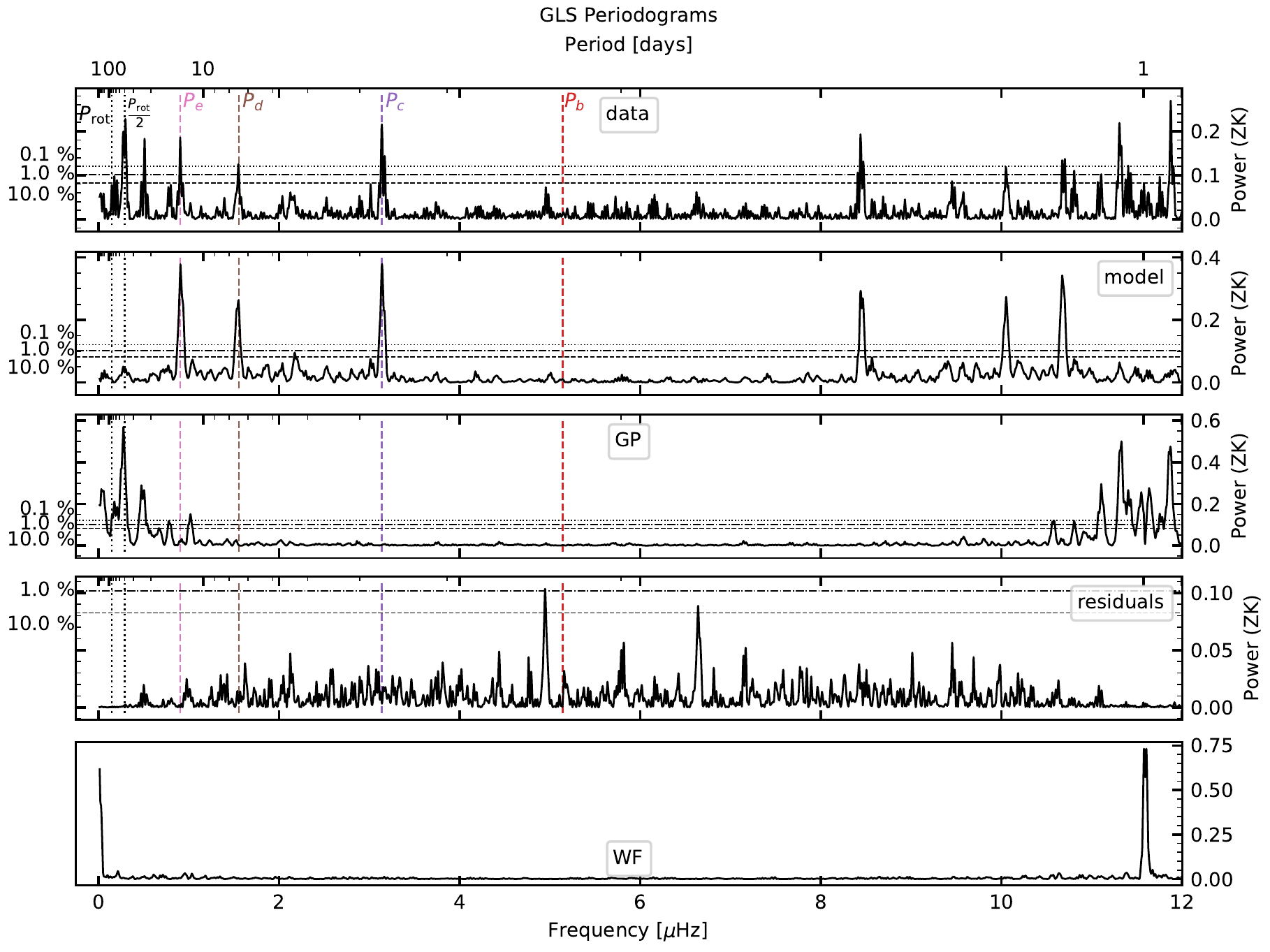}}  \caption{\figlabel{4plGLS} Outcome of the fit of the four-planet model: \glsp s of the \rv\ time series (top) and of the planetary (second) and stellar activity (third) models sampled at the same time as the \rv\ data, \glsp\ of the time series of the residuals (fourth) and the window function (bottom). The vertical lines on the \glsp s correspond to the orbital periods of planets b, c, d, and e, half and the full rotation period (estimated at 80 days) from right to left.}
\end{figure*}

Extensive outputs are shown and discussed in \app{modelselect}. From the fit of the three planets model (see Figs. \figref{3plTS} and \figref{3plGLS}), the \glsp\ of the residuals displays a narrow peak at 13 days, which we consider to be a strong insight for the presence of a fourth planet in the L 98-59 system at this period.
For the analysis with four planets, we adopted a noninformative \prior\ for the orbital period of the potential fourth planet (see \tab{syspar}). However, to speed up convergence, we drew its initial values from a Gaussian distribution with a mean of 13 days and a standard deviation of 1 day.
As shown in \fig{4plGLS}, the \glsp\ of the residuals of the four-planet model shows two narrow peaks around 1.743 and 2.341 days. These two peaks are aliases of one another. Because transit signals in the \tess\ \lc\ are absent at these periods, we did not explore the possibility of a planet at these periods. However, the peak at 23 days in the \glsp\ of the \rv s appears to be absorbed by the stellar activity model despite the absence of signal at 23 days in the \glsp s of the \fwhm\ and other activity indicators. We thus performed another analysis with five planets. We again set a noninformative \prior\ for the orbital period of the potential fifth planet (see \tab{syspar}), but we drew its initial values from a Gaussian distribution with a mean of 23 days and standard deviation of 1 day. The fit converged toward a significant detection of the semi-amplitude of a fifth Keplerian signal.

\tab{modcomp} regroups the Bayesian information criterion (\bic) values computed for all the models we tested.
However, the \bic\ is not necessarily adapted for our analysis because our models are nonlinear and our \prior s uninformative but relatively complex (see \app{priors}). Consequently, we also computed the Bayesian evidence ($\mathcal{Z}$) of our models using the Perrakis algorithm \citep{perrakis2014} using the Python implementation \texttt{bayev}\footnote{\texttt{bayev} is available at \url{https://github.com/exord/bayev}.} \citep{diaz2014}. We computed the logarithm of $\mathcal{Z}$ based on 5000 sets of parameters values and repeated the process 150 times. From these 150 computations, we extracted the median and the 68\,\% confidence interval (using the 16th and 84th percentiles) and report these values in \tab{modcomp}.
The Bayesian evidence agrees with the \bic\ values. According to both criteria, the four-planet model is favored and obtains the best values (minimum for the \bic\ and maximum for the Bayesian evidence). The only difference is in the absolute difference between the four- and the five-planet models. The \bic\ values of the five-planet model is significantly higher ($\Delta \bic = 3$ for the \rv+\fwhm\ analysis), while the Bayesian evidence of these two models is very similar ($\Delta \ln \mathcal{Z} = 0.4$).

We thus conclude that our additional \espresso\ \rv\ campaign allows us to identify one additional planet in the L\,98-59 system: a fourth planet, hereafter planet \object{L 98-59 e}, with an orbital period of 12.80 days. We also identify a planetary candidate, a potential fifth planet, hereafter planet 5, with an orbital period of 23.2 days. 
We show in \sect{noadditionaltransits} that these two additional planets do not transit.

Finally, retrieving the relevant information on L\,98-59 from the new Gaia Early Data Release 3 (\textsc{EDR3}), we note that an astrometric excess noise of 0.171 mas is reported, and the reduced unit weight error (RUWE) statistics has a value of 1.27. At G = 10.6 mag. The star is not so bright as to be strongly affected by unmodeled systematics due to the limited calibration. The Gaia EDR3 astrometry information (particularly RUWE) can thus be interpreted as providing weak evidence for the possible existence of an unresolved, massive outer companion \citep[e.g.,][]{belokurov2020,penoyre2020}. However, no long-term trend is observed in our \rv\ analysis.

\begin{table}[!htb]
\tiny
\caption{\tablabel{modcomp}Comparison of different models of the \rv s of the L 98-59 system.}
\raggedright
\begin{tabular}{p{0.10\columnwidth}P{0.20\columnwidth}P{0.20\columnwidth}P{0.20\columnwidth}}
\hline\noalign{\smallskip} 
 Nb planets & Types of data modeled & $\Delta \textsc{bic}$ & $\Delta \ln \mathcal{Z}$ \\
\hline\noalign{\smallskip} 
3          & \rv  &  0 & $0^{0.23}_{-0.18}$ \\ 
4          & \rv  & -12.0 & $7.08^{0.20}_{-0.15}$ \\ 
5          & \rv  & -6.2 & $6.27^{0.48}_{-0.43}$\\ 
\noalign{\smallskip}\hline\noalign{\smallskip} 
3          & \rv\ + \fwhm  &   0 & $0^{0.19}_{-0.13}$  \\ 
4          & \rv\ + \fwhm  &  -24.6 & $11.9^{0.25}_{-0.17}$ \\ 
5          & \rv\ + \fwhm  &  -21.6 & $11.5^{0.64}_{-0.29}$ \\ 
\noalign{\smallskip} 
\hline
\end{tabular}
\tablefoot{$\Delta \bic$ and $\Delta \ln \mathcal{Z}$ indicate the difference between a given model and the value of the three-planet model. For $\Delta \ln \mathcal{Z}$, our value for the three-planet model is 0 and is affected by error bars because our evidence estimates have quantified uncertainties and we use the best value of the three-planet model to derive the difference.}
\end{table}

\subsection{Joint analysis of \rv\ and photometry data}\sectlabel{finalana}

For the joint analysis of the \rv\ and photometry data, we only fit the best model identified by the \rv-only analysis due to the much higher
computational time associated with the data of the nine \tess\ sectors: The 
four planets plus stellar activity model on the \rv\ and \fwhm\ data sets.

The model of the \rv , \fwhm,\ and \lc\ data as well as the inference process is similar to the models used in Sects. \sectref{rvonly} and \sectref{lconly}.
The \prior\ \pdf\ assumed for the parameters is given in \tab{syspar} and discussed in \app{priors}. The initial values were drawn from the \prior\ \pdf s with a few exceptions. For $P$, $t_{\mathrm{ic}}$, ${R_p / R_*}$ , and $\cos i_{p}$ of the three transiting planets, we used the \post\ \pdf\ obtained by \kostov\ to draw the initial values. For $P$ of the two exterior planets, we used Gaussian \prior s with a standard deviation of 1 day and a mean value of 13 and 23 days for planet e and planetary candidate 5, respectively.

From our \textsc{mcmc} exploration, we extracted the estimates of the model parameters using the median of the converged iterations as best model values and their 16th and 84th percentiles as the boundaries of the 68\,\% confidence level intervals. We also derived estimates for secondary parameters.
As opposed to the model parameters (also called main or jumping parameters) described in the previous sections, secondary parameters are not used in the parameterization chosen for our modeling and are not necessary to perform the \textsc{mcmc} exploration. However, they provide quantities that can be computed from values of the main parameters and are of interest for describing the system. The secondary parameters that we computed are ${\Delta F/ F}$ the transit depth, $i$ the orbital inclination, $e$ the eccentricity, $\omega$ the argument of periastron, $a$ the orbital semimajor axis, $M_{\textrm{ref}}$ the mean anomaly at a given reference time (set as BTJD = 1354, the time of the first \tess\ measurement), $b$ the impact parameter, $D14$ the outer transit duration (duration between $\text{first}^{\textrm{}}$ and $\text{fourth}^{\textrm{}}$ contact), $D23$ the inner transit duration (duration between  $\text{second}^{\textrm{}}$ and $\text{third}^{\textrm{}}$ contact), $R_p$ the planetary radius, $M_p$ the planetary mass, $F_{i}$ the incident flux at the top of the planetary atmosphere, and $T_{\textrm{eq}}$ the equilibrium temperature of the planet (assuming an albedo of 0).
After the full \textsc{mcmc} analysis, we drew for each iteration a mass, a radius, and an effective temperature value for the star using Gaussian distributions whose mean and standard deviation were set according the results of our stellar analysis (see \sect{star} and \tab{syspar}). We then consistently computed the value of all the secondary parameters at each iteration of the \texttt{emcee} exploration, which provided us with chains for the secondary parameters. Finally, we estimated their best model values and 68\,\% confidence intervals with the same method as for the main parameters.

\subsubsection{Dynamical stability and parameters of the L 98-59 system}\sectlabel{stabilityandfinalparam}

In compact multiplanetary systems such as L 98-59, the assumption of long-term stability of the system can bring strong constraints on the planetary masses and orbital properties. Both \kostov\ and \cloutier\ performed N-body dynamical simulations with the objective of constraining the orbital eccentricity of the planets in this system. Both studies provide compatible conclusions: The eccentricity of planets c and d should be 0.1 or lower. As only the three inner planets were known at the time, the discovery of a fourth planet in this system requires revisiting this question. To do this, we used the framework implemented in the \texttt{spock} Python package \citep{tamayo2021,tamayo2020,tamayo2016}. \texttt{spock} has been developed specifically to assess the stability of compact multiplanetary systems. It performs a short, and thus relatively inexpensive, N-body simulations ($10^{4}$ orbits of the inner planet) using the Python package \texttt{rebound} \citep{rein2012b}. This simulation is then used to compute metrics based on established stability indicators \citep[see][and references therein]{tamayo2020}. These metrics are then provided to a machine-learning algorithm that estimates the probability that the simulated system is stable on the long term (typically $10^{9}$ orbits of the inner planet). According to \texttt{spock}, the probability that the system described by the best model parameters inferred from our joint analysis of the \rv , \fwhm,\ and photometry data is stable is 0. This means that the simulated system becomes unstable during the short N-body simulation (within $10^{4}$ orbits of the inner planet). This stresses the importance of considering the dynamical stability for this system.

Following the procedure described in \citet{tamayo2021}, we used \texttt{spock} to compute the probability of stability of the $10^{5}$ versions of the L\,98-59 systems described by the last $10^{5}$ converged \textsc{mcmc} iterations of our joint analysis. For these computations, we used the \texttt{WHFast} symplectic integrator \citep{rein2015} of \texttt{rebound}. We set a maximum distance of 0.4 AU ($\sim 6$ times the semimajor axis of planet e), meaning that all simulations that led to one of the planets traveling 0.4 AU away from the barycenter of the system were stopped and their probability of stability was set to 0. For each \textsc{mcmc} iteration we considered, we provided for the N-body simulation the mass of the star, the masses of the planets, and their orbital elements orbital period, semimajor axis, inclination, eccentricity, argument of periastron passage, mean anomaly at the beginning of the simulation (set as 1354 BTJD, the time of the first \tess\ measurement where BTJD = BJD - 2,457,000), and the longitude of ascending node. All these quantities, except for the longitude of the ascending node, are either main or secondary parameters of the model (see \sect{finalana}). Their values were thus taken directly from the \textsc{mcmc} chains or from their associated secondary parameters chains. For the longitudes of the ascending node, we drew values from a uniform distribution between 0 and 2 $\pi$.

With the probability of long-term stability estimated for the last $10^{5}$ iterations of our \textsc{mcmc} analysis of the joint fit of the data, we selected the iterations for which the probability of stability is higher than 40\% \citep[as in][]{tamayo2020}. This left us with only 1588 iterations. From these iterations and using their probability of stability as weight, we computed the weighted median and the weighted 16th and 84th percentile that we used as the best model values and the boundaries of the 68\,\% confidence interval, respectively, as suggested by \citet[][]{tamayo2021}. These estimates now describe a system with a high probability of long-term stability, and they are reported in \tab{syspar}. The phase-folded data (\rv\ and photometry) and the best model are displayed in Figs. \figref{datacompRV} and \figref{datacompLC}. The main impact of the long-term dynamical stability condition is on the eccentricity of planet c, which decreases from $0.147_{-0.048}^{+0.044}$ to $0.103_{-0.058}^{+0.045}$. The eccentricities of the other planets remain unchanged or decrease slightly, but they are well within $1\,\sigma$ of the previous estimates. The other parameters of the system are all compatible with their previous estimates at better than $1\,\sigma$. With these updated estimates, the eccentricities of the three transiting planets satisfy the constraints derived by both \kostov\ and \cloutier\ from their respective N-body simulations.

Finally, in order to assess whether planets c and d are in mean motion resonance, we performed an additional N-body simulation for each iteration of the system with a probability of long-term stability higher than 40\,\%. For each iteration, we started the simulation using the parameter values found in the \textsc{mcmc} chains or the associated secondary parameters chains, as before. We used \texttt{rebound} and the \texttt{WHFast} symplectic integrator with a time step of $10^{-4}\,\mathrm{year / 2\pi}$ (which corresponds to $\sim 10^{3}$ time steps per orbits of planet c). We integrated each simulation for the duration of our observations, 560 days between the beginning of the \tess\ observations and the last \espresso\ point. For each time step, we calculated the 2:1 resonant angles ($\theta$) of planet c and d, whose equation is \citep[e.g.,][Eq. 1]{quillen2014}
$$\theta_{i} = 2 \lambda_{d} - \lambda_{c} - \omega_{i}\,,\ i \in [\mathrm{c, d}],$$
where $\lambda$ is the mean longitude. As explained in \citet[][]{delisle2017}, if planets c and d are in mean motion resonance, their resonant angles should librate around a constant value. Following a procedure already used by \citet[][]{hara2020}, we computed the derivative of the resonant angles using the finite difference approximation and averaged their value over the duration of the simulation. The normalized histogram of the 1588 values of the average derivatives of the resonant angles obtained is not compatible with zero and indicates that planets c and d are not in mean motion resonance.

   \begin{figure*}[!htb]
      \resizebox{\hsize}{!}{\includegraphics{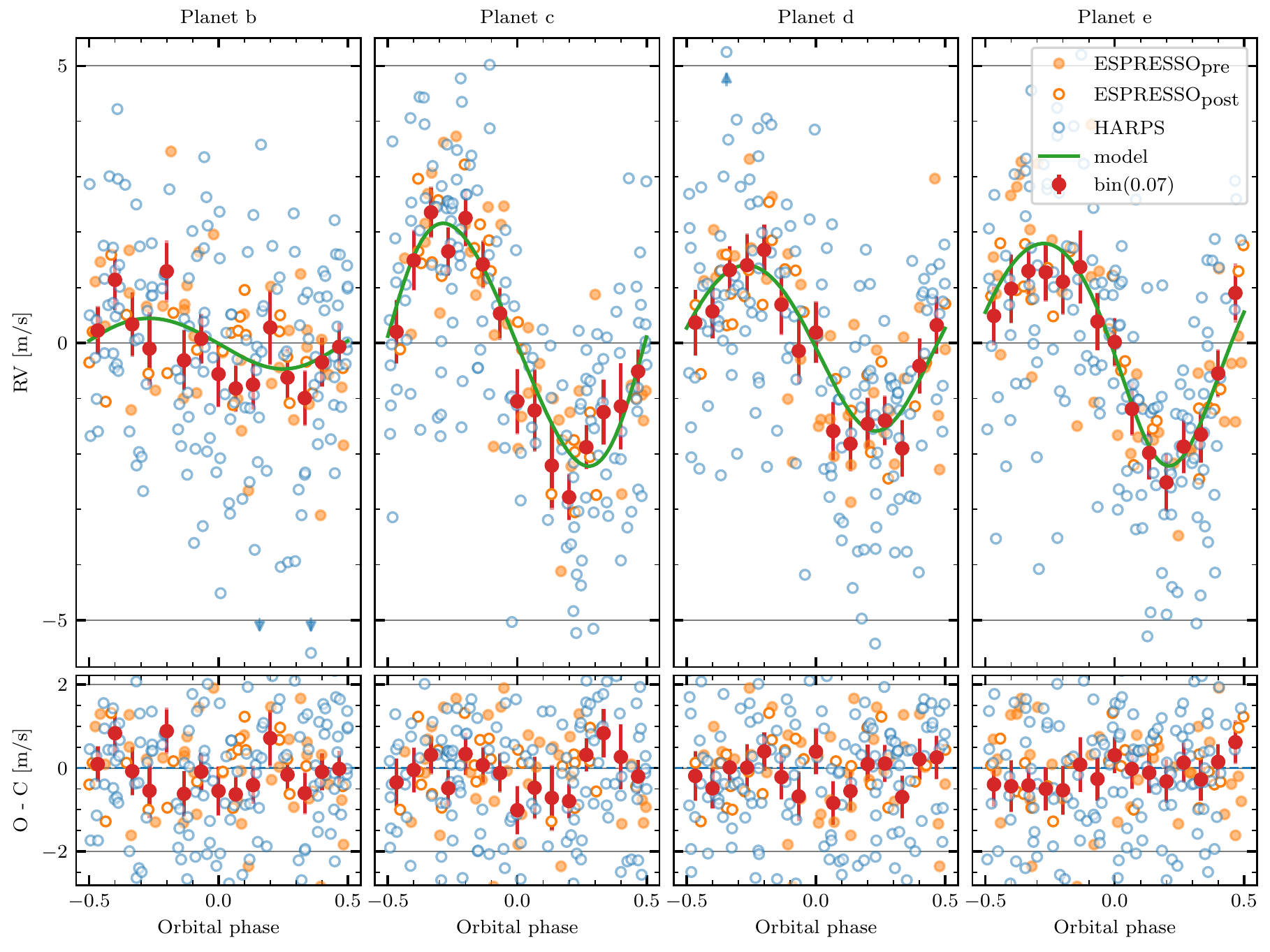}}
      \caption{\figlabel{datacompRV}Phase-folded \harps\ and \espresso\ \rv s, best model (top) and residuals (bottom) for the four planets. The \harps\ data, presented in \sect{harpsdata}, are displayed with empty blue circles, and the \espresso\ data, presented in \sect{espdata}, are displayed with orange circles. The filled orange circles are for the data taken before the fiber change of \espresso. The empty orange circles are for data taken after the change. For clarity, the error bars of the \harps\ and \espresso\ data points are not displayed. For this plot, the stellar activity model has been subtracted from each data point. The points with error bars in red correspond to averages of the data within evenly spaced bins in orbital phase, whose size is 0.07 orbital period. The best model is shown with a green line. Before the subtraction of the stellar activity model the \textsc{rms} of the \rv\ data is 3.5, 3.4, and 3.2 $\ms$ for \harps, \espresso$_{\textrm{pre}}$ , and \espresso$_{\textrm{post}}$ , respectively. After the subtraction of the stellar activity model, it is 2.9, 2.5, and 2.3 $\ms$ for \harps, \espresso$_{\textrm{pre}}$ , and \espresso$_{\textrm{post}}$ , respectively. Finally, after subtraction of the planetary model,  the \textsc{rms} of the residuals is 1.8, 1.2, and 0.7 $\ms$ for \harps, \espresso$_{\textrm{pre}}$ , and \espresso$_{\textrm{post}}$ , respectively.}
  \end{figure*}

  \begin{figure*}[!htb]
      \resizebox{\hsize}{!}{\includegraphics{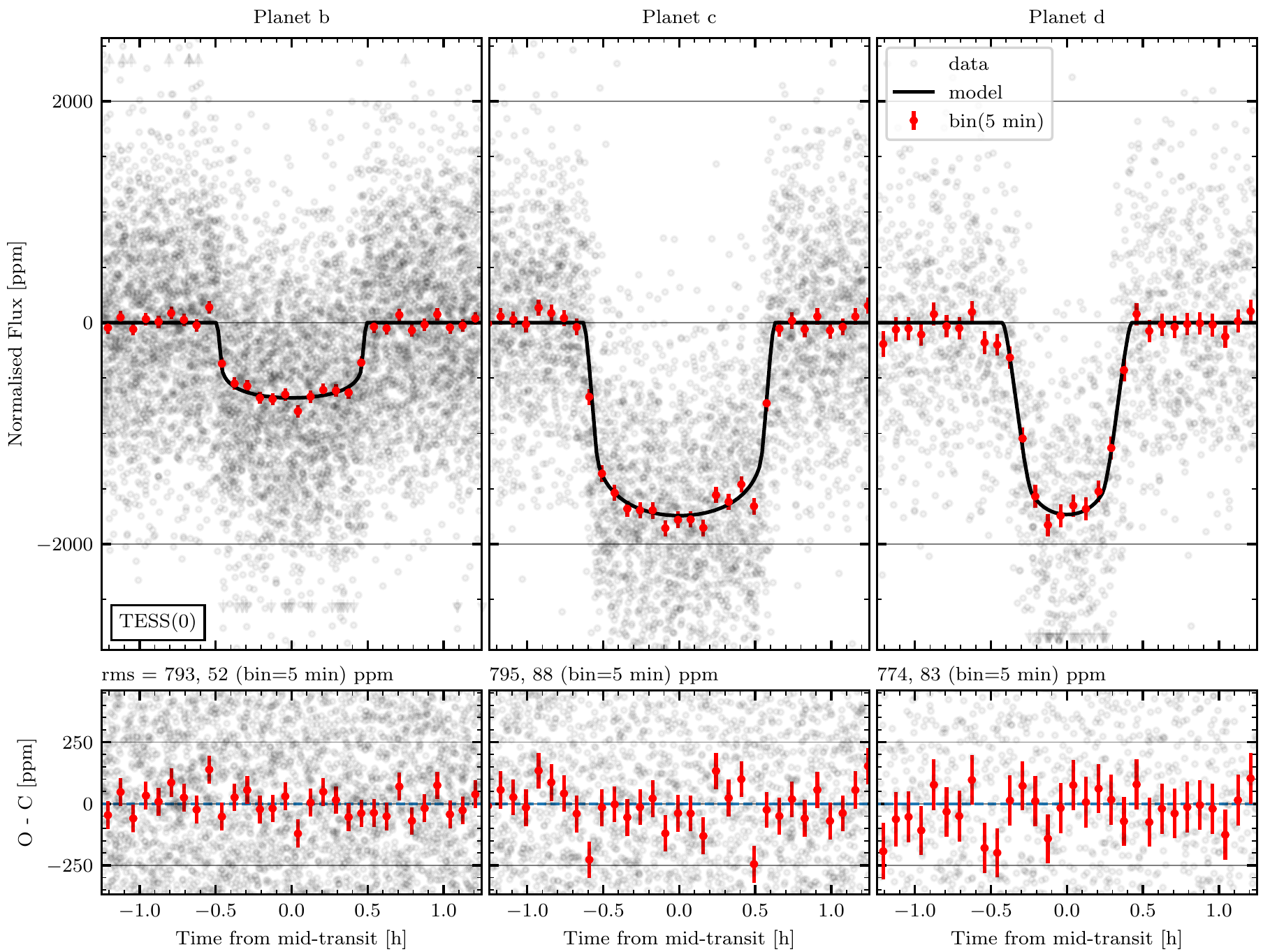}}
      \caption{\figlabel{datacompLC}Phase-folded \tess\ \lc, best model (top) and residuals (bottom) for the three transiting planets. The data presented in \sect{tessdata} are displayed in black. For clarity, the error bars are not displayed. The points with error bars in red correspond to averages of the data within evenly spaced bins in orbital phase whose size corresponds to 5 min. The best model is shown with a black line. The standard deviation of the raw and binned residuals is indicated above each residual plot.}
  \end{figure*}

\subsection{Three transiting planets}\sectlabel{noadditionaltransits}

Our \rv\ analysis (see \sect{modselect}) concluded with the existence of a fourth planet and a planetary candidate. They have not been reported before. Assuming that all planets in the system are coplanar, we can infer an orbital inclination of $88.21_{-0.27}^{+0.35}$ degrees and predict the impact parameter of planet e ($1.47_{-0.30}^{+0.27}$) and candidate 5 ($2.27_{-0.43}^{+0.46}$). From these impact parameter distributions, we estimate a probability of 4.8\% and 0.11\,\% that planet e and planetary candidate 5 transit their host star, respectively.
Using the nine \tess\ sectors and the best ephemerides inferred from our analysis, we do not detect any sign of transit from either planet e or planetary candidate 5 (see \fig{notransit} and \app{transitsearch} for more details of the analysis we performed).

\section{Discussion}\sectlabel{disc}

\subsection{Stellar activity modeling and mitigation}

Stellar activity mitigation is a current focus of the exoplanet community due to its impact on the detection and characterization of low-mass planets, both in \rv\ \citep[e.g.,][]{dumusque2017} and transit photometry \citep[e.g.,][]{barros2020}. For this analysis, we used a \gp\ with a quasi-periodic kernel to account for the important stellar activity imprint on the \rv\ data that were already identified by \cloutier. We analyzed the data with two slightly different approaches (see \sect{rvonly}): one used a \gp\ on the \rv\ data alone, and the other used the time series of a stellar activity indicator (here the \fwhm) that was fit simultaneously with the \rv. The motivation for the second approach is to place stronger constraints on the hyperparameters of the \gp .
In the case of L 98-59, we have shown in \sect{modselect} that the two approaches provide similar answers for the preferred model. A comparison of the \post\ \pdf\ of all common parameters to the two approaches shows that they also provide compatible estimates (within $1\,\sigma$).

\subsection{Four-planet system hosting the smallest planet measured through \rv}\sectlabel{disc:4planets}

The additional six sectors we analyzed compared to \kostov\  improved the characterization of the three transiting planets presented by \kostov\ and \cloutier\ (see \tab{syspar}). The ephemerides of the three planets are improved by factors $\sim 2$ and $\sim 10$ for the time of transit and the orbital period, respectively. The relative precisions of the radius ratios (${R_p / R_*}$) are also improved by a factor $\sim 2$ for the two inner planets and by a factor $\sim 4$  for planet d.

We also improved the mass determinations for these three planets. We derived the mass of planet b with 40\,\% relative precision (\cloutier\ only provided an upper limit). With an \rv\ semi-amplitude of $0.46_{-0.17}^{+0.20},\ms$ and a mass of $0.40^{+0.16}_{-0.15}$\,\MEarth (half the mass of Venus), \object{L 98-59 b} currently is the lowest-mass exoplanet measured through \rv\footnote{Confirmed planets with lower masses that can be found in \href{exoplanet.eu}{exoplanet.eu}
 and the \href{exoplanetarchive.ipac.caltech.edu}{NASA exoplanet archive} were all measured through transit-timing variations.
}. It represents a new milestone that illustrates the capability of \espresso\ to yield the masses of planets with \rv\ signatures of about 10\,cm.s$^{-1}$ in multiplanetary systems even when there is stellar activity.
The relative precision of the \rv\ semi-amplitude of the other two previously known planets is also improved by a factor $\sim 1.5$ for planet c and by a factor of $\sim 2$ for planet d. 
We obtain a relative mass precision of 11\%\ and 14 \% for planets c and d, respectively, which is the best precision for the mass measurement of super-Earths around M dwarfs that can currently be achieved \citep{suarezmascareno2020,lillo-box2020}.

For the three transiting planets, we achieve bulk densities with relative precision of 46, 21, and 24\,\% for planets b, c, and d, respectively. Considering the size and mass of these planets and the difficulties associated with a precise characterization of the mass and radius of M dwarfs, these density measurements are references for the field. Figure \figref{mrplot} shows these three planets in the mass-radius diagram and in the context of the known exoplanet population. These three planets are located below the radius gap \citep[][]{fulton2017,fulton2018b,cloutier2020} and appear to be mostly rocky (see \sect{intcomp}).

\begin{figure}[!htb]
      \includegraphics[]{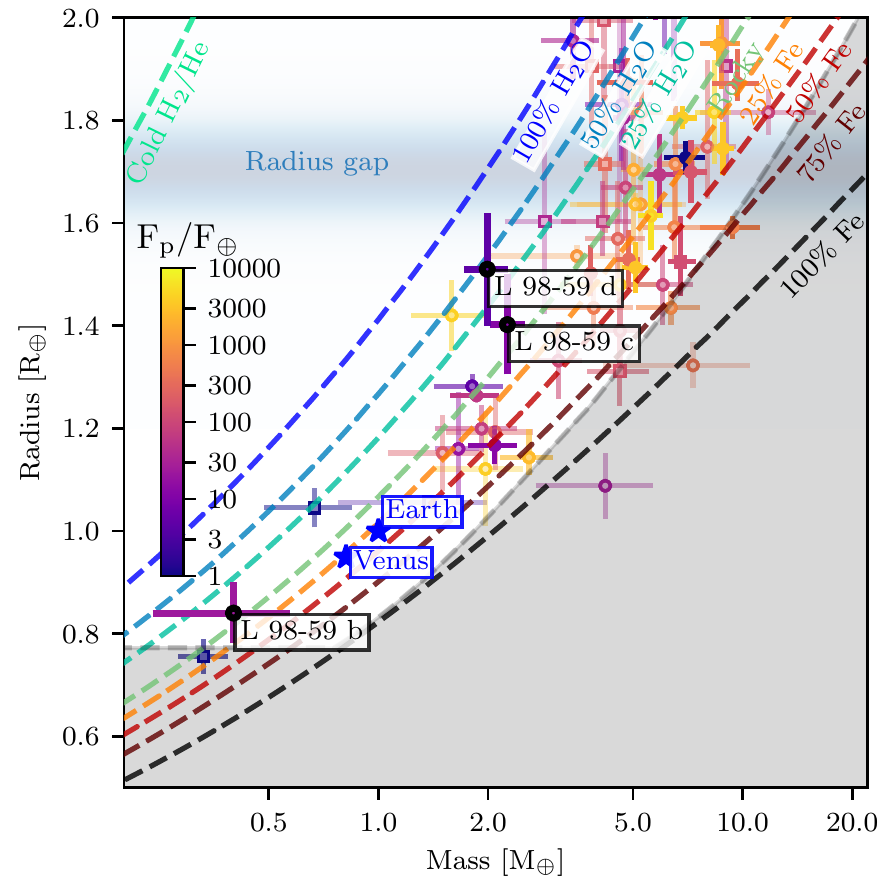}
      \caption{\figlabel{mrplot}Mass-radius diagram of the small-planet population. Each point represents a confirmed exoplanet for which the mass and radius are measured with a relative precision better than 50\,\%. These data have been extracted from exoplanet.eu \citep{schneider2011}. The shape of the points indicates the technique used to measure the mass of the planet: circles for \rv,\ and squares for transit timing variations. The color of the point reflects the intensity of the incident flux. The level of transparency of the error bars indicates the relative precision of the planetary bulk density. The better the precision, the more opaque the error bars. The three transiting planets in the L\,98-59 system are labeled and appear circled in black. The labeled blue stars indicate the Solar System planets. The colored dashed lines are the mass-radius models from \citet{zeng2016}. The gray region indicates the maximum collision stripping of the mantle. The shaded horizontal blue line represent the radius gap \citep{fulton2017}. L\,98-59\,b is in a sparsely populated region of the parameter space and currently the lowest-mass planet with a mass measured through \rv. Lower planetary masses have all been measured through transit-timing variation, e.g., for  Trappist-1\,h \citep{gillon2017a} to the left of L\,98-59\,b. This plot has been produced using the code available at \url{https://github.com/odemangeon/mass-radius_diagram}.}
\end{figure}

We also expand the view of this system with the discovery of a fourth planet and a planetary candidate. These planets do not transit, but with minimum masses of $3.06_{-0.37}^{+0.33}$ and $2.46_{-0.82}^{+0.66}$\,\MEarth, they are probably both rocky planets or water worlds \citep[also called ocean worlds, e.g.,][]{adams2008}. With an equilibrium temperature of $285_{-17}^{+18}$\,K, the planetary candidate 5, if confirmed, would orbit in the habitable zone of its parent star.

\subsection{Internal composition of the three transiting super-Earths}\sectlabel{intcomp}

We performed a Bayesian analysis to determine the \post\ distribution of the internal structure parameters of the planets. The method follows \citet{Dorn15} and \citet{Dorn17}, and has been used in \citet{mortier2020}, \citet{leleu2021}, and \citet{delrez2021}.
The model consists of two parts. The first is the forward model, which provides the planetary radius as a function of the internal structure parameters (iron molar fraction in the core,  Si and Mg molar fraction in the mantle, mass fraction of all layers,  age of the planet, and irradiation from the star), and the second part is the Bayesian analysis, which provides the \post\ distribution of the internal structure parameters based on the observed radii, masses, and stellar parameters (in particular, its composition). The details of the analysis performed along with additional outputs are provided in \app{intcomp}.

\begin{figure}[!htb]
 \centering
 \subfloat[Planet b]{\includegraphics[width=0.8 \hsize]{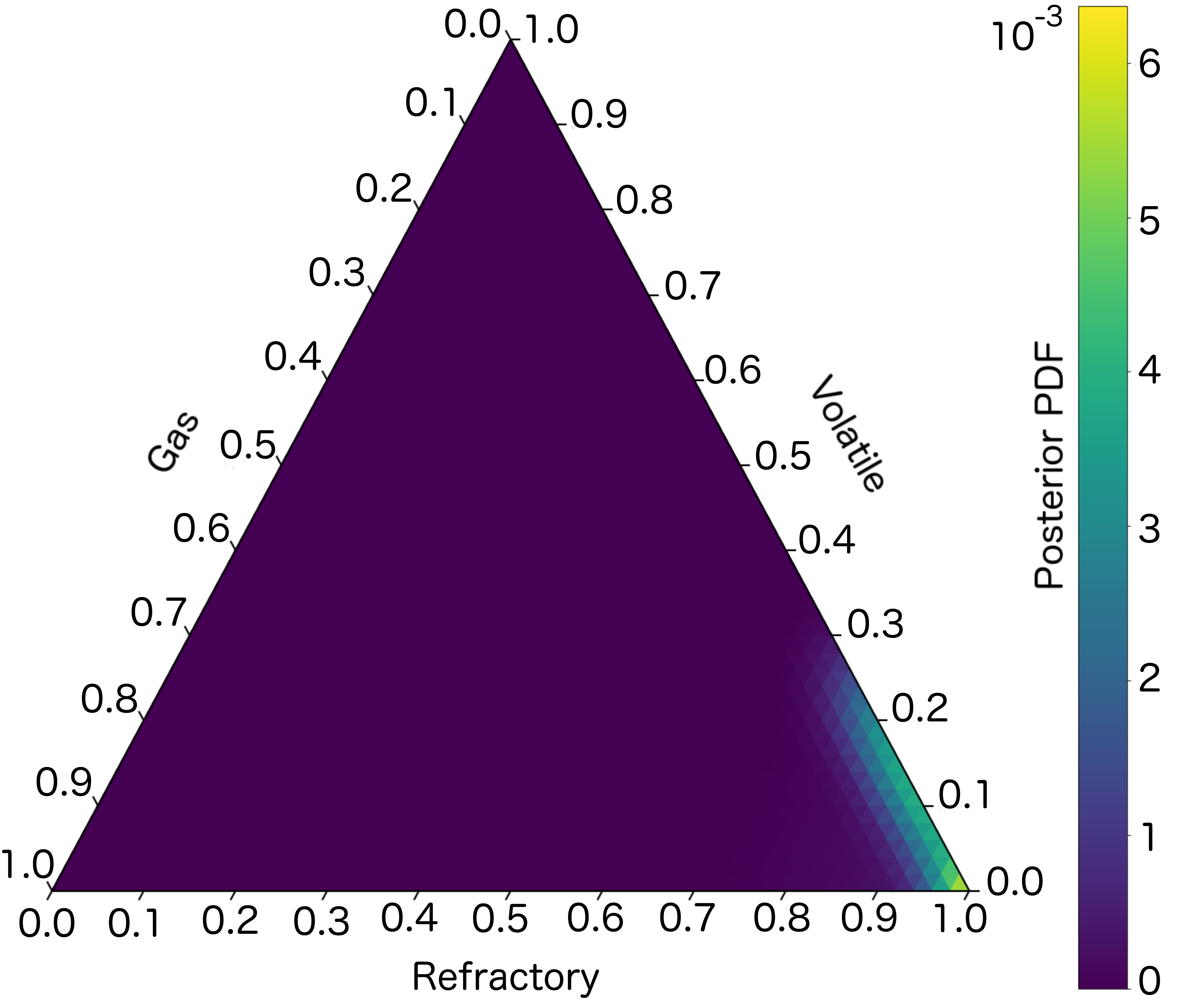}}
 \quad
 \subfloat[Planet c]{\includegraphics[width=0.8 \hsize]{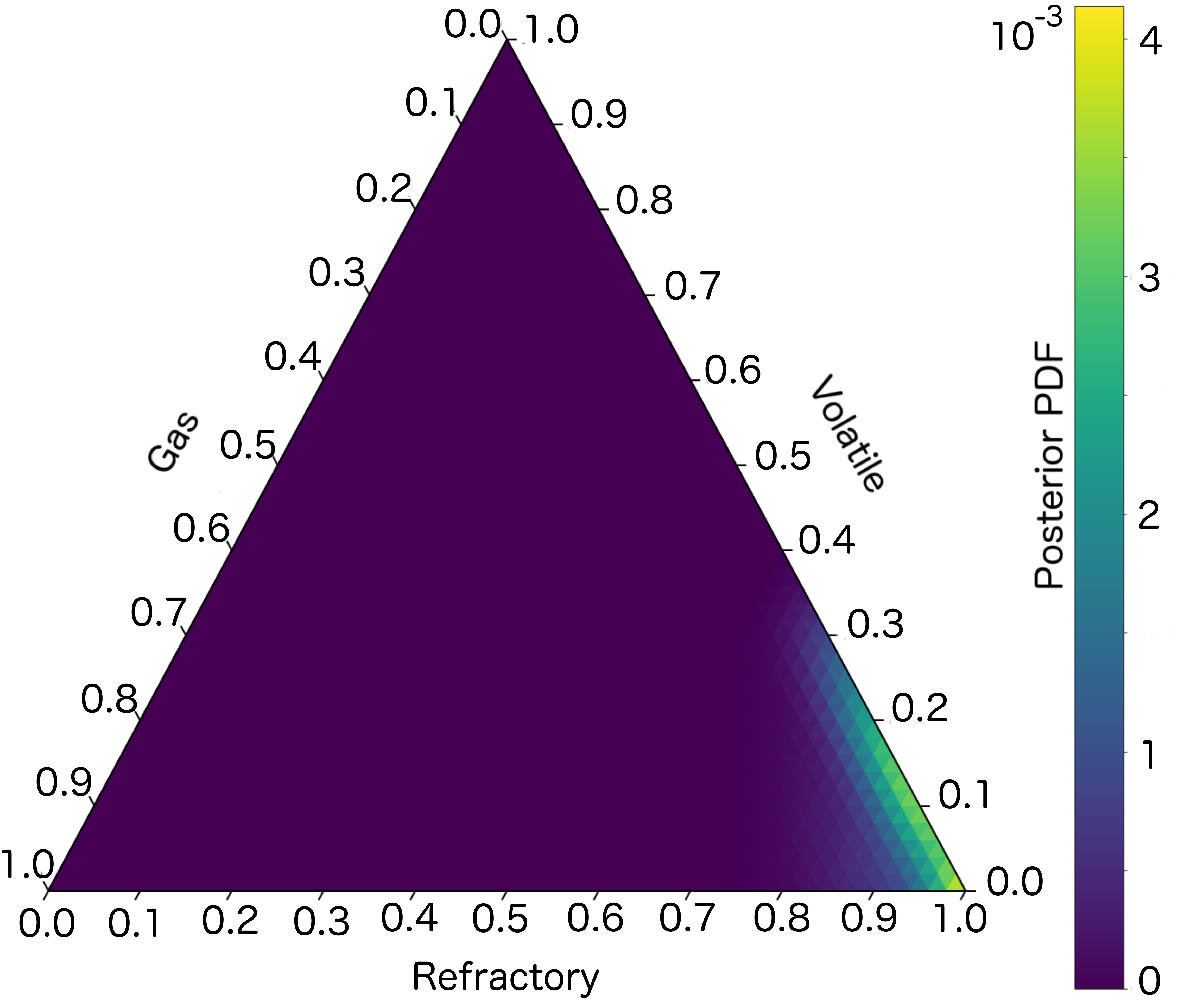}}
  \quad
  \subfloat[Planet d]{\includegraphics[width=0.8 \hsize]{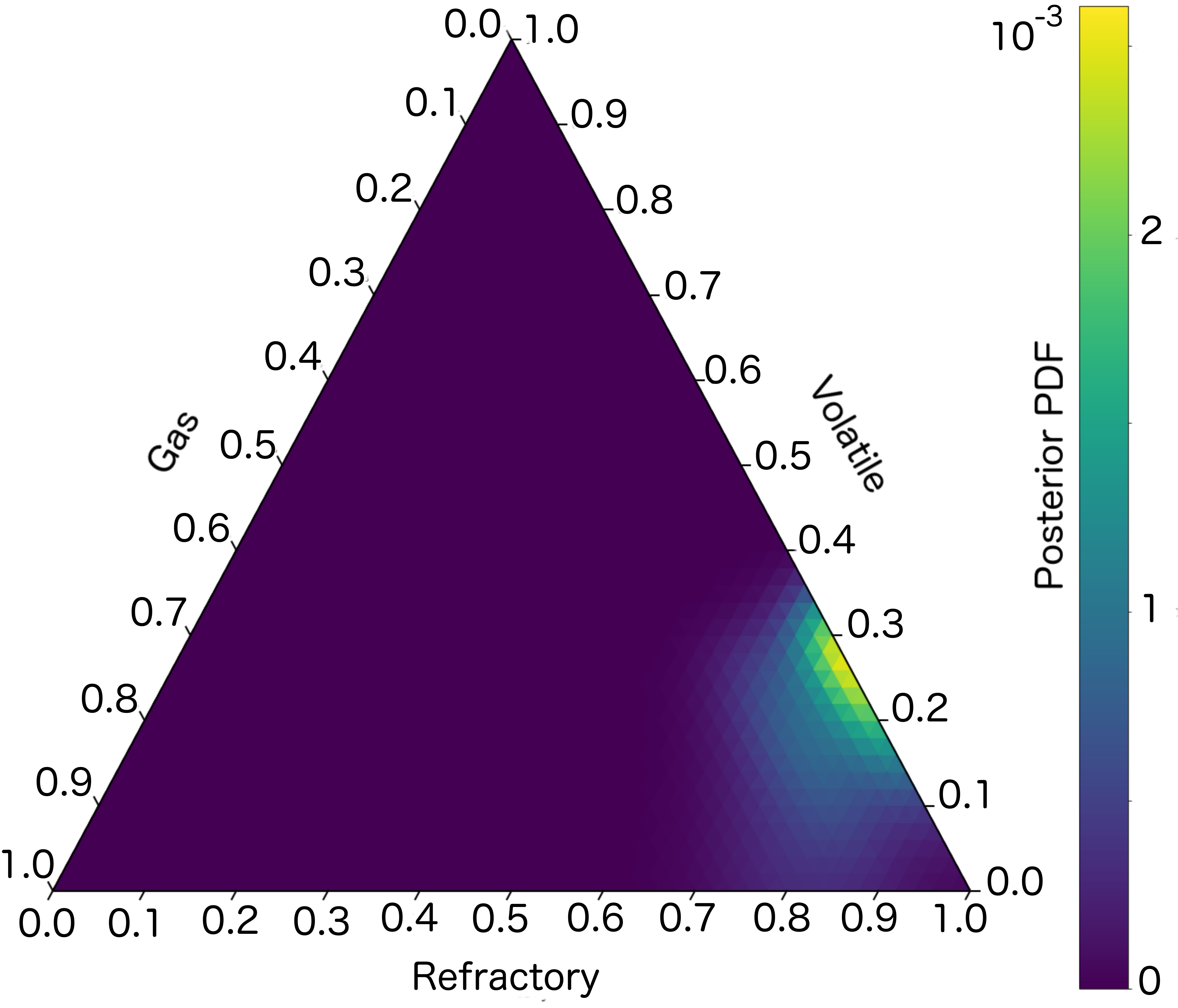}}
  \caption{\figlabel{triangles}Ternary diagrams showing the internal composition (mass fractions of the gas (H and He), the volatile (water) and the refractory elements) for the three transiting planets in system L\,98-59.}
\end{figure}

Figure \figref{triangles} provides the ternary diagrams representing the \post\ distributions of the composition of the three transiting planets in system L\,98-59. Furthermore, Figs. \figref{cornerb} to \figref{cornerd} in \app{intcomp} provide the detailed \post\ distributions of the most important parameters (mass fractions and composition of the mantle) of each planet.
The three planets are characterized by small iron cores (12 to 14 \% in mass), which reflects the small iron abundance (compared to Si and Mg) in the star. According to the Bayesian analysis, the two innermost planets are likely to have a small mass fraction of water (the mode of the distribution is at 0) and a low gas mass, if they have any gas at all. Interestingly, the internal structure parameters of L\,98-59\,d according to the Bayesian analysis are substantially different: the mode of the water-mass fraction distribution is at $\sim$ 0.3, whereas the mode of the gas mass peaks at  $\sim 10^{-6} M_\oplus$. Because the Bayesian analysis provides the joint distribution of all planetary parameters, we can easily compute the probability that the mass fraction of gas and water is higher in L\,98-59\,d than in L\,98-59\,b and L\,98-59\,c. Based on our model, the values are 79.3 \% and 72.0 \% for gas and water, respectively, for planet d \textit{\textup{versus}} planet b. These values are 79.6 \% and 79.1 \% for gas and water, respectively, for planet d \textit{\textup{versus}} planet c. Planet d therefore appears to be likely richer in gas and water. On the other hand, planets b and c are very similar in composition. We emphasize finally that these numbers result from the Bayesian analysis, and they therefore depend on the assumed \prior s that we took to be as uninformative as possible.

Our modeling favors a dry and hydrogen- and helium-free model for planet b and c. The \post\ distributions of their gas and water content peak at 0, but the $3\,\sigma$ confidence interval still allows for a water mass fraction of up to $\sim\, 25\,\%$  (see Figs. \figref{cornerb} and \figref{cornerc}). In order to understand how promising planets b, c, and even d are for atmospheric characterization, we need to understand whether these warm planets ($T_{\mathrm{eq}}$ between $\sim\,400$ and $\sim\,600$\,$\mathrm{K}$) could retain a water-dominated atmosphere. Providing a robust answer to this question requires modeling the complex phase diagram of water \citep[e.g.,][]{french2009,mousis2020,turbet2020}, the radiative transfer in a water-dominated atmosphere irradiated by an M star including potential runaway greenhouse effects \citep[e.g.,][]{arnscheidt2019}, and the hydrodynamic escape of water potentially assisted by ultraviolet photolysis \citep[e.g.,][]{bourrier2017a}. This analysis is beyond the scope of this paper. However, we can study the example of the TRAPPIST-1 system \citep{gillon2017a,luger2017a} for comparison. \citet{turbet2020} stressed the impact of irradiation on a water-dominated atmosphere. If the received irradiation is higher than the runaway greenhouse irradiation threshold \citep[e.g.,][]{kasting1993}, which should be the case for TRAPPIST-1\,b to d \citep[][]{wolf2017}, water should be in a steamed phase instead of a condensed phase, as classically assumed. In this case, the estimated water content of the planets decreases by several orders of magnitude. The authors further concluded that planets with masses lower than $0.5\,\MEarth$ that are more irradiated than the runaway greenhouse irradiation threshold are probably unable to retain more than a few percent of water by mass due to an efficient hydrodynamic escape. For the TRAPPIST-1 system, \citet{bourrier2017a} followed a theoretical study from \citet[][]{bolmont2017}, however, in the attempt to assess the water loss experienced by the planets during their lifetime. The authors concluded that planets g and planets closer in could have lost up to 20 Earth oceans through hydrodynamic escape. However, they noted that depending on the exact efficiency of the photolysis, even TRAPPIST-1\,b and c could still harbor significant amounts of water. 

L\,98-59\,b is similar in mass and radius to TRAPPIST-1\,d. However, it is significantly more irradiated \citep[$T_{\mathrm{eq}} = 288 \pm 5.6$\,K for TRAPPIST-1\,d][]{gillon2017a}. L\,98-59\,b might thus undergo or have undergone efficient hydrodynamic escape. \object{L 98-59 c} and d are more massive than any of the TRAPPIST-1 planets, but they are also more irradiated \citep[$T_{\mathrm{eq}} = 400.1 \pm 7.7$\,K for TRAPPIST-1\,b][]{gillon2017a}, and the comparison is thus less pertinent. They are likely to have undergone runaway greenhouse effect, but their higher masses might inhibit atmospheric escape. A more detailed study and observational evidence are thus required to reliably assess the nature and content of the atmosphere of the transiting planets in the L\,98-59 system.

\section{Conclusion: L\,98-59, a benchmark system for super-Earth comparative exoplanetology around an M dwarf}\sectlabel{conc}

Multiplanetary systems are ideal laboratories for exoplanetology because they offer the unique possibility of comparing exoplanets that formed in the same protoplanetary disk and are illuminated by the same star. According to the exoplanet archive\footnote{\url{https://exoplanetarchive.ipac.caltech.edu/}} \citep{akeson2013}, we currently know 739 multiplanetary systems. A large fraction of them ($\sim 60\,\%$) were discovered by the Kepler survey \citep{borucki2010,lissauer2011a}. From a detailed characterization and analysis of the properties of the Kepler multiplanetary systems, \citet[][hereafter \weiss]{weiss2018} extracted the so-called "peas in a pod" configuration. The authors observed that consecutive planets in the same system tend to have similar sizes. The planets also appear to be preferentially regularly spaced. The authors also noted that the smaller the planets, the tighter their orbital configuration. For \tab{peasinpod} we computed the metrics identified by \weiss\ for the L\,98-59 system along with the distributions of these metrics derived by the authors from their sample. From \tab{peasinpod} we conclude that system L\,98-59 closely follows the peas in a pod configuration. Most systems in the \weiss\ sample have FGK host stars. For example, none of the 51 host stars that host four or more planets have a mass lower than $0.6 M_{\Sun}$. The fact that the L\,98-59 system also follows the peas in a pod configuration thus further strengthens the universality of this configuration and the constraints that it brings on planet formation theories. The only trend observed by \weiss\ that the L\,98-59 system does not display is the positive correlation between the equilibrium temperature difference of consecutive planets and their radius ratio. Furthermore, assuming a $v\sin{i_{*}}$ of 1 km/s, the semi-amplitude of the expected Rossiter-McLaughling effect \citep[e.g.,][]{queloz2000} is 40 cm/s, 1 m/s, and 37 m/s for planets b, c, and d, respectively. These amplitudes might at least for planets c and d be within the reach of high-resolution spectrographs such as \espresso. This would give us access to the spin-orbital angle in this system, which would further constrain its architecture and the possible mechanisms of its formation and migration.

\begin{table}[!htb]
\caption{\tablabel{peasinpod} Peas in a pod statistics in system L\,98-59 }
\begin{tabular}{p{0.5\columnwidth}p{0.4\columnwidth}} 
\hline \noalign{\smallskip} 
Metric from the L\,98-59 system & \weiss\ distribution \\      
\hline \noalign{\smallskip}
$R_c / R_b = 1.669$ & $1.14 \pm 0.63$\\ \noalign{\smallskip}  
$R_d / R_c = 1.077$ & (mean = 1.29) \\ 
\noalign{\smallskip} \hline \noalign{\smallskip}

$(P_d / P_c) / (P_c / P_b) = 1.232 $ & \multirow{2}{0.40\columnwidth}{$1.00 \pm 0.27$}\\  
$(P_e / P_d) / (P_d / P_c) = 0.851 $ & \\
$(P_{05} / P_e) / (P_e / P_d) = 1.053 $ & (mean = 1.03)\\
\noalign{\smallskip} \hline \noalign{\smallskip}
$\Delta(c, b) = 15.260$ & \multirow{4}{0.40\columnwidth}{Mode between 10 and 20 with long tail towards high values for 4+ planets systems}\\
$\Delta(d, c) = 18.414$ & \\
$\Delta(e, d) = 13.569$ & \\
$\Delta(05, e) = 14.389$ & \\
\noalign{\smallskip} \hline \noalign{\smallskip}
$T_{\textrm{eq}, b} - T_{\textrm{eq}, c} = 49$ K & \multirow{2}{0.40\columnwidth}{$T_{\textrm{eq}, i} - T_{\textrm{eq}, i+1}$ positively correlated with $R_{i + 1} / R_{i}$} \\
$T_{\textrm{eq}, c} - T_{\textrm{eq}, d} = 152$ K & \\
\noalign{\smallskip} \hline \hline           
\end{tabular}
\tablefoot{- $\Delta(i, j)$ is the separation in mutual Hill radii (see Eq. 5 in \weiss)\\
When the notation $x \pm y$ is used in the column "\weiss\ distribution",  $x$ is the median of the observed distribution, and $y$ is its standard deviation.}
\end{table}

The fact that L\,98-59\,A is an M dwarf sets this system apart among multiplanetary systems. According to the exoplanet archive and the recent literature, only seven multiplanetary systems are currently confirmed (including L-98-59) around M dwarfs for which the planetary masses and radius of at least two planets have been measured. The other six are TRAPPIST-1 \citep{gillon2017a}, LTT-3780 \citep{cloutier2020a}, TOI-1266 \citep{demory2020}, LHS-1140 \citep{lillo-box2020}, K2-146 \citep{hamann2019}, and Kepler-138 \citep{jontof-hutter2015}. With a V magnitude of 11.7 and a distance of 10.6 pc, L\,98-59 is the brightest and closest of these systems.

Finally, according to the transmission spectrum metric \citep[\textsc{tsm},][]{kempton2018}, with values of 49, 37, and 255 for planets b, c, and d, respectively, the three transiting planets in system L\,98-59 are comfortably above the thresholds proposed by \citet[][]{kempton2018} for a super-Earth atmospheric characterization with the \jwst. This threshold is 12 for planets with radii smaller than 1.5 \REarth\ , like planets b and c, and 92 for planets with radii between 1.5 and 2.75 \REarth\ , like planet d.
Figure \figref{tsmplot} shows the \textsc{tsm} values for the well-characterized small-planet population. L\,98-59\,b and c are the two planets with the highest \textsc{tsm} value below $1.5\,\mathrm{R_{\Earth}}$ , and L\,98-59\,d has the second highest above this value. These three planets thus belong to the most favorable warm to temperate ($T_{\mathrm{eq}} < 650$ K) super-Earths ($R_p < 1.5 R_{\Earth}$) for an atmospheric characterization. Furthermore, L\,98-59 is located at the border of the continuous viewing zone ($\sim 200$ days per year) of the \jwst,\ making it a golden system for atmospheric characterization and comparative planetology.
Even if the \textsc{tsm} is specifically tailored to the \jwst , these planets are also suitable for transmission spectroscopy with other facilities such as \espresso, the HST \citep[][]{sirianni2005}, NIRPS \citep{bouchy2017a}, or Ariel \citep{tinetti2016}.   

\begin{figure}[!htb]
      \includegraphics[]{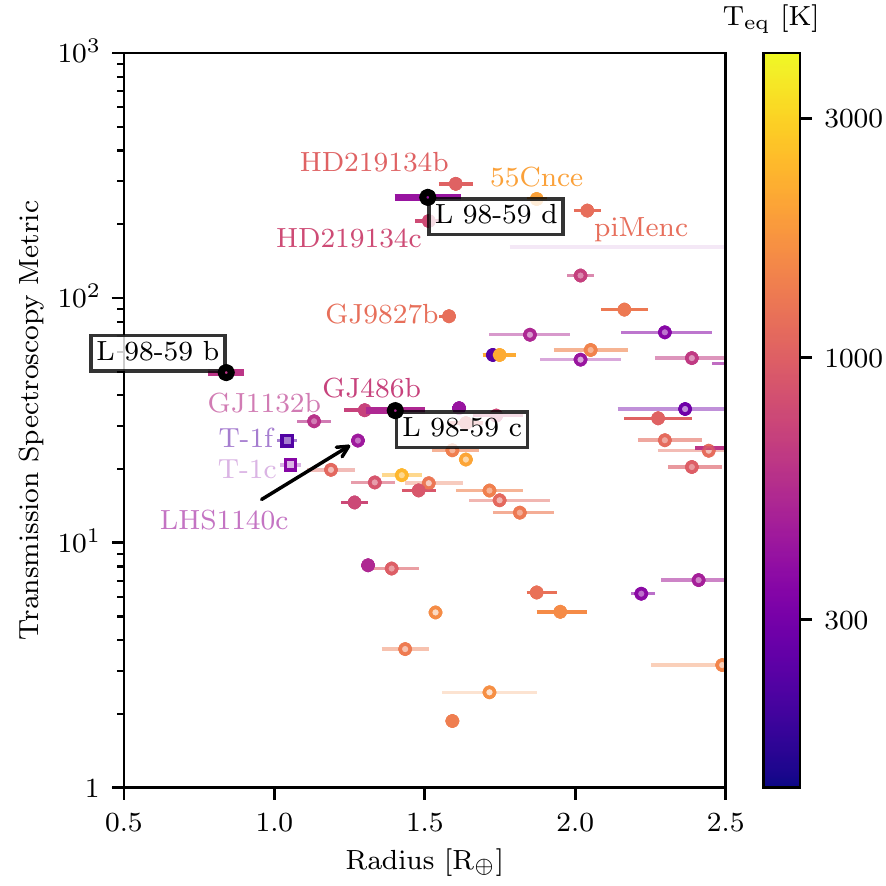}
      \caption{\figlabel{tsmplot} Transmission spectrum metric vs. planetary radius diagram of the small-planet population. Each point represents a confirmed exoplanets with mass and radius measured with a relative precision better than 50\,\%. These data have been extracted from the \href{https://exoplanetarchive.ipac.caltech.edu/}{exoplanet archive}. The shape of the points indicates the technique used to measure the mass of the planet: circles for \rv,\ and squares for transit timing variations. The color of the point reflects the equilibrium temperature of the planet. The level of transparency of the error bars indicates the relative precision of the planetary bulk density. The better the precision, the more opaque the error bars. The three transiting planets in system L\,98-59 are labeled and appear circled in black. We also display the names of the other planets with the highest transmission spectrum metrics.
      This plot has been produced using the code available at \url{https://github.com/odemangeon/mass-radius_diagram}.}
\end{figure}

\begin{acknowledgements}
The authors acknowledge the ESPRESSO project team for its effort and dedication in building the ESPRESSO instrument.

This research has made use of the NASA Exoplanet Archive, which is operated by the California Institute of Technology, under contract with the National Aeronautics and Space Administration under the Exoplanet Exploration Program.

This work has made use of data from the European Space Agency (ESA) mission
 (\href{https://www.cosmos.esa.int/gaia}{\it Gaia}),
 processed by the {\it Gaia}
Data Processing and Analysis Consortium (\href{https://www.cosmos.esa.int/web/gaia/dpac/consortium}{\textsc{dpac}}).
Funding for the \textsc{dpac}
has been provided by national institutions, in particular the institutions
participating in the {\it Gaia} Multilateral Agreement.

This research has made use of the NASA/IPAC Infrared Science Archive, which is funded by the National Aeronautics and Space Administration and operated by the California Institute of Technology.

This publication makes use of \vosa, developed under the Spanish Virtual Observatory project supported by the Spanish MINECO through grant AyA2017-84089.
\vosa\ has been partially updated by using funding from the European Union's Horizon 2020 Research and Innovation Programme, under Grant Agreement nº 776403 (EXOPLANETS-A)

This work was supported by FCT - Funda\c{c}\~{a}o para a Ci\^{e}ncia - through national funds and by FEDER through COMPETE2020 - Programa Operacional Competitividade e Internacionalização by these grants: UID/FIS/04434/2019; UIDB/04434/2020; UIDP/04434/2020; PTDC/FIS-AST/32113/2017 \& POCI-01-0145-FEDER-032113; PTDC/FIS-AST/28953/2017 \& POCI-01-0145-FEDER-028953; PTDC/FIS-AST/28987/2017 \& POCI-01-0145-FEDER-028987; PTDC/FIS-AST/30389/2017 \& POCI-01-0145-FEDER-030389.

O.D.S.D.~is supported in the form of work contract (DL 57/2016/CP1364/CT0004) funded by FCT.

J.I.G.H. acknowledges financial support from Spanish Ministry of Science and Innovation (MICINN) under the 2013 Ram\'on y Cajal  programme RYC-2013-14875.
J.I.G.H., A.S.M., R.R., and C.A.P. acknowledge financial support  from the Spanish MICINN AYA2017-86389-P.
A.S.M. acknowledges financial support from the Spanish Ministry of Science and Innovation (MICINN) under the 2019 Juan de la Cierva Programme.

This work has been carried out with the support of the framework of the National Centre of Competence in Research PlanetS supported by the Swiss National Science Foundation (SNSF). The authors acknowledge the financial support of the SNSF and in particular YA and JH acknowledge the SNSF for supporting research through the grant 200020\_19203.

NJN acknowledges support form the following projects: CERN/FIS-PAR/0037/2019, PTDC/FIS-OUT/29048/2017.

R. A. is a Trottier Postdoctoral Fellow and acknowledges support from the Trottier Family Foundation. This work was supported in part through a grant from FRQNT.

A.A.K. and S.G.S. acknowledge the support from FCT in the form of the exploratory projects with references IF/00028/2014/CP1215/CT0002, IF/00849/2015/CP1273/CT0003.

A.S. acknowledges support from the Italian Space Agency (ASI) under contract 2018-24-HH.0. The financial contribution from the agreement ASI-INAF n.2018-16-HH.0 is gratefully acknowledged.

This project has received funding from the European Research Council (ERC) under the European Union’s Horizon 2020 research and innovation programme (project {\sc Four Aces}; grant agreement No 724427).

Used Simbad, Vizier, exoplanet.eu.

Most of the analyses presented in this paper were performed using the Python language (version 3.6) available at \href{http://www.python.org}{python}
and several scientific packages: Numpy \citep{vanderwalt2011}, Scipy \citep{virtanen2020}, Pandas \citep{mckinney2010}, Ipython \citep{perez2007}, Astropy \citep{astropycollaboration2013, astropycollaboration2018} and Matplotlib \citep{hunter2007}.

\end{acknowledgements}

\bibliographystyle{aa}
\bibliography{bibliography}

\begin{appendix}

\section{Characterization of the M dwarf L 98-59 A}\applabel{stelchar}

\subsection{Atmospheric parameters of L\,98-59\,A: Detailed description of the different methods}\applabel{atmostelchar}

In addition to the derivation made by \kostov, we applied three different methods to derive the \teff, \logg\ and \feh\ of L\,98-59\,A.

\subsubsection{Spectral synthesis with {\scshape SteParSyn}}\applabel{specsyn}

We employed the BT-Settl model grid \citep{allard2012}, the radiative transfer code {\tt turbospectrum} \citep{ple12}, and a VALD3-based line list \citep{vald3}. The stellar atmospheric parameters of our selected set of synthetic spectra span between 2600 and 4500 K in $T_{\rm eff}$, 4.0 to 6.0 dex in \logg, and -1 to +0.5 dex in [Fe/H]. In addition, we took the instrumental broadening into account by means of a Gaussian kernel (R~$=$~140\,000). 
We used the latest version of the {\scshape SteParSyn} code \citep[][]{tab18,tab20b} to infer the stellar parameters. We fit the combined spectrum of L\,98-59\,A that was constructed using 61 \espresso\ spectra ($\snr = 1063$ at 7580 \AA). We selected the TiO band system at 7050\,\AA{} together with some \ion{Fe}{i} and \ion{Ti}{i} lines \citep[see ][]{mar20b} to fit the observations. The latest version of {\sc SteParSyn} relies on {\tt emcee} \citep{emcee}, a Markov chain Monte Carlo (\textsc{mcmc}) method used to fully sample the underlying distribution of the stellar parameters of L\,98-59. In addition to the \teff, \feh\ , and \logg\ values shown in \tab{specpar}, the method also provides an estimate for the quadratic sum of the macroturbulence ($\zeta$) and the stellar equatorial spin velocity projected on the plane of the sky ($\varv \sin i$):  $\sqrt{\zeta^2 +  (\varv \sin i)^2} = 3.78 \pm 0.44 $ \kms.

\subsubsection{Machine-learning regression with \textsc{odusseas}}\applabel{machinelearning}

The \textsc{odusseas} software \citep{antoniadis-karnavas2020} receives a 1D spectrum and its resolution as input. The pseudo-equivalent widths are measured and used as input for a supervised machine-learning algorithm (ridge regression model) that is used to derive the spectroscopic parameters \teff\ and \feh. The implementation relies on the machine-learning Python package \texttt{scikit learn}. The training and testing sets were taken from a reference sample of 65 \harps\ spectra with associated \teff\ and \feh\ derived by \citet{casagrande2008} and \citet{neves2012}. When the spectra provided as input did not have the same resolution as the \harps\ spectra from the reference sample, the spectra with the highest resolution were degraded (by convolution) to the lowest of the two resolutions.
The estimates of \teff\ and \feh\ result from the average of 100 determinations obtained by randomly shuffling and splitting the training and testing groups. The reported uncertainties are the wide uncertainties of the machine-learning models at this resolution, after taking the intrinsic uncertainties of the reference sample parameters during the machine-learning process into consideration. The estimates provided by this method are also reported in \tab{specpar}.

\subsubsection{Spectral energy distribution fitting with \textsc{vosa}}\applabel{vosa}

The \vosa\ \citep[][]{bayo2008} online tool estimates the \teff, \feh, \logg, extinction ($A_{\textrm{V}}$) , and alpha enhancement by fitting the photometric \sed\ with theoretical models. It also computes the total flux ($F_{\textrm{tot}}$) by integrating over the best template and then uses the distance to infer the luminosity ($L$). \vosa\ offers a wide variety of stellar models. We chose the BT-Settl model \citep{allard2012} for its treatment of dust and clouds, which is important for low-mass stars. Because of the small distance of 10.6194 pc \citep[inferred from GAIA parallaxes, ][]{bailer-jones2018}, we fixed the extinction to 0. The photometric measurements we used for the photometric \sed\ are listed in \tab{sed}. The \teff, \feh\ , and \logg\ provided by this analysis are provided in \tab{specpar}. Additionally, the fitting procedure inferred an alpha-element enhancement ([$\alpha$/Fe]) of $-0.03^{+0.16}_{-0.13}$ dex and a luminosity of $L = 0.01128 \pm 0.00042\,\mathrm{L}_{\sun}$.

\begin{table}[!htb]
\tiny
\caption{\tablabel{sed} Broadband photometry of L 98-59}
\center
\begin{tabular}{lcccc}
\hline
Filter ID                       & Observed Flux                      \\
                                        & $[\mathrm{erg/s/cm^2}/$\AA] \\
\hline\\[-5pt]
APASS.B                 & $3.139\cdot10^{-14} \pm 7.8\cdot10^{-16}$ \\
SLOAN/SDSS.g       & $5.208\cdot10^{-14} \pm  9.1\cdot10^{-16}$ \\
GAIA/GAIA2.Gbp     & $6.7184044482566\cdot10^{-14} \pm                      0$ \\
APASS.V                 & $7.91\cdot10^{-14} \pm    1.2\cdot10^{-15}$ \\
SLOAN/SDSS.r        & $1.08\cdot10^{-13} \pm    4.4\cdot10^{-15}$ \\
GAIA/GAIA2.G         & $1.4670979389511\cdot10^{-13} \pm                      0$ \\
GAIA/GAIA2.Grp      & $2.1402666745903\cdot10^{-13} \pm                      0$ \\
WISE/WISE.W1       & $1.376\cdot10^{-14} \pm    7.9\cdot10^{-16}$ \\
WISE/WISE.W2       & $4.744\cdot10^{-15} \pm   9.2\cdot10^{-17}$ \\
AKARI/IRC.S9W      & $4.95\cdot10^{-16} \pm   2.1\cdot10^{-17}$ \\
WISE/WISE.W3       & $1.357\cdot10^{-16} \pm   2.0\cdot10^{-18}$ \\
WISE/WISE.W4       & $1.190\cdot10^{-17} \pm   5.2\cdot10^{-19}$ \\
\hline \hline
\end{tabular}
\end{table}

\subsubsection{\kostov\ approach}\applabel{kostovatmopar}

\kostov\ estimated \teff\ and \logg\ from two mostly independent derivations. \teff\ was derived using the Stefan-Boltzman law. The required bolometric luminosity was estimated from V- and K-band photometry using empirical bolometric correction relations \citep[][erratum]{pecaut2013a, mann2015a}. For the radius, they used $0.312 \pm 0.014\,\mathrm{R}_{\sun}$ derived from the mass-luminosity relation of \cite{benedict2016} and the mass-radius relation of \citet{boyajian2012}. \feh\ was derived from \textsc{sed} fitting \citep{stassun2017, stassun2016}. This procedure also yielded an estimate of \teff\  that was compatible within $1\,\sigma$ with the previous one, but was not preferred by the authors.

\subsubsection{Choice of the adopted set of atmospheric parameters}\applabel{choicespecpar}

\begin{table}[!htb]
\tiny
\caption{\tablabel{specpar}Different approaches to the spectroscopic parameters of L 98-59}
\raggedright
\begin{tabular}{p{0.30\columnwidth}P{0.15\columnwidth}P{0.18\columnwidth}P{0.16\columnwidth}}
\hline\noalign{\smallskip} 
& \teff\ [K] & \feh\ [dex] & \logg \\
\noalign{\smallskip} \hline\noalign{\smallskip} 
{\sc SteParSyn}         & $3415 \pm 60$                 & $-0.46 \pm 0.26$         & $4.86 \pm 0.13$ \\
\noalign{\smallskip}
{\sc ODUSSEAS}  & $3280 \pm 65$                 & $-0.34 \pm 0.10$      & -- \\
\noalign{\smallskip}
\textsc{vosa}             & $3362^{+140}_{-47}$ & $-0.24 \pm 0.51$      & $4.88 \pm 0.64$ \\
\noalign{\smallskip}
Stefan-Boltzman law + \textsc{SED} fitting ({\tiny \kostov}) &$3367 \pm 150$ & $-0.5 \pm 0.5$ & -- \\
\noalign{\smallskip}\hline
\end{tabular}
\tablefoot{The adopted estimates are provided in \tab{syspar}.\\
-- indicates that \logg\ is not estimated by these methods.}
\end{table}

\tab{specpar} compiles the four estimates of the spectroscopic parameters of L\,98-59\,A obtained with the four approaches presented above. It makes sense to separate them into two groups: the \vosa\ and \kostov\ estimates, which rely on the photometric \sed, on one side and the spectral synthesis and machine-learning estimates, which rely on the high-resolution \espresso\ spectra, on the other.
For \teff, the \sed\ based estimates are similar in terms of best values and uncertainties. They are both compatible within $1\,\sigma$ with the two \espresso- based estimates. However, the latter are 2.5 times more precise. The \espresso- based estimates provide similar uncertainties, but are only compatible at $2.25\,\sigma$. We do not currently know of any study that demonstrates the higher accuracy of one of the two \espresso- based approaches for M stars. Consequently, we did not exclude any of these estimates as an obvious outlier. However, judging from the data, the spectral synthesis and machine-learning uncertainties appear to be underestimated.
For \feh, the four estimates are compatible within $1.6\,\sigma$. As expected, the spectral synthesis and machine-learning methods provide more precise estimates with uncertainties up to five times better.
Finally, the two \logg\ estimates provided by the spectral synthesis and \vosa\ approaches are compatible within $1\,\sigma$. The spectral synthesis method provides a more accurate estimate because it uses data with high spectral resolution.

In this paper, which focuses on the characterization of the planets in system L 98-59, we need to conclude with one final set of \teff, \feh\ , and \logg\ estimates. In order to keep a physically self-consistent set of estimates, we decided to use the best values inferred by one method as the final best values for the three spectroscopic parameters. The use of high-resolution spectroscopy data, which offers the possibility of directly characterizing the chromospheric lines, is clearly an asset for inferring \feh\ and \logg\ compared to the use of the photometric \sed. The larger wavelength coverage toward the infrared offered by the \sed\ can provide important constraints for the inference of \teff. However, the \teff\ estimates provided by the \sed- based methods include the estimates of the methods based on high spectral resolution within $1\,\sigma$. We thus decided to use one of the two methods based on high spectral resolution to obtain our set of best values. We chose the spectral synthesis method because of the lack of a benchmark analysis demonstrating the accuracy of the relatively recent machine-learning approach and because it does not provide an estimate for \logg. The only exception was the uncertainties on the \teff\ , which we identified as underestimated. We chose to enlarge this uncertainty to encompass the best values provided by the other three methods within $1\,\sigma$, leading to the adopted values and uncertainties provided in \tab{syspar}.

\subsection{Stellar modeling: Mass, radius, and age}\applabel{starmod}

The derivation of the radius of L\,98-59\,A is already presented in detail in \sect{starmod}, but for the derivation of its mass, we again used several methods. The first method relies on our estimate of \logg\ (see \sect{spectro}), which combined with our radius estimates provides a mass of $0.241^{+0.097}_{-0.069}\,\MSun$.
The second method relies on the stellar density retrieved by \kostov\ from the fit of the transits of the three transiting planets. Combined with our radius estimates, it provides a mass of $0.311^{+0.10}_{-0.081}\,\MSun$.
Our third method relies on the mass-luminosity relation in K band of \citet{mann2019}. From the absolute K magnitude of $6.970 \pm 0.019\,\mathrm{mag}$, obtained from the observed magnitude provided by the 2MASS catalog \citep{cutri2003} and the distance provided by the Gaia collaboration \citep{bailer-jones2018}, we obtain a mass of $0.290 \pm 0.020\,\MSun$. 
The fourth approach is based on the recently published studies of M dwarfs by \citet[][see in particular Table 6]{cifuentes2020}. The authors performed a comprehensive analysis of 1843 nearby bright low-mass star using \sed\ photometry. They derived bolometric luminosities, effective temperature, radius, and mass for this sample. The masses are based on \citet{schweitzer2019}. They thus provide an equivalence between absolute bolometric luminosity, effective temperature, radius, and mass. Our bolometric luminosity estimate would indicate a radius of $0.343 \pm 0.082\,\RSun$ and a mass of $0.338 \pm 0.087 \,\MSun$. Our effective temperature estimate would indicate a radius of $0.433 \pm 0.086\,\RSun$ and a mass of $0.432 \pm 0.090\,\MSun$.
For our fifth approach, we used the \vosa\ online tools that are described in \app{vosa}. \vosa\ derives the stellar mass of $0.273 \pm 0.030 \,\MSun$ by comparing the measured \teff\ and bolometric luminosity to evolutionary tracks (BT-Settl model \citet{allard2012} for consistency with our analysis of the photometric \sed).
Finally, \kostov\ provided an estimate of the mass of $0.313 \pm 0.014\,\MEarth$ using the mass-luminosity relation for M dwarfs of \citet{benedict2016}. They derived the luminosity from K-band observations.

\begin{table}[!htb]
\tiny
\caption{\tablabel{physpar}Mass, radius, and density of L 98-59 derived with different approaches}
\raggedright
\begin{tabular}{p{0.30\columnwidth}P{0.22\columnwidth}P{0.22\columnwidth}P{0.08\columnwidth}}
\hline\noalign{\smallskip} 
Method & $M_*$  & $R_*$  & $\rho_*$\\
 & [\MSun] & [\RSun] & [$\rho_\sun$] \\
\hline\noalign{\smallskip} 
Stefan-Boltzmann law            &                                               & $0.303^{+0.026}_{-0.023}$       &  \\   
\noalign{\smallskip}
$\logg + R_*$                           & $0.241^{+0.097}_{-0.069}$     & //                                      &  $8.5^{+4.1}_{-2.1}$\\
\noalign{\smallskip}
$\rho_* + R_*$                          & $0.311^{+0.10}_{-0.081}$      & //                                      &  $11.2^{+2.1}_{-1.7}$\\
\noalign{\smallskip}
\textsc{vosa}                           & $0.273 \pm 0.030$                     & --                                              &  $9.8^{+3.1}_{-1.5}$\\
\noalign{\smallskip}
Cifuentes+20 ($f(L)$)           & $0.338 \pm 0.087$             & $0.343 \pm 0.082$                       & $8.2^{+11}_{-3.1}$ \\ 
\noalign{\smallskip}
Cifuentes+20 ($f(\teff)$)               & $0.432 \pm 0.090$             & $0.433 \pm 0.086$                        & $5.3^{+5.2}_{-1.7}$ \\ 
\noalign{\smallskip}
mass-lum \citep{mann2019}       & $0.290 \pm 0.020$                     & --                                              & $10.4^{+3.1}_{-1.5}$\\        
\noalign{\smallskip}    
\kostov                                         & $0.313 \pm 0.014$             & $0.312 \pm 0.014$               & $10.3^{+1.6}_{-0.89}$ \\
\noalign{\smallskip}\hline
\end{tabular}
\tablefoot{The adopted estimates are provided in \tab{syspar}.\\
// indicates that the radius estimate used as input of the method was provided by the Stefan-Boltzmann law.\\
-- indicates that the radius is not estimated by the method and that we used the estimate provided by the Stefan-Bolztmann law to compute the stellar density.}
\end{table}

\tab{physpar} gathers all these estimates of the radius and mass of L 98-59 A. From these, we also computed the resulting stellar densities through Monte Carlo simulations. We drew 100,000 samples of stellar mass and radius from normal distributions with mean and standard deviation as provided by the estimates from the corresponding row of \tab{physpar}. When the error bars were asymmetric, we used the average of the upper and lower uncertainties as standard deviation. From these 100,000 samples, we computed 100,000 stellar density values. We then computed the estimate of the stellar density using the 50th, 16th and 84th percentiles.
The relative precision on the stellar density provides us with a lower limit on the relative precision that we can achieve for the planetary density (see \tab{syspar}). The absolute value of the stellar density will also impact the measured planetary densities and thus is of particular interest for modeling their interior (see \sect{intcomp}). All stellar density estimates agree within $1\,\sigma$. However, the dispersion of best values shows that the one inferred from \citet{cifuentes2020} using the \teff\ is clearly off. The associated mass and radius are also significantly above all others. This might be due to the scale of the \citet{cifuentes2020} study. The table from which we derive our estimates is a summary of the properties of around 2000 stars, which might be relevant for a large sample, but might fail to accurately represent a specific case such as that of L 98-59 A. We thus discarded this estimate. We also note that the \citet{cifuentes2020} estimates that are based on the bolometric luminosity (instead of the \teff) agree well with the others.

The remaining radius estimates agree within $1\,\sigma$, but their uncertainties vary by a factor of up to $\sim 6$ between the \kostov\ estimate and the one from \citet{cifuentes2020} based on the bolometric luminosity. As already mentioned in the previous paragraph, because of the scale of the \citet{cifuentes2020} study, their uncertainties are probably overestimated. The uncertainties of the other two estimates differ by less than a factor two. We adopted the values derived from the Stefan-Boltzmann law because they are based on first principles.

The mass estimates also agree within $1\,\sigma$, but their uncertainties vary by a factor of up to $\sim 10$.
Compared with the dispersion of the best values, the \kostov\ uncertainty appears to be underestimated. The \logg\ and stellar density based values and the \citet{cifuentes2020} uncertainties, in contrast, appear to be overestimated. In between the two remaining estimates, \vosa\ and \citet{mann2019}, we adopted the estimate derived with \vosa. The \vosa\ tools provided \teff, \feh\ , and \logg\ values in good agreement with the value we adopted. We also used \vosa\ to derive the bolometric luminosity used to derive L\,98-59\,A radius. The \vosa\ mass estimate thus provides a physically consistent set of stellar parameters. The final set of adopted values and uncertainties is provided in \tab{syspar}.

To determine the age of L\,98-59\,A, we used the accurate photometry and distance provided by Gaia. We constructed the color-magnitude diagram shown in \fig{colormag}, where we also depict the well-known empirically determined mean sequences of stellar members of the $\beta$ Pictoris moving group ($\sim$20 Myr, \citealt{miret-roig2020}), the Tucana-Horologium moving group ($\sim$45 Myr, \citealt{bell2015}), the Pleiades open cluster ($\sim$120 Myr, \citealt{gossage2018}), and the field (possible ages in the range 0.8--10 Gyr). These sequences were taken from \citet{luhman2018} and \citet{cifuentes2020} and were derived by employing Gaia data; therefore the direct comparison with L\,98-59\,A is feasible without any systematic effect. From its location in the Gaia color-magnitude diagram, we infer that L\,98–59 likely has an age that is consistent with that of the field (our target lies below the mean field sequence of M dwarfs). We did not correct L\,98–59\,A data for interstellar extinction because based on its optical and infrared photometry (\tab{sed}) and optical HARPS and ESPRESSO spectroscopy, there is no evidence of strong or anomalous absorption. The field age is consistent with the measured mass and radius of the star, and the actual position of L\,98-59\,A below the bottom borderline of the $1\,\sigma$ dispersion of the field sequence also agrees with a slightly subsolar metallicity.
Finally, the kinematics of L\,98-59\,A can also provide indications about its age. Using the \rv\ systemic velocity, the Gaia parallax, the \textsc{ra}/\textsc{dec} coordinates and proper motions, we derived the UVW velocities of L\,98-59\,A (see \tab{kinematics}). L\,98-59\,A appears to belong to the thin disk and does not belong to any know young moving group. Therefore it is kinematically older than the oldest moving group currently known, that is, its age is older than 800 Myr.

\begin{table}[!htb]
\tiny
\caption{\tablabel{kinematics}Kinematics of L\,98-59\,A}
\raggedright
\begin{tabular}{p{0.45 \columnwidth} p{0.45 \columnwidth}}
\hline\noalign{\smallskip} 
U & $15.42 \pm 0.22$ \kms \\
V  & $10.31 \pm 1.06$ \kms \\
W  & $-2.59 \pm 0.34$ \kms \\
P(thick)  &  2 \% \\
P(thin)    &  98 \% \\
P(halo)  &  0 \% \\
Group membership & Thin disk \\
\noalign{\smallskip}\hline
\end{tabular}
\tablefoot{U, V, W are the three velocity components in the solar reference frame. P(thin), P(thick), and P(halo) are the probability of L\,98-59\,A of belonging to the thin disk, the thick disk, and the galactic halo, respectively.
}
\end{table}

\section{Measuring radial velocities and activity indicators}\applabel{rvs}

\tab{rvs} provides the measurements of the \rv s and activity indicators from the \espresso\ spectrograph used in this paper. For the \rv s and activity indicators measurements from the \harps\ spectrograph, we refer to \cloutier.

\begin{table*}[!htb]
\tiny
\centering
\caption{\tablabel{rvs} \espresso\ \rv, \fwhm, \bis, contrast, $S_{\textrm{index}}$, \Ha , \NaD, and \berv\\ measurements for L 98-59}
\begin{tabular}{lcccccccccc}
\hline
$\textrm{BJD}_{\textrm{TDB}}$ & \rv & $\sigma_{\rv}$ & \fwhm & $\sigma_{\fwhm}$ & \bis & $\sigma_{\bis}$ & Contrast & $\sigma_{\textrm{Contrast}}$ & \dots & Inst. \\
- 2 400 000                   & & & & & & & & & & \\
days                          & \ms & \ms & \kms & \kms & \ms & \ms & \dots & & & \\
\hline\\[-5pt]
58436.80567402998 & -5573.322521 & 0.803703 & 4499.159260 & 1.607407 & 20.138559 & 1.607407 & 42.799022 & 0.015291 & \dots & Pre \\
58444.83918777015 & -5576.670284 & 0.792801 & 4498.863898 & 1.585603 & 19.679371 & 1.585603 & 42.598489 & 0.015014 & \dots & Pre \\
58463.82528164983 & -5579.975907 & 0.646632 & 4507.424364 & 1.293264 & 20.459179 & 1.293264 & 42.763800 & 0.012270 & \dots & Pre \\
58470.77212886    & -5578.914555 & 0.657531 & 4503.113471 & 1.315063 & 16.920078 & 1.315063 & 42.714825 & 0.012474 & \dots & Pre \\
\multicolumn{11}{l}{The full table is available in electronic form at the CDS ....} \\
\hline \hline
\end{tabular}
\tablefoot{$\sigma_{X}$ represents the $1\,\sigma$ error bar measured for the quantity $X$.\\
Inst. stands for instrument and indicates whether a measurement has been taken before or after the technical intervention on \espresso\ (see \sect{espdata}).}
\end{table*}

\section{Rotational modulation in photometric time series}\applabel{protfromphotometry}

In order to address the presence of stellar activity induced modulation in the \tess\ data, we first attempted to fit the \lc\ with \gp\ and mean offsets for each sector.  
The \gp\ was implemented with the \texttt{celerite} Python package, as in \sect{tessdata}, but this time, the functional form of the kernel was designed to model quasi-periodic signal. Its equation is

\begin{equation}\eqlabel{rotkernel}
    k(\tau) = \frac{B}{2 + C} e^{-\tau / L} \left[\cos\left(\frac{2 \pi\,\tau}{P_{\mathrm{rot}}}\right) + ( 1 + C)\right],
\end{equation}

and it is taken from \citet[][eq. 56]{foreman-mackey2017}. $P_{\mathrm{rot}}$ is an estimator of the stellar rotation period, $L$ is the correlation timescale, $B$ is a positive amplitude term, and $C$ is a positive factor. We performed the fit by maximizing the log likelihood with \texttt{emcee}. We used 32 walkers. For each walker, we first maximized the log likelihood using the L-BFGS-B algorithm \citep{morales2011,zhu1997,byrd1995} implemented in the \texttt{scipy.optimize} Python package. Then we performed a first exploration of 5000 iterations followed by a second exploration of 10\,000 iterations starting at the last position of the first iteration.

The \post\ \pdf\ of the main hyperparameters ($B$, $L$ and $P_{rot}$) is presented in \fig{corner_gp_param}. The rotation period is poorly constrained ($190^{+189}_{-134}$ days). It is also worth noting that the retrieved amplitude and timescales are low and short: $0.11^{+0.05}_{-0.01}$ ppm for the amplitude, and $6.3^{+2.6}_{-1.0}$ days for the timescales. In particular, $\frac{L}{P_{rot}}$ appears too low to be physical because the timescale is expected to be on the same order of magnitude as or higher than the rotation period \citep{angus2018, haywood2014}. The timescale and the amplitude are strongly correlated. The short timescale and low amplitude can thus be tentatively explained by this degeneracy, which would result in a strong underestimation of both quantities. 

\begin{table}[!htb]
\caption{\tablabel{tessoffsets} Photometric offset derived for each \tess\ sector}
\begin{tabular}{lc}
\hline
Sector & Offset [\%] \\
\hline\\[-5pt]
2        &      $1.771^{+0.021}_{-0.022}$\\
5        &      $0.774^{+0.021}_{-0.022}$\\
8        &      $0.303^{+0.021}_{-0.020}$\\
9        &      $-0.007^{+0.023}_{-0.020}$\\
10       &      $-0.976^{+0.019}_{-0.019}$\\
11       &      $-2.170^{+0.017}_{-0.020}$\\
12       &      $-0.186^{+0.020}_{-0.020}$\\
28       &      $0.736^{+0.019}_{-0.022}$\\
29       &      $-0.847^{+0.021}_{-0.023}$\\
\hline \hline
\end{tabular}
\end{table}

Because the rotation period is only poorly determined, we used the \glsp\ as a more model-independent approach to determining the rotational modulation in the \tess\ \lc . \tess\ is designed for high-precision relative photometry (as opposed to high-precision absolute photometry). The photometry can thus show offsets between each sector that would strongly affect the \glsp\ of the \lc . We thus used the offsets derived from our \gp\ fit (see \tab{tessoffsets}) to realign the different sectors before we computed the \glsp. As the retrieved offsets are several orders of magnitude higher than the amplitude of the \gp\ signal, we can assume that they are independent of the exact model used to describe the stellar activity (hyperparameters and choice of kernel). The result of the \glsp\ analysis is presented in \fig{glsptess} and discussed in \sect{stelact}.

\begin{figure*}[!htb]
  \resizebox{\hsize}{!}{\includegraphics[]{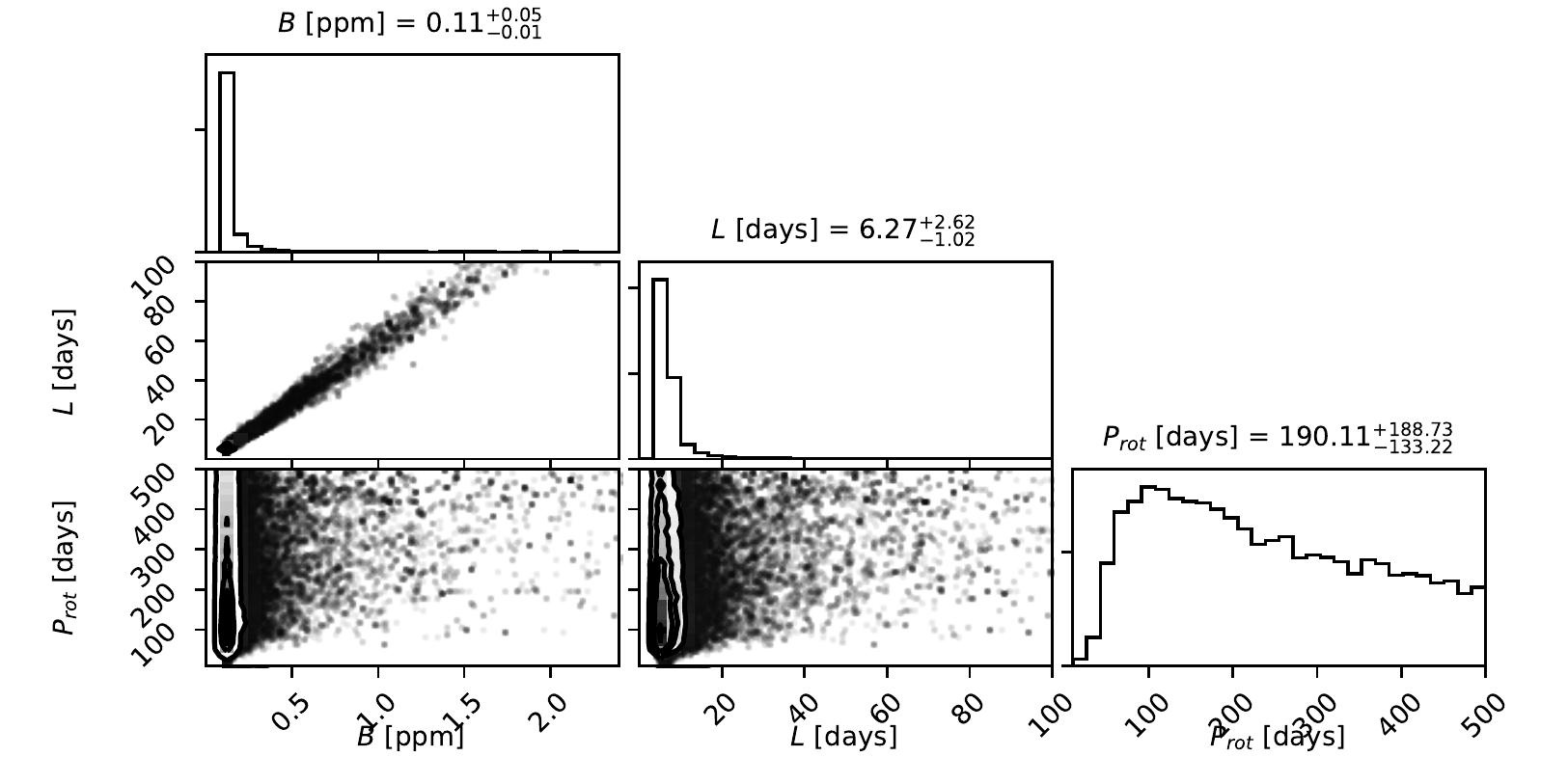}}
\caption{\figlabel{corner_gp_param}\post\ distributions of the main hyperparameters of the rotational kernel (\Eq{rotkernel})}
\end{figure*}

\section{Choice of priors}\applabel{priors}

The \prior\ \pdf\ used for the analyses described in Sects. \sectref{lconly}, \sectref{rvonly} and \sectref{finalana} are provided in \tab{syspar} (column \prior ). In this appendix, we explain the reasons for the choice of each \prior .

\subsection{\textit{Prior}s used for the \tess\ \lc\ analysis (\sect{lconly})}\applabel{priorslconly}

For the instrumental \prior , the \tess\ additive jitter term ($\sigma_{\textrm{\tess}}$), we adopted a uniform distribution between zero and five times the median value of the reported error bars.
The orbital parameters $e\cos\omega$ and $e\sin\omega$ were assigned a joint \prior . A joint \prior\ consists of a transformation between two sets of parameters to define the \prior\ on the new set of parameters instead. In this case, $e\cos\omega$ and $e\sin\omega$ were converted into $e$ and $\omega$. For the \prior\ \pdf\ of $e$, as recommended by \citet{kipping2013b}, we used a beta distribution with the following values for the two shape parameters: $a = 0.867$ and $b = 3.03$. For the \prior\ \pdf\ of $\omega$, we used a uniform distribution between $-\pi$ and $\pi$.
The remaining planetary parameters, $P$, $t_{\textrm{ic}}$, ${R_p / R_*}$ , and $\cos i_{p}$ , were also assigned a joint \prior . This joint \prior , which we call transiting \prior , also includes the stellar density $\rho_*$. Its main objective is to exclude regions of the parameter space where the three transiting planets are not transiting. It performs two changes of coordinates. It first computes the impact parameter ($b$) from $P$, $\rho_*$ , and  $\cos i_{p}$ (assuming a circular orbit), effectively converting the parameter $\cos i_{p}$ into $b$. Then it computes the orbital phase ($\phi$) from $P$ and  $t_{\textrm{ic}}$. For this conversion, we need to define a reference time that corresponds to $\phi = 0$. We chose this reference time to be the floored value of the first \espresso\ observation, $\mathrm{t_{ref}} = 1436$ BTJD. Then $t_{\mathrm{ic}} = \mathrm{t_{ref}} + P \phi$. We thus transformed the set of parameters $\rho_*$, $P$, $t_{\textrm{ic}}$, ${R_p / R_*}$ , and $\cos i_{p}$ into the new set of parameters $\rho_*$, $P$, $\phi$, ${R_p / R_*}$ , and $b$. To $\rho_*$, we assigned as \prior\ the \post\ of the \kostov\ analysis. To $P$, we assigned a Jeffreys distribution between 0.1 day and the time span of the \rv\ observations ($\sim520$ days). To avoid degenerate values of $t_{\mathrm{ic}}$ separated by a multiple of the period, we chose as \prior\ a uniform distribution between zero and one for $\phi$. For ${R_p / R_*}$, we assigned a uniform distribution between $10^{-3}$ and 1. For the \prior\ of $b$, we used a uniform distribution between 0 an 2 in order to allow grazing transiting, but we imposed the condition that $b < 1 + {R_p / R_*}$ to ensure that the configuration is transiting.

Finally, for the \prior\ on the limb-darkening coefficients, we used Gaussian \pdf s whose first two moments were defined using the Python package \texttt{ldtk}\footnotemark[11] \citep{parviainen2015a}.
Using a library of synthetic stellar spectra, it computes the limb-darkening profile of a star that is observed in a given spectral bandpass (specified by its transmission curve), and defined by its \teff, \logg\ , and \feh.
Provided the values and error bars for these stellar parameters (see \sect{spectro}) and the spectral bandpass of \tess , \texttt{ldtk} uses an \textsc{mcmc} algorithm to infer the mean and standard deviation of the Gaussian \pdf s for the coefficients of a given limb-darkening law (nonlinear in our case).
\texttt{ldtk} relies on the library of synthetic stellar spectra generated by \citet{husser2013}.
It covers the wavelength range from $500\,\mbox{\AA}$ to 5.5\,$\mu$m  and the stellar parameter space delimited by $2\,300\,\mathrm{K} \leq \teff \leq 12\,000\,\mathrm{K}$, $0.0 \leq \logg \leq +6.0$, $-4.0 \leq \feh \leq +1.0$, and $-0.2 \leq \afe \leq +1.2$. This parameter space is well within the requirements of our study (see \tab{specpar}).

\subsection{\textit{Prior}s used for the \rv\ analysis (\sect{rvonly})}\applabel{priorsrvonly}

Regarding the instrumental \prior s, the \prior\ \pdf\ of the offsets between the \rv\ instruments ($\Delta \mathrm{RV}_{\harps/\mathrm{pre}}$, $\Delta \mathrm{RV}_{\mathrm{post/pre}}$) are Gaussian distributions with means equal to the difference of the median values of the data sets and variances equal to the sum of their variances. The \prior\ \pdf s of the constant levels of the \fwhm\ ($C_{\mathrm{pre}}$, $C_{\mathrm{post}}$, and $C_{\harps}$) are Gaussian distributions with means equal to the median values of each data set and variances equal to their variances. The \prior\ \pdf\ of the additive jitter parameters ($\sigma_{\rv, \mathrm{pre}}$, $\sigma_{\rv, \mathrm{post}}$, $\sigma_{\rv, \harps}$, $\sigma_{\fwhm, \mathrm{pre}}$, $\sigma_{\fwhm, \mathrm{post}}$, and $\sigma_{\fwhm, \harps}$) are uniform distributions between zero and five times the median values of the reported error bars for each data set.

Regarding the star related \prior s, the \prior\ \pdf\ of the systemic velocity ($v_0$) is a Gaussian with the mean equal to the median value of the \rv\ data taken by \espresso\ before the fiber change and a variance equal to its variance. The other parameters are the hyperparameters of the quasi-periodic kernels. The \prior\ \pdf s of the two amplitudes ($A_{\rv}$, $A_{\fwhm}$) are uniform between zero and the maximum of the peak-to-peak values of the joint data sets taken by the three instruments. For the period of recurrence (\Prot), the \prior\ \pdf\ chosen is a Jeffreys distribution between 5 days and the time span of our observations ($\sim520$ days). Considering the age and the spectral type of L\,98-59, 5 days appears to be a good lower limit for the rotation period.
This \prior\ comfortably encompasses the estimate of $\sim 80$ days made by \cloutier\ based on the periodogram of the $H_{\alpha}$ measurements.
For the decay timescale ($\tau_{\mathrm{decay}}$), we chose a Jeffreys distribution between 2.5 days and five times the time span of observations. This upper limit was set to prevent the \gp\ from producing stellar activity models that would be completely coherent over the time span of our observations. In other words, we imposed that the stellar activity signal is quasi-periodic and not periodic. The objective was to avoid that the \gp\ reproduces planetary signals. 
Furthermore, we imposed that the decay timescale was superior to half of the period of recurrence. This condition, suggested by \citet{angus2018} and \citet{haywood2014}, prevents the \gp\ from producing stellar activity signals that are too incoherent and thus close to white noise. In these cases, the \gp\ signal and the additive jitter terms start to become degenerate.
The \prior\ \pdf\ of the periodic coherence scale ($\gamma$) was uniform between 0.05 and 5. The typical value for $\gamma$ in the literature is thought to be 0.5 \citep{dubber2019}. This \prior\ is designed to explore one order of magnitude below and above this typical values.
{\symbolfootnotes
Regarding the planetary \prior s, the \prior\ \pdf\ of $K$ is uniform between 0 and the maximum of the peak-to-peak values of the \rv\ data sets taken by the three instruments. 
For the ephemeris parameters ($P$ and $t_{\mathrm{ic}}$) and for the three known transiting planets, we used as \prior s the \post s of our analysis of the \tess\ \lc\ (see notes \footnotemark[3] at the end of \tab{syspar}).
}

For the nontransiting planets that we identified in the \glsp, we used a joint \prior. This joint \prior\ converts $P$ and $t_{\mathrm{ic}}$ into $P$ and $\phi$ , similarly to what was done within the transiting joint \prior\ in \app{priorslconly}. The reference time used, which corresponds to $\phi = 0$, is the same ($\mathrm{t_{ref}} = 1436$ BTJD). We chose a uniform distribution between zero and one for $\phi$.
For $P$, we used a Jeffreys distribution between 0.1 day and the time span of the \rv\ observations ($\sim520$ days).
Finally, the last two parameters are $e\cos\omega$ and $e\sin\omega$. We used the same joint \prior\ as in \app{priorslconly} , which results in a beta distribution with shape parameters $a = 0.867$ and $b = 3.03$ for the \prior\ \pdf\ of $e$ \citep{kipping2013b}, and a uniform distribution between $-\pi$ and $\pi$ for $\omega$.

\subsection{\textit{Prior}s used for the joint analysis of the \rv\ and \lc\ data (\sect{finalana})}

The \prior s used for this analysis are the same as were used for the analysis of the \tess\ \lc\ (see \app{priorslconly}). For the parameters that are not present in this analysis, we used the same \prior s as we used for our analysis of the \rv\ data (see \app{priorsrvonly}). All \prior s are mentioned in \tab{syspar}.

\section{Searching for the transits of planet e and planetary candidate 5}\applabel{transitsearch}

\begin{figure}[!htb]
    \centering
    \subfloat[Planet e]{\includegraphics[width=\columnwidth]{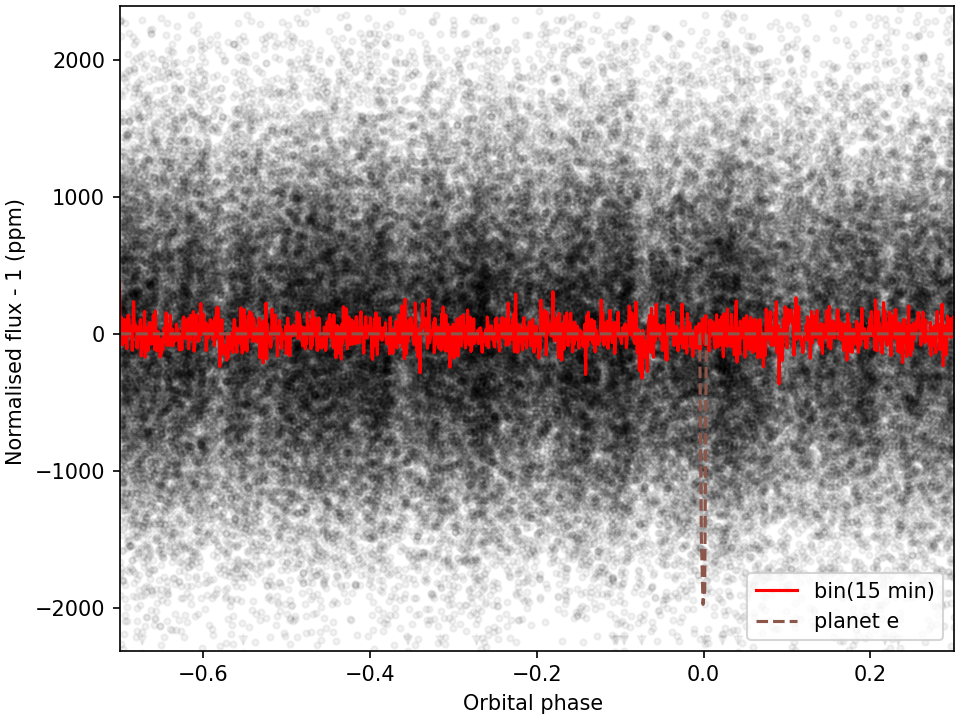}}
    \quad
    \subfloat[Candidate 5]{\includegraphics[width=\columnwidth]{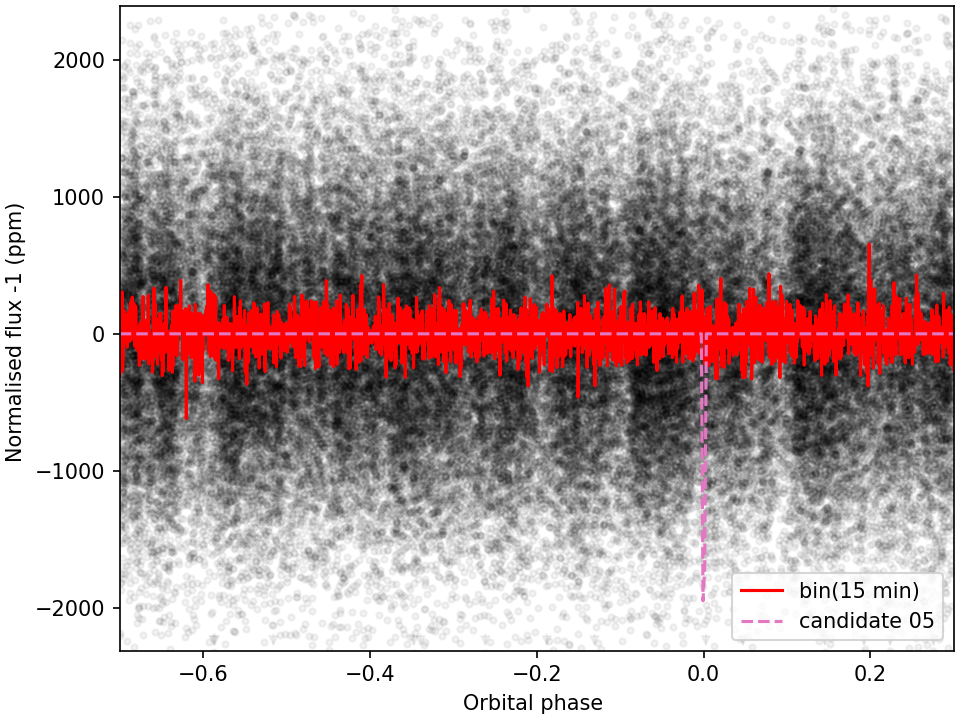}} 
    \caption{\figlabel{notransit}Phase-folded \tess\ \lc\ assuming the best model ephemerides of planet e (a) and planetary candidate 5 (b). The black points are the \tess\ data points at the original cadence. The red line is the data binned in phase using bins of 15 min. The dashed pink and brown lines are the expected transit signal assuming that the planets have the same radius as planet d (see \tab{syspar}).}
\end{figure}

We searched the \tess\ data for previously unreported planetary transit signal including planet e and planetary candidate 5. We used a procedure similar to \citet{barros2016}.
For this analysis, we did not use the \lc\ detrended with a \gp\ described in \sect{tessdata} because the flexibility of the \gp\ might alter the transit signals. Instead we detrended each sector separately by dividing the \lc\ by a spline interpolation of third degree. We used a knot every 0.5 days. Combined with an iterative $3\,\sigma$ clipping to identify outliers, this allowed us to better preserve unidentified transits signals in the detrended \lc\ \citep{barros2016}.
Then we removed the transits of the three known transiting planets by cutting out data within a window of two transit durations centered on the predicted transit time. The additional 0.5 transit duration before and after transit allows accounting for errors in the ephemerides or unknown transit-timing variations. We then performed a box least-squares (BLS) search \citep{kovacs2002} to find periodicities between 0.5 and 40 days.
The resulting periodogram is shown in \fig{blsperiodo} , with the highest peak corresponding to 1.049 days, which is probably due to aliases linked to the rotation of Earth. Phase-folding the \lc\ at this period does not show a typical transit signature. No other significant peaks are seen in the BLS periodogram, including at the periods of the candidate planets detected in RV (see \sect{modselect}). 
We also performed a transit search using the \textsc{tls} software \citep{hippke2019} and obtained the same conclusion.

To confirm the absence of transit signal for planet e and planetary candidate 5, we phase-folded the TESS \lc\ using the ephemeris of \tab{syspar}. We do not observe any transit signal in either case. We show that if the planetary radii are similar to those of the other transiting planets, the transit signal would have been clear in the \tess\ \lc.

\begin{figure}[!htb]
    \resizebox{\hsize}{!}{\includegraphics[]{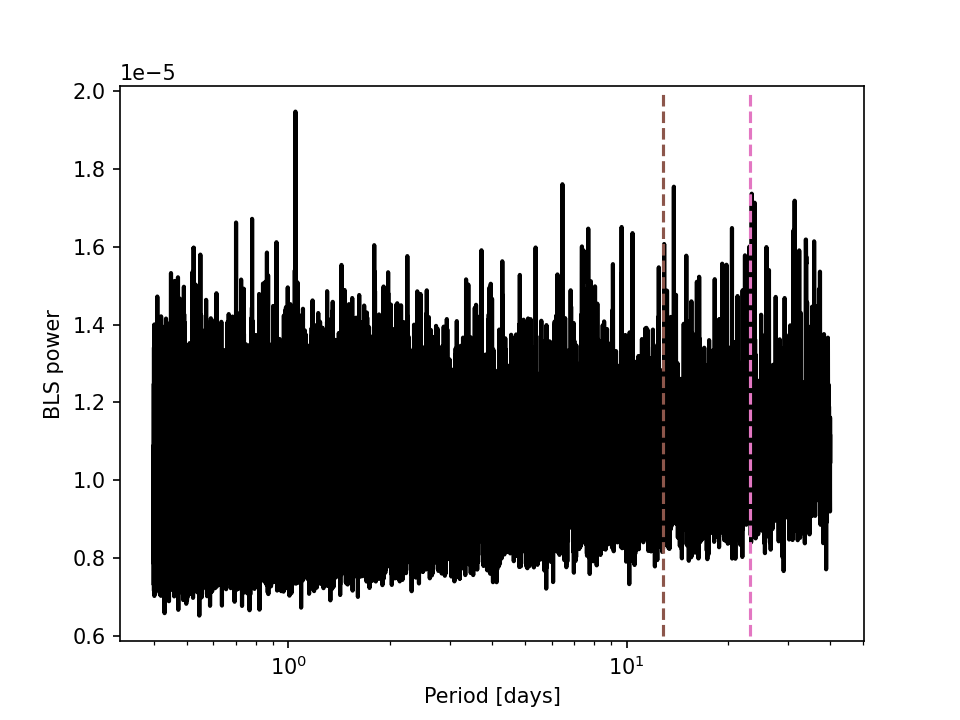}}
    \caption{\figlabel{blsperiodo}Periodogram provided by the BLS search in the \tess\  data. The dashed vertical pink and brown links indicate the orbital period of planet e and planetary candidate, respectively. There is no significant power at these periods.}
\end{figure}

\section{Evidence for additional planets in the L 98-59 system}\applabel{modelselect}

\begin{figure*}[!htb]
  \resizebox{0.9 \hsize}{!}{\includegraphics[]{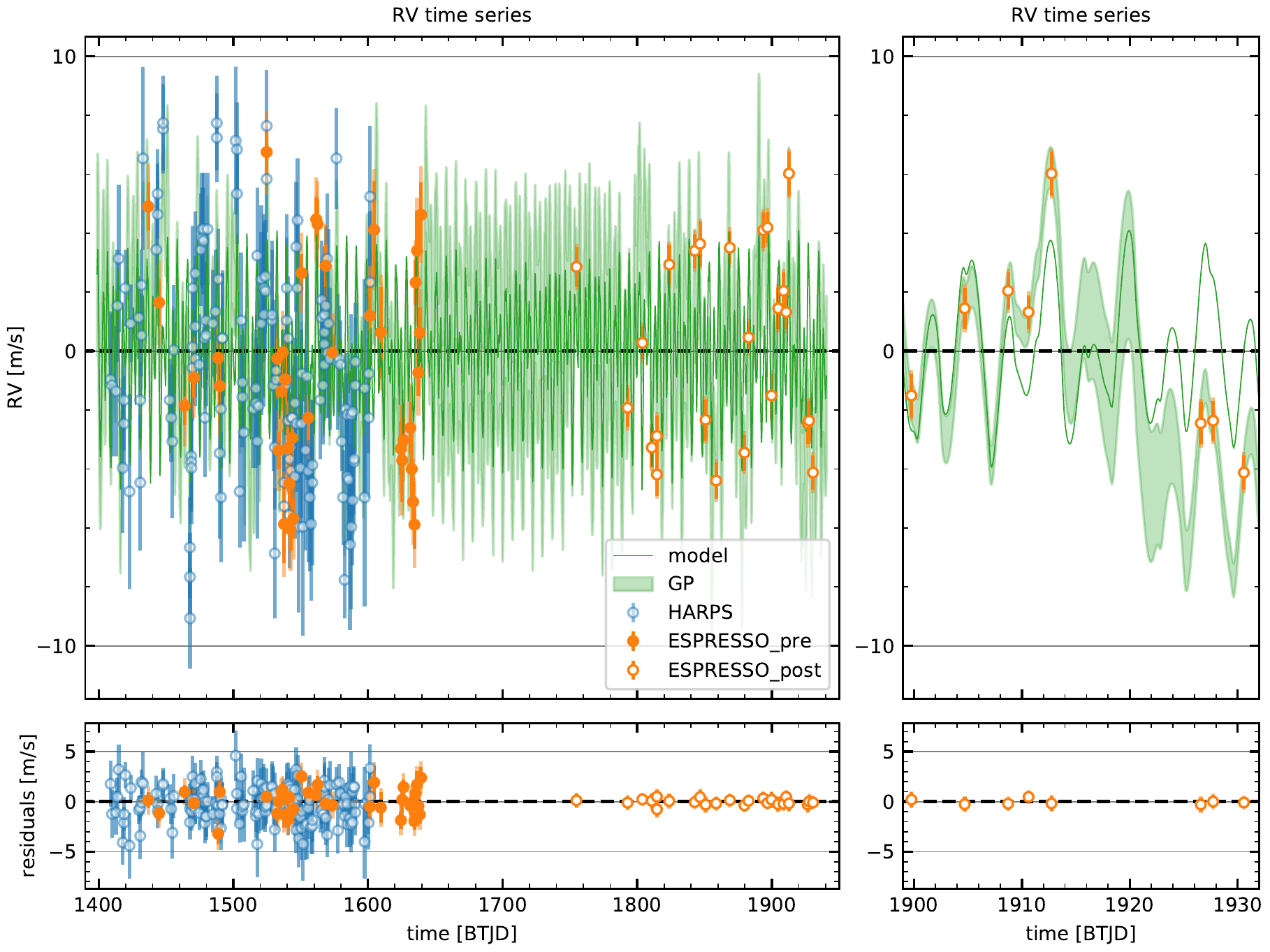}}
  \caption{\figlabel{3plTS} Outcome of the fit of the three-planet model: The format of this figure is identical to that in \fig{4plTS}, but is described again here for convenience. (Top left) \rv\ time series along with the best model (solid green line) that includes the planetary signals and best prediction from the \gp\ stellar activity model. The $1\,\sigma$ uncertainties from the \gp\ prediction are also displayed (shaded green area). For this plot, we subtracted  the systemic velocity and the instruments offsets from the \rv\ data (see values in \tab{syspar}). (Bottom left) Time series of the residuals of the best model. (Right) Zoom on a small portion of the time series to better visualize the short-timescale variations.}
\end{figure*}

\begin{figure*}[!htb]
  \resizebox{0.9 \hsize}{!}{\includegraphics[]{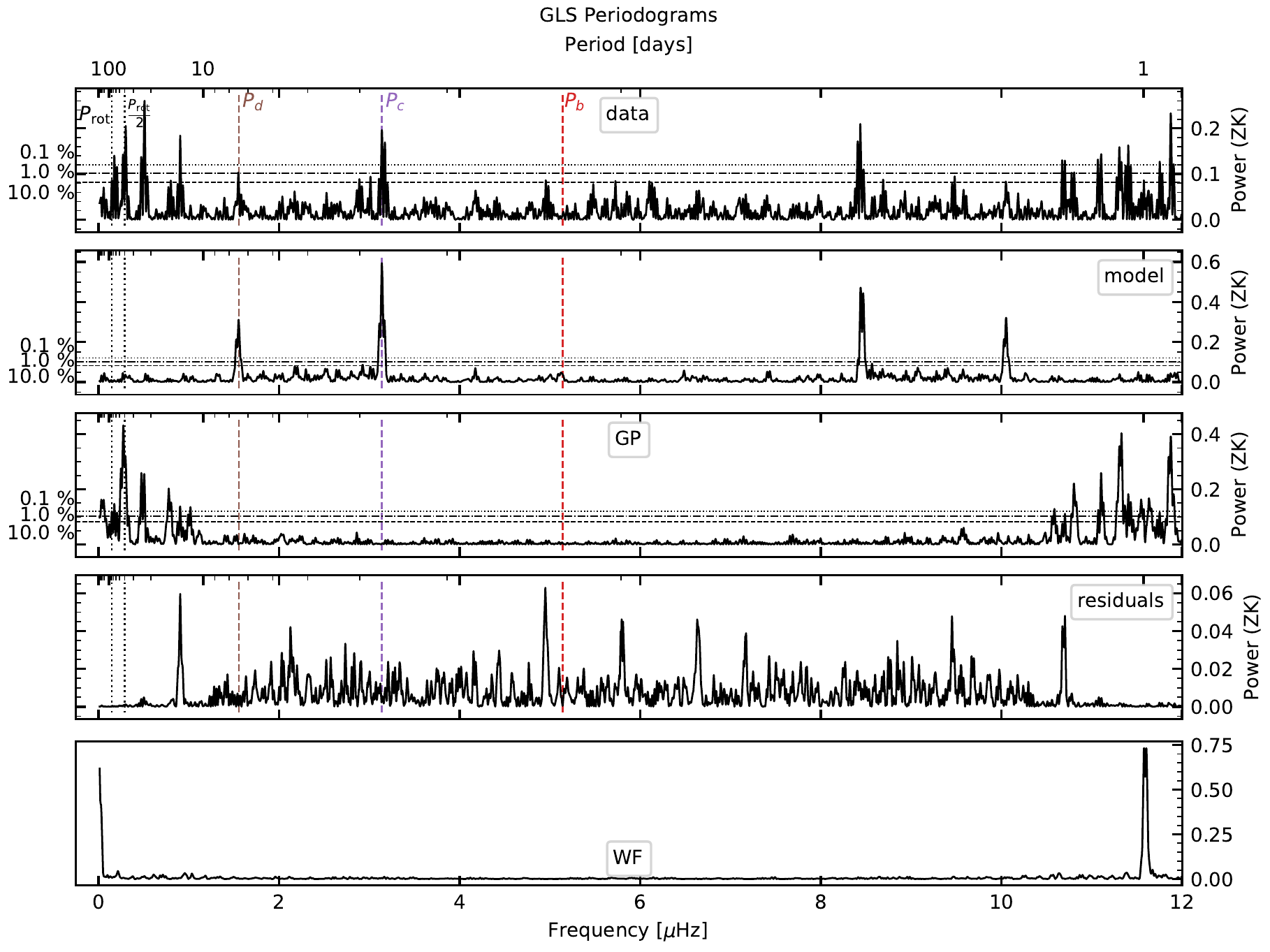}}
  \caption{\figlabel{3plGLS} Outcome of the fit of the three-planet model: The format of this figure is identical to that in \fig{4plGLS}, but is described again here for convenience. \glsp s of the \rv\ time series (top) and of the planetary (second) and stellar activity (third) models sampled at the same times as the \rv\ data. \glsp\ of the time series of the residuals (fourth) and the window function (bottom). The vertical lines on the \glsp s correspond to the orbital periods of planets b, c, and d, half and the full rotation period (estimated at 80 days) from right to left.}
\end{figure*}

As mentioned in \sect{modselect}, in order to assess the presence of additional planets in the L\,98-59 system, we first performed the two analyses that included only the three previously known planets. Figures \figref{3plTS} and \figref{3plGLS} follow the same format as Figs. \figref{4plTS} and \figref{4plGLS}. Figure \figref{3plTS} shows the \rv\ time series including the data from both instruments, the best three planets plus activity model, and the residuals of this fit. Figure \figref{3plGLS} displays the \glsp\ of the combined \rv\ data and the residuals, the \glsp s of the planetary and stellar activity model sampled at the same times as the \rv\ time series, and the \wf. 
The \glsp\ of the combined \rv s in \fig{3plGLS} shows two narrow peaks with an \fap\ below 0.1 \% at the two periods, 13 and 23 days, which were previously identified as potential additional planetary signals. The \glsp\ of the residuals displays a narrow peak at 13 days.

\begin{figure*}[!htb]
  \resizebox{0.9 \hsize}{!}{\includegraphics[]{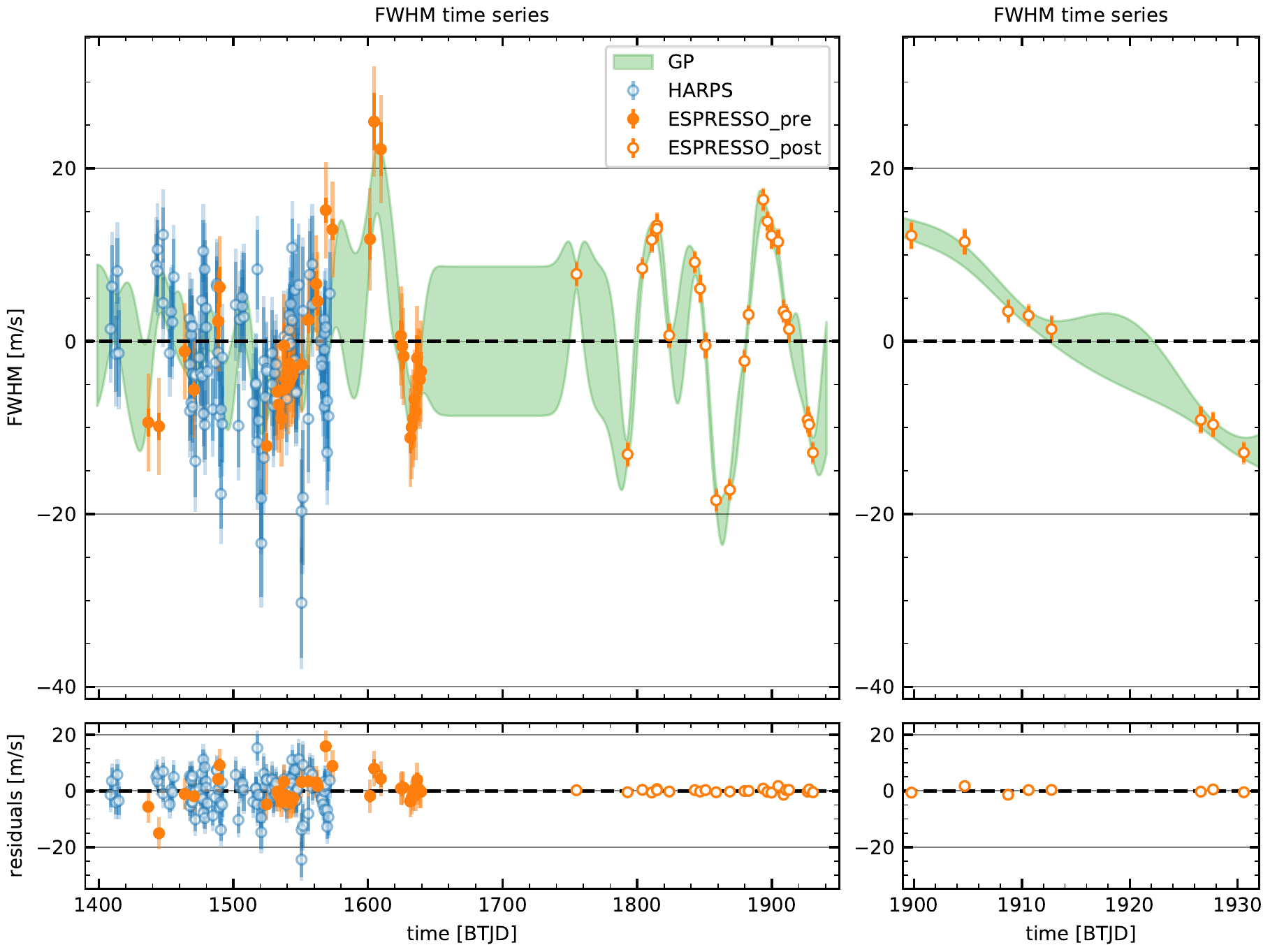}}
  \caption{\figlabel{4plTSFWHM} Outcome of the fit of the four-planet model regarding the \fwhm: The structure of this figure is similar to that of \fig{4plTS} or \fig{3plTS} , but the \fwhm\ data and model are displayed instead of the \rv\ data and model.}
\end{figure*}

\begin{figure*}[!htb]
  \resizebox{0.9 \hsize}{!}{\includegraphics[]{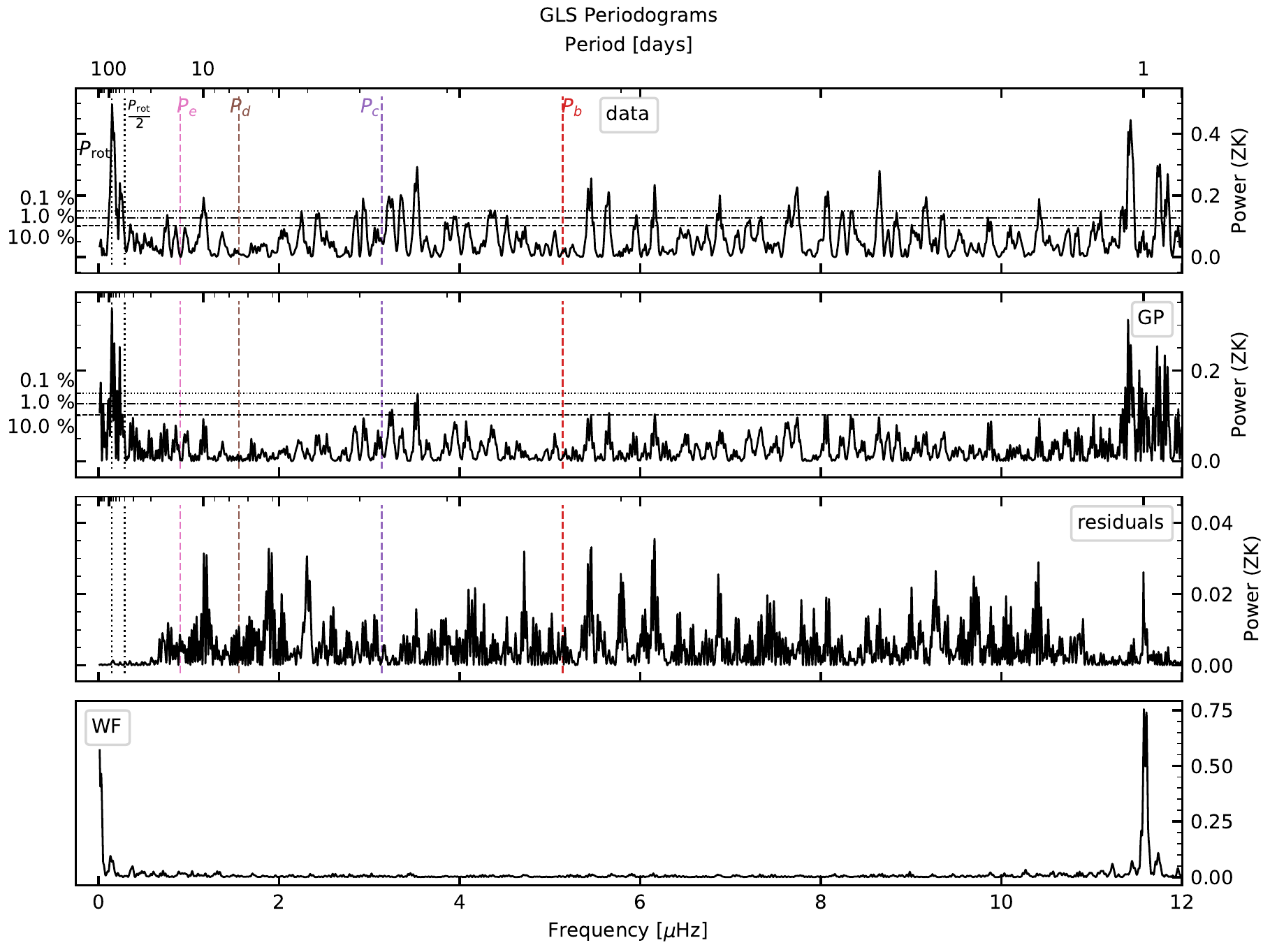}}
  \caption{\figlabel{4plGLSFWHM} Outcome of the fit of the four-planet model regarding the \fwhm: The structure of this figure is similar to that of \fig{4plGLS} or \fig{3plGLS} , but the \fwhm\ data and model are displayed instead of the \rv\ data  and model.}
\end{figure*}

The analyses with four planets converges toward a significant detection of the semi-amplitude of a fourth Keplerian signal. Similarly to \fig{3plTS},  \fig{4plTS} shows the time series and the best model, and \fig{4plGLS} shows the \glsp s. We also performed an iterative \glsp\ analysis in \fig{4plIGLS}. This allows showing the peak on the \glsp\ that corresponds to planet b, which is invisible in other figures. The \glsp\ of the residuals after the subtraction of the model for planet b (also shown in \fig{4plGLS}) shows two peaks around 1.743 and 2.341 days and no peak around 23 days. The analysis of the \tess\ \lc\ did not show transit signals at 1.743 or 2.341 days, so that we did not pursue a planetary origin for these peaks. However, the \glsp\ of the activity model does show a peak around 23 days. This indicates that the signal at 23 days might be generated by stellar activity. Based on our stellar activity model, which analyzes the \fwhm\ data simultaneously with the \rv\ data, we can also analyze the behavior of this activity indicator. Figures \figref{4plTSFWHM} and \figref{4plGLSFWHM} show similar information to that in Figs. \figref{4plTS} and \figref{4plGLS}, but for the \fwhm\ data. There is no significant power around 23 days in the \glsp\ of the combined \fwhm\ data or in those of the stellar activity model and the residuals. Similarly, the \glsp s of all the other activity indicators (see Figs. \fig{actindgls} and \fig{glsptess}) do not display significant power around 23 days. The analysis of the activity indicator does not confirm the stellar activity origin of the 23-day signal.

\begin{figure*}[!htb]
  \resizebox{0.9 \hsize}{!}{\includegraphics[]{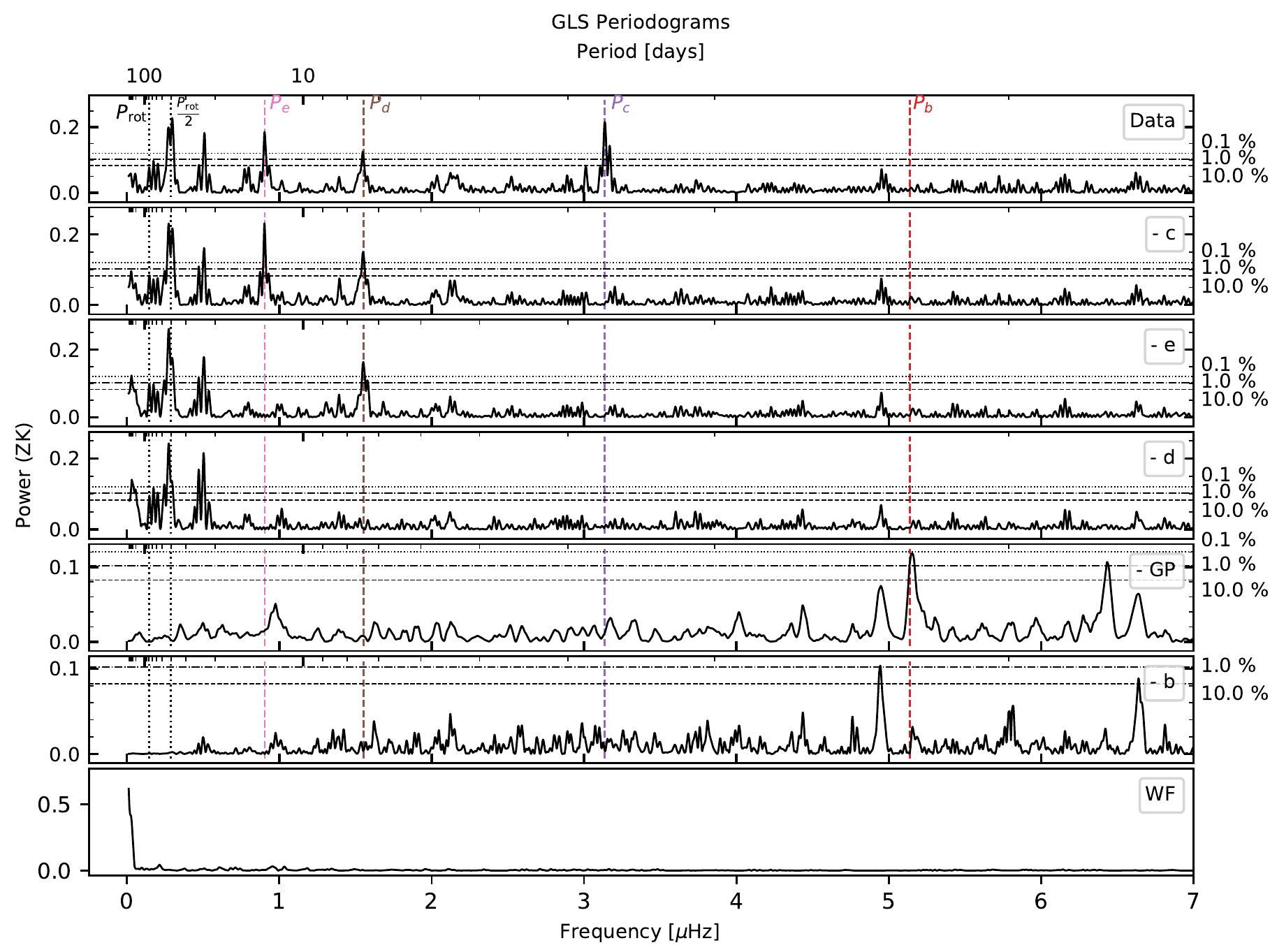}}
  \caption{\figlabel{4plIGLS} Iterative \glsp\ for the four planets model: \glsp\ of the \rv\ data (top) and the window function (bottom). The \glsp s of the data shown in the previous row minus the model for planets c, e, and d, the stellar activity model, and the model for planet b are displayed in the second, third, fourth, fifth, and sixth row, respectively.}
\end{figure*}

Consequently, we performed other analyses with five planets. The fits converge toward a significant detection of the semi-amplitude of a fifth Keplerian signal. Figures \figref{5plTS} and \figref{5plGLS} show the time series, the best model, and the \glsp s. The \glsp\ of the stellar activity model still displays power around 23 days, but it is less significant and has a much more flattened profile compared with the four-planet analyses (\fig{4plGLS}).

\begin{figure*}[!htb]
  \resizebox{0.9 \hsize}{!}{\includegraphics[]{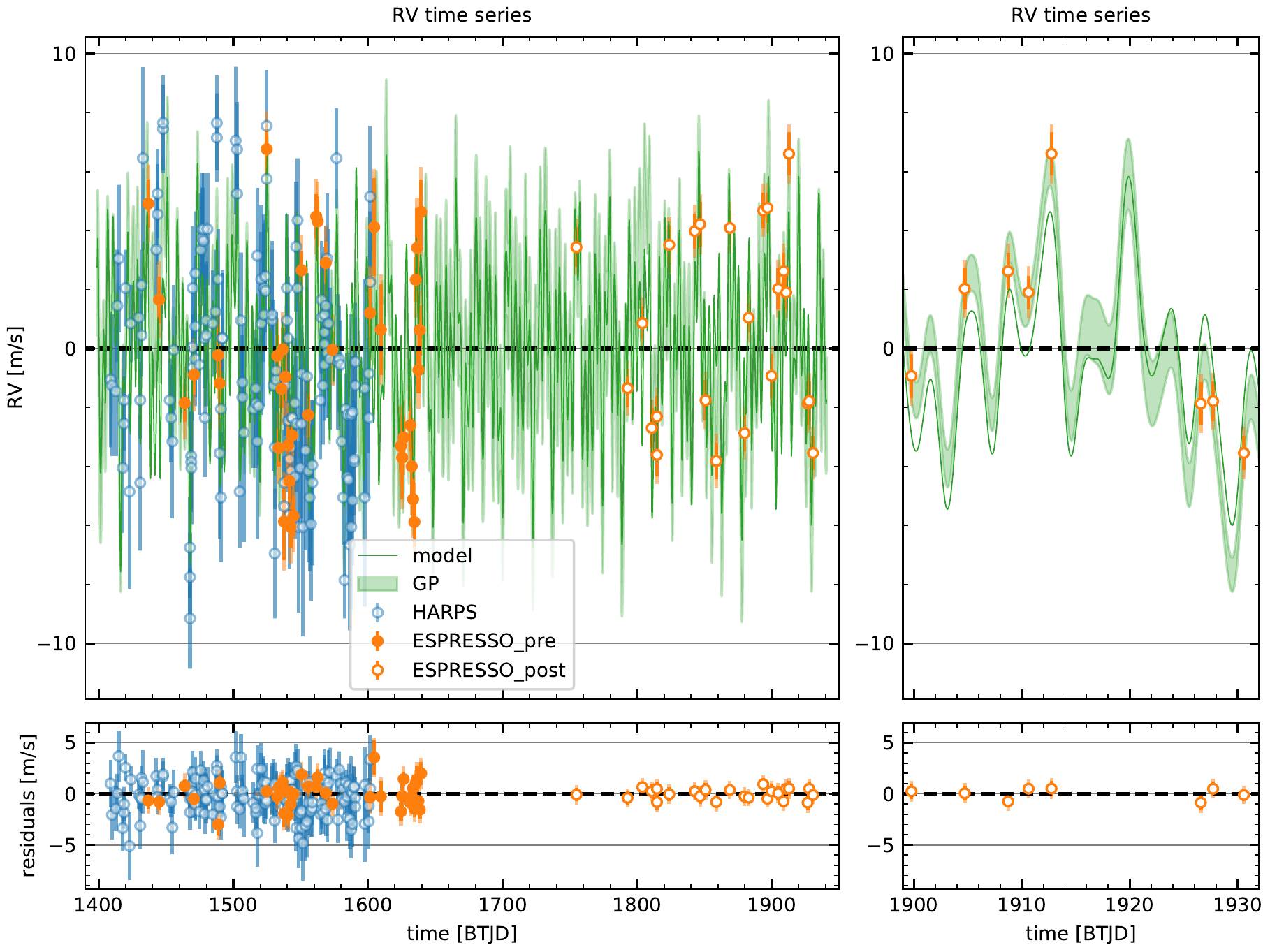}}
  \caption{\figlabel{5plTS} Outcome of the fit of the five-planet model: The format of this figure is identical to that in Figs. \figref{4plTS} and \figref{3plTS}.}
\end{figure*}

\begin{figure*}[!htb]
  \resizebox{0.9 \hsize}{!}{\includegraphics[]{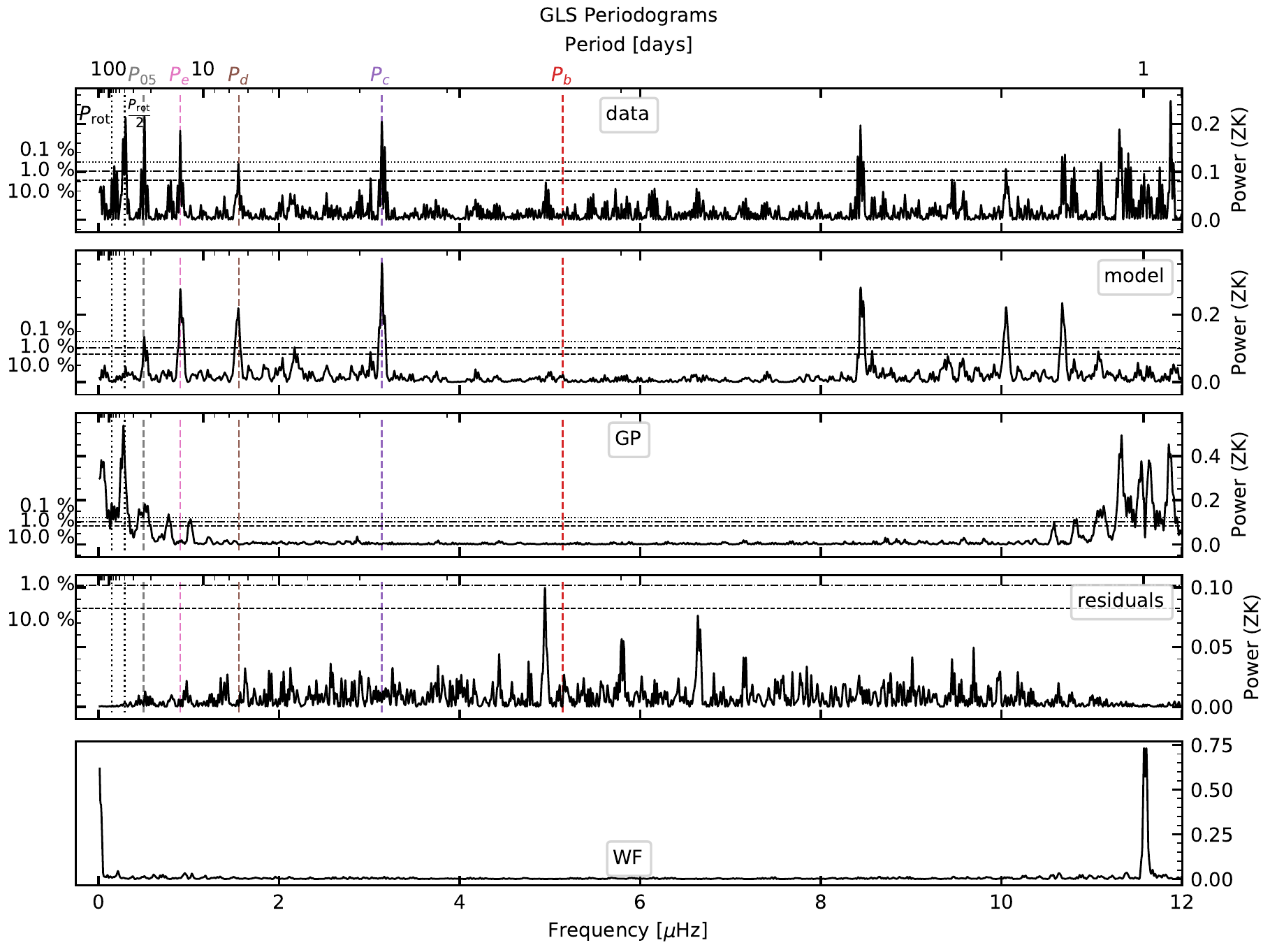}}
  \caption{\figlabel{5plGLS} Outcome of the fit of the five-planet model: The format of this figure is identical to that in Figs. \figref{4plGLS} and \figref{3plGLS}.}
\end{figure*}

\section{Internal composition of three transiting super-earths}\applabel{intcomp}

As explained in \sect{intcomp}, our framework for modeling the interior of the three transiting planets is composed of a forward model and a Bayesian retrieval.
In the forward model, each planet is made of four layers: an iron or sulfur inner core, a mantle, a water layer, and a gas layer. We used for the core the equation of state (EOS) of \citet{Hakim}, for the silicate mantle, the EOS of \citet{Sotin}, and the water EOS was taken from \citet{Haldemann_20}. These three layers constitute the solid part of the planets. The thickness of the gas layer (assumed to be made of pure H and He) was computed as a function of the stellar age, mass, and radius of the solid part and irradiation from the star using the formulas of \citet{lopez2014}. 

In the Bayesian analysis part of model, we proceeded in two steps. We first generated 150000 synthetic stars, their mass, radius, effective temperature,  age, and composition ([Si/H], [Fe/H] , and [Mg/H]), as well as the associated error bars, which were taken at random following the stellar parameters quoted above. For each of these stars, we generated 1000 planetary systems for which we varied the internal structure parameters of all planets, and we assumed that the bulk Fe/Si/Mg molar ratios are equal to the stellar ratios. We then computed the transit depth and RV semi-amplitude for each of the planets and retained models that fit the observed data within the error bars. 
With this procedure, we included the fact that all synthetic planets orbit a star with exactly the same parameters. Planetary masses and radii are correlated by the fact that the fitted quantities are the transit depth and RV semi-amplitude, which depend on the stellar radius and mass. In order to take  this correlation into account, it is therefore important to fit the planetary system at once, and not each planet independently.

The \prior s used in the Bayesian analysis are the following: The mass fraction of the gas envelope is uniform in log, the mass fraction (relative to the solid planet, i.e., excluding the mass of gas) of the inner core, mantle, and water layer are uniform on the simplex (the surface on which they add up to one). Finally, we constrain the mass fraction of water to be 50 $\%$ at most \citep{Thiabaud,Marboeuf}. The molar fraction of iron in the inner core is uniform between 0.5 and 1, and the molar fraction of Si, Mg, and Fe in the mantle is uniform on the simplex (they add up to one). 

The \post\ distributions of the most important parameters (mass fractions and composition of the mantle) of each planet in L\,98-59 are shown in Figs. \figref{cornerb} to \figref{cornerd}.

 \begin{figure*}[!htb]
  \resizebox{\hsize}{!}{\includegraphics[]{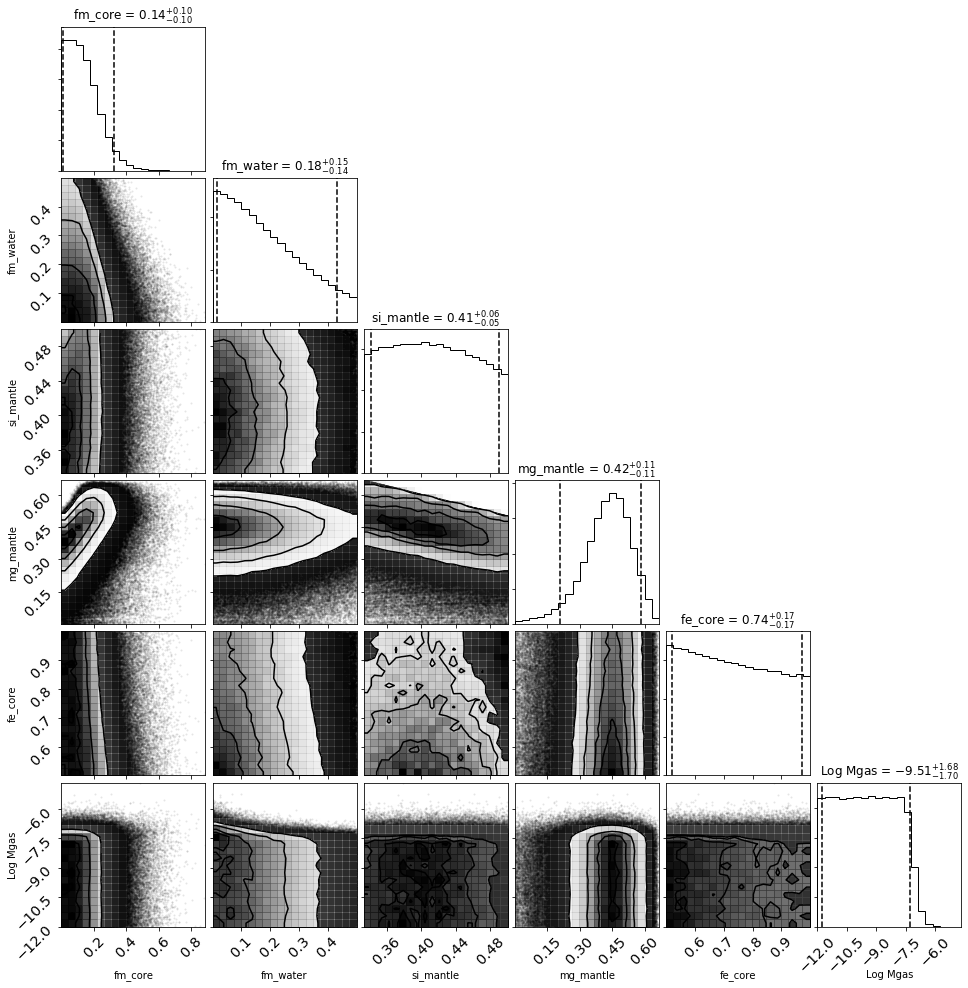}}
  \caption{Corner plot showing the main internal structure parameters of L\,98-59\,b. We show the mass fraction of the inner core, the mass fraction of water, the Si and Mg mole fraction in the mantle, the Fe mole fraction in the inner core, and the mass of gas (log scale). The values at the top of each column are  the mean and 5\% and 95\% quantiles.}
  \figlabel{cornerb}
\end{figure*}

\begin{figure*}[!htb]
  \resizebox{\hsize}{!}{\includegraphics[]{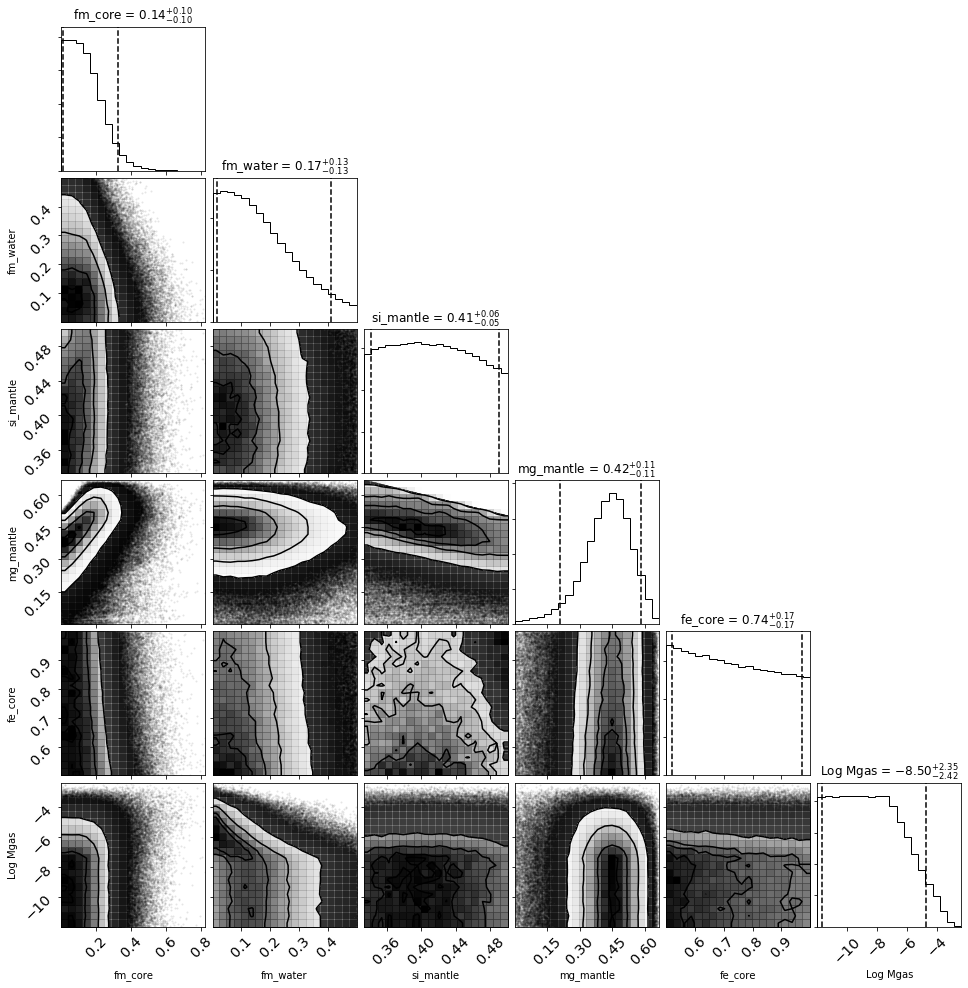}}
  \caption{Same as Fig. \ref{fig:cornerb} for L\,98-59\,c.}
  \figlabel{cornerc}
\end{figure*}

\begin{figure*}[!htb]
  \resizebox{\hsize}{!}{\includegraphics[]{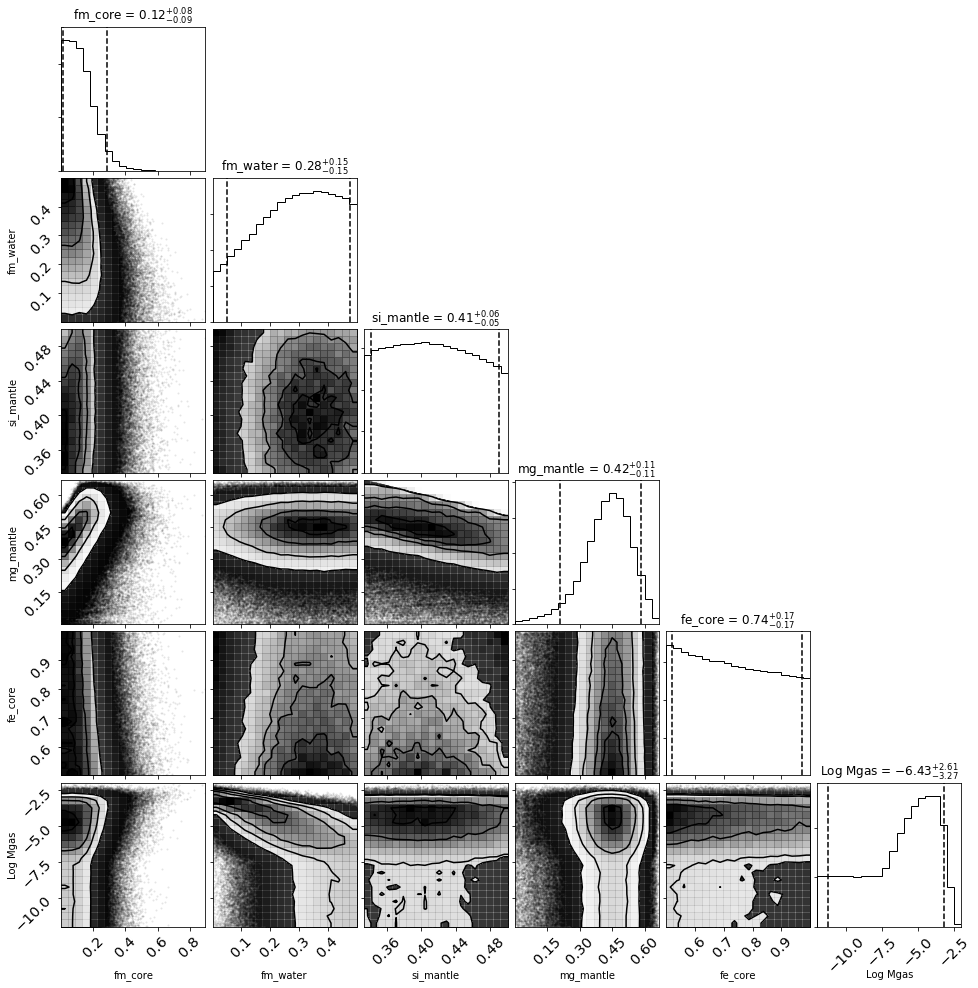}}
  \caption{Same as Fig. \ref{fig:cornerb} for L\,98-59\,d.}
  \figlabel{cornerd}
\end{figure*}

\end{appendix}

\clearpage
\onecolumn
\setcounter{table}{2}
\begin{symbolfootnotes}
\begin{raggedleft}
\begin{longtable}{p{0.25\textwidth}P{0.30\textwidth}L{0.25\textwidth}L{0.15\textwidth}}
\caption{\tablabel{syspar} \textbf{Parameter estimates of the planetary system L 98-59}}\\
\hline\noalign{\smallskip}
 & Posterior & Prior & Source\\
\noalign{\smallskip}\hline\noalign{\smallskip}
\endfirsthead
\multicolumn{4}{c}%
{\tablename\ \thetable\ -- \textit{Continued from previous page}} \\
\hline\noalign{\smallskip}
 & Posterior & Prior & Source\\
\noalign{\smallskip}\hline\noalign{\smallskip}
\endhead
\hline \multicolumn{4}{r}{\textit{Continued on next page}} \\
\endfoot
\hline\\
\endlastfoot

\multicolumn{4}{l}{\textit{Planetary parameters}} \\
\noalign{\smallskip}\hline\hline\noalign{\smallskip}
                                                            & \multicolumn{3}{l}{\textit{Planet b}} \\
$M_p$ [\MEarth]                                         & $0.40_{-0.15}^{+0.16}$ &  & \\ \noalign{\smallskip}
$R_p$ [\REarth]                                         & $0.850_{-0.047}^{+0.061}$ &  & \\ \noalign{\smallskip}
$\rho_p$ [$\mathrm{g.cm^{-3}}$]                         & $3.6_{-1.5}^{+1.4}$ &  & \\ \noalign{\smallskip}
$T_{\textrm{eq}}$ [K]                                   & $627_{-36}^{+33}$  &   &  \\ \noalign{\smallskip}
${P}^{\ \bullet}$\ [days]                               & $2.2531136_{-1.5e-06}^{+1.2e-06}$  & $\mathrm{JP}_{\textrm{transiting}}(P: \mathcal{J}(0.1, 520))$  & \\ \noalign{\smallskip}
${t_{\textrm{ic}}}^{\bullet}$\ [BJD$_{\mathrm{TDB}}$ - 2\,457\,000]   & $1366.17067_{-0.00033}^{+0.00036}$ & $\mathrm{JP}_{\textrm{transiting}}(\phi: \mathcal{U}(0, 1))$ & \\ \noalign{\smallskip}
$a$ [AU]                                                & $0.02191_{-0.00084}^{+0.00080}$ &   & \\ \noalign{\smallskip} 
$e$                                                     & $0.103_{-0.045}^{+0.117}$ &    & \\ \noalign{\smallskip}
$\omega_*$ [$^\circ$]                                   & $192_{-155}^{+70}$ &  &  \\ \noalign{\smallskip}
$M_{\textrm{ref}}$ \footnotemark[5] [radians]           & $2.7_{-1.7}^{+1.9}$ &  &  \\ \noalign{\smallskip}
$i_p$ [$\deg$]                                                          & $87.71_{-0.44}^{+1.16}$ &  &  \\ \noalign{\smallskip}
${e\cos \omega_*}^{\bullet}$                            & $-0.027_{-0.144}^{+0.099}$  &  $\mathrm{JP}_{e\cos \omega_*, e\sin \omega_*}(e:\beta(0.867, 3.03), $  &  \\ \noalign{\smallskip} 
${e\sin \omega_*}^{\bullet}$                            & $-0.028_{-0.072}^{+0.090}$ &  {$\omega_*:\mathcal{U}(-\pi, \pi))$}  & \\  \noalign{\smallskip} 
${K}^{\bullet}$ [\ms]                                   & $0.46_{-0.17}^{+0.20}$ & $\mathcal{U}(0, 17)$  & \\ \noalign{\smallskip}
${R_p / R_*}^{\bullet}$                                             & $0.02512_{-0.00064}^{+0.00072}$ & $\mathrm{JP}_{\textrm{transiting}}({R_p / R_*}: \mathcal{U}(10^{-3}, 1))$ & \\ \noalign{\smallskip} 
${\cos i_p}^{\bullet}$                                              & $0.0400_{-0.0203}^{+0.0076}$ & $\mathrm{JP}_{\textrm{transiting}}(b: \mathcal{U}(0, 2))$ & \\ \noalign{\smallskip}
$a / R_*$                                               & $15.0_{-1.0}^{+1.4}$ &   & \\ \noalign{\smallskip}
$b$                                                                         & $0.53_{-0.22}^{+0.14}$  &   & \\ \noalign{\smallskip}
$D14$ [h]                                                                   & $0.992_{-0.032}^{+0.090}$ &   & \\ \noalign{\smallskip}
$D23$ [h]                                                                   & $0.928_{-0.032}^{+0.075}$ &   & \\ \noalign{\smallskip}
$F_{i}$ [$F_{i, \oplus}$]                               & $24.7_{-4.1}^{+5.0}$ &   & \\ \noalign{\smallskip} 
$H$ [km]                                                & $430_{-110}^{+290}$ &   & \\ \noalign{\smallskip} 
                                                            & \multicolumn{3}{l}{\textit{Planet c}} \\
$M_p$ [\MEarth]                                         & $2.22_{-0.25}^{+0.26}$ &   & \\ \noalign{\smallskip} 
$R_p$ [\REarth]                                         & $1.385_{-0.075}^{+0.095}$ &  & \\ \noalign{\smallskip} 
$\rho_p$ [$\mathrm{g.cm^{-3}}$]                         & $4.57_{-0.85}^{+0.77}$ &  & \\ \noalign{\smallskip}
$T_{\textrm{eq}}$ [K]                                   & $553_{-26}^{+27}$ &   & \\ \noalign{\smallskip} 
${P}^{\ \bullet}$\ [days]                               & $3.6906777_{-2.6e-06}^{+1.6e-06}$ &  $\mathrm{JP}_{\textrm{transiting}}(P: \mathcal{J}(0.1, 520))$  & \\ \noalign{\smallskip}
${t_{\textrm{ic}}}^{\bullet}$\ [BJD$_{\mathrm{TDB}}$ - 2\,457\,000]   & $1367.27375_{-0.00022}^{+0.00013}$ & $\mathrm{JP}_{\textrm{transiting}}(\phi: \mathcal{U}(0, 1))$ & \\ \noalign{\smallskip} 
$a$ [AU]                                                & $0.0304_{-0.0012}^{+0.0011}$ &   & \\ \noalign{\smallskip} 
$e$                                                     & $0.103_{-0.058}^{+0.045}$ &   & \\ \noalign{\smallskip}  
$\omega_*$ [$^\circ$]                                   & $261_{-10}^{+20}$ &   & \\ \noalign{\smallskip}
$M_{\textrm{ref}}$ \footnotemark[5] [radians]           & $5.83_{-0.65}^{+0.19}$ &  &  \\ \noalign{\smallskip}
$i_p$ [$\deg$]                                                          & $88.11_{-0.16}^{+0.36}$ &  &  \\ \noalign{\smallskip}
${e\cos \omega_*}^{\bullet}$                            & $-0.014_{-0.022}^{+0.027}$  & $\mathrm{JP}_{e\cos \omega_*, e\sin \omega_*}(e:\beta(0.867, 3.03), $  & \\ \noalign{\smallskip} 
${e\sin \omega_*}^{\bullet}$                            & $-0.099_{-0.046}^{+0.056}$ &  {$\omega_*:\mathcal{U}(-\pi, \pi))$}  & \\ \noalign{\smallskip}
${K}^{\ \bullet}$ [\ms]                                 & $2.19_{-0.20}^{+0.17}$ & $\mathcal{U}(0, 17)$ & \\ \noalign{\smallskip} 
${R_p / R_*}^{\bullet}$                                                 & $0.04088_{-0.00056}^{+0.00068}$ & $\mathrm{JP}_{\textrm{transiting}}({R_p / R_*}: \mathcal{U}(10^{-3}, 1))$ & \\ \noalign{\smallskip} 
${\cos i_p}^{\bullet}$                                                  & $0.0330_{-0.0062}^{+0.0028}$ & $\mathrm{JP}_{\textrm{transiting}}(b: \mathcal{U}(0, 2))$ & \\ \noalign{\smallskip} 
$a / R_*$                                               & $19.00_{-0.80}^{+1.20}$ &   & \\ \noalign{\smallskip}
$b$                                                                     & $0.601_{-0.066}^{+0.081}$ &   & \\ \noalign{\smallskip}
$D14$ [h]                                                                   & $1.346_{-0.069}^{+0.122}$  &   & \\ \noalign{\smallskip} 
$D23$ [h]                                                               & $1.167_{-0.050}^{+0.125}$  &   & \\ \noalign{\smallskip}
$F_{i}$ [$F_{i, \oplus}$]                               & $12.8_{-2.1}^{+2.6}$ &   & \\ \noalign{\smallskip}
$H$ [km]                                                & $184_{-23}^{+43}$ &   & \\ \noalign{\smallskip}
                                                            & \multicolumn{3}{l}{\textit{Planet d}} \\
$M_p$ [\MEarth]                                         & $1.94_{-0.28}^ {+0.28}$ &  &  \\ \noalign{\smallskip}
$R_p$ [\REarth]                                         & $1.521_{-0.098}^{+0.119}$ &  & \\ \noalign{\smallskip}
$\rho_p$ [$\mathrm{g.cm^{-3}}$]                         & $2.95_{-0.51}^{+0.79}$ &  & \\ \noalign{\smallskip}
$T_{\textrm{eq}}$ [K]                                   & $416_{-20}^{+20}$ &   & \\ \noalign{\smallskip} 
${P}^{\ \bullet}$\ [days]                               & $7.4507245_{-4.6e-06}^{+8.1e-06}$ & $\mathrm{JP}_{\textrm{transiting}}(P: \mathcal{J}(0.1, 520))$ &  \\ \noalign{\smallskip}
${t_{\textrm{ic}}}^{\bullet}$\ [BJD$_{\mathrm{TDB}}$ - 2\,457\,000]   & $1362.73974_{-0.00040}^{+0.00031}$ & $\mathrm{JP}_{\textrm{transiting}}(\phi: \mathcal{U}(0, 1))$ & \\ \noalign{\smallskip} 
$a$ [AU]                                                & $0.0486_{-0.0019}^{+0.0018}$ &   & \\ \noalign{\smallskip} 
$e$                                                     & $0.074_{-0.046}^{+0.057}$ &   & \\ \noalign{\smallskip} 
$\omega_*$ [$^\circ$]                                   & $180_{-50}^{+27}$ &  &  \\ \noalign{\smallskip} 
\noalign{\smallskip}
$M_{\textrm{ref}}$ \footnotemark[5] [radians]           & $3.76_{-0.61}^{+0.66}$ &  &  \\ \noalign{\smallskip}
$i_p$ [$\deg$]                                                          & $88.449_{-0.111}^{+0.058}$ &  &  \\ \noalign{\smallskip}
${e\cos \omega_*}^{\bullet}$                            & $-0.062_{-0.061}^{+0.057}$ & $\mathrm{JP}_{e\cos \omega_*, e\sin \omega_*}(e:\beta(0.867, 3.03), $  & \\ \noalign{\smallskip} 
${e\sin \omega_*}^{\bullet}$                            & $0.000_{-0.026}^{+0.032}$ &   {$\omega_*:\mathcal{U}(-\pi, \pi))$}  & \\ \noalign{\smallskip}
${K}^{\ \bullet}$ [\ms]                                 & $1.50_{-0.19}^{+0.22}$ & $\mathcal{U}(0, 17)$  & \\ \noalign{\smallskip} 
${R_p / R_*}^{\bullet}$                                         & $0.0448_{-0.0010}^{+0.00106}$ &  $\mathrm{JP}_{\textrm{transiting}}({R_p / R_*}: \mathcal{U}(10^{-3}, 1))$ & \\ \noalign{\smallskip}
${\cos i_p}^{\bullet}$                                                      & $0.0271_{-0.0010}^{+0.0019}$ & $\mathrm{JP}_{\textrm{transiting}}(b: \mathcal{U}(0, 2))$  & \\ \noalign{\smallskip} 
$a / R_*$                                               & $33.7_{-1.7}^{+1.9}$ &   & \\ \noalign{\smallskip} 
$b$                                                                         & $0.922_{-0.059}^{+0.059}$ &   & \\ \noalign{\smallskip}
$D14$ [h]                                                                   & $0.84_{-0.20}^{+0.15}$  &   & \\ \noalign{\smallskip}  
$D23$ [h]                                                                   & $0.51_{-0.18}^{+0.23}$  &   & \\ \noalign{\smallskip} 
$F_{i}$ [$F_{i, \oplus}$]                               & $5.01_{-0.83}^{+1.02}$ &   & \\ \noalign{\smallskip}
$H$ [km]                                                & $195_{-37}^{+37}$ &   & \\ \noalign{\smallskip}
                                                            & \multicolumn{3}{l}{\textit{Planet e}}  \\
$M_p \sin i$ [\MEarth]                                  & $3.06_{-0.37}^{+0.33}$  &   & \\ \noalign{\smallskip} 
$T_{\textrm{eq}}$ [K]                                   & $342_{-18}^{+20}$  &   & \\ \noalign{\smallskip}
${P}^{\ \bullet}$\ [days]                               & $12.796_{-0.019}^{+0.020}$  & \multirow{2}{0.25\textwidth}{\raggedleft$\mathrm{JP}_{P, t_{\textrm{ic}}}\left(P:\mathcal{N}(12.8, 1), \phi:\mathcal{U}(0, 1) \right)$}  & \\ \noalign{\smallskip}
${t_{\textrm{ic}}}^{\bullet}$\ [BJD$_{\mathrm{TDB}}$ - 2\,457\,000]   & $1439.40_{-0.36}^{+0.37}$  &  & \\ \noalign{\smallskip}
$a$\footnotemark[1] [AU]                                & $0.0717_{-0.0048}^{+0.0060}$  &   & \\ \noalign{\smallskip}
$e$                                                     & $0.128_{-0.076}^{+0.108}$  &   & \\ \noalign{\smallskip}
$\omega_*$ [$^\circ$]                                   & $165_{-29}^{+40}$ &   & \\ \noalign{\smallskip}
$M_{\textrm{ref}}$ \footnotemark[5] [radians]           & $1.07_{-0.49}^{+2.1}$ &   &  \\ \noalign{\smallskip}
${e\cos \omega_*}^{\bullet}$                            & $-0.106_{-0.095}^{+0.095}$  & $\mathrm{JP}_{e\cos \omega_*, e\sin \omega_*}(e:\beta(0.867, 3.03), $  & \\ \noalign{\smallskip} 
${e\sin \omega_*}^{\bullet}$                            & $0.023_{-0.070}^{+0.056}$  &  {$\omega_*:\mathcal{U}(-\pi, \pi))$}  & \\ \noalign{\smallskip} 
${K}^{\ \bullet}$ [\ms]                                 & $2.01_{-0.20}^{+0.16}$  & $\mathcal{U}(0, 17)$  &  \\ \noalign{\smallskip}
$a / R_*$                                               & $49.8_{-3.8}^{+3.9}$  &   & \\ \noalign{\smallskip}
                                                            & \multicolumn{3}{l}{\textit{Planetary candidate 05}} \\
$M_p \sin i$ [\MEarth] \footnotemark[4]                 & $2.46_{-0.82}^{+0.66}$  &    &\\ \noalign{\smallskip} 
$T_{\textrm{eq}}$ [K] \footnotemark[4]                  &  $285_{-17}^{+18}$  &   & \\ \noalign{\smallskip}
${P}^{\ \bullet}$\ [days] \footnotemark[4]              & $23.15_{-0.17}^{+0.60}$ & \multirow{2}{0.25\textwidth}{\raggedleft$\mathrm{JP}_{P, t_{\textrm{ic}}}\left(P:\mathcal{N}(22.8, 1), \phi:\mathcal{U}(0, 1) \right)$}   & \\ \noalign{\smallskip}
${t_{\textrm{ic}}}^{\bullet}$\ [BJD$_{\mathrm{TDB}}$ - 2\,457\,000] \footnotemark[4]  & $1435.4_{-2.5}^{+2.5}$  &   &\\ \noalign{\smallskip}
$a$\footnotemark[1] [AU] \footnotemark[4]                            & $0.1034_{-0.0044}^{+0.0042}$  &   & \\ \noalign{\smallskip}
$e$ \footnotemark[4]                                                 & $0.21_{-0.11}^{+0.17}$  &   & \\ \noalign{\smallskip}
$\omega_*$ [$^\circ$] \footnotemark[4]                               & $-23_{-76}^{+85}$  &   & \\ \noalign{\smallskip}
${e\cos \omega_*}^{\bullet}$ \footnotemark[4]                         & $0.08_{-0.16}^{+0.15}$  & $\mathrm{JP}_{e\cos \omega_*, e\sin \omega_*}(e:\beta(0.867, 3.03), $  & \\ \noalign{\smallskip}
${e\sin \omega_*}^{\bullet}$ \footnotemark[4]                         & $-0.04_{-0.16}^{+0.17}$  &  {$\omega_*:\mathcal{U}(-\pi, \pi))$}  & \\ \noalign{\smallskip}
${K}^{\ \bullet}$ [\ms] \footnotemark[4]                             & $1.37_{-0.43}^{+0.33}$  & $\mathcal{U}(0, 17)$   & \\ \noalign{\smallskip}
$a / R_*$ \footnotemark[4]                                           & $73.3_{-6.4}^{+7.3}$  &  &  \\ \noalign{\smallskip} 

\\[-3pt]
\multicolumn{4}{l}{\textit{Stellar parameters}} \\
\noalign{\smallskip}\hline\hline\noalign{\smallskip}
\textsc{ra}$^{\textsc{gaia-crf2}}$  [hh:mm:ss.ssss]  & 08:18:07.89  &  & GAIA-DR2 \\ \noalign{\smallskip}   
\textsc{dec}$^{\textsc{gaia-crf2}}$ [dd:mm:ss.ss]    & -68:18:52.08  &  & GAIA-DR2 \\  \noalign{\smallskip} 
Sp. Type                                            & M3V &   & \kostov \\ \noalign{\smallskip}
V mag                                                & $11.685 \pm 0.02$ &  & APASS DR9 \\ \noalign{\smallskip}
Ks mag                                          & $7.101 \pm 0.018$ & & 2MASS \\ 
J mag                           & 7.9 & & 2MASS \\ \noalign{\smallskip}
parallax [mas]                                                  & $94.1385 \pm 0.0281$ &  & GAIA-DR2 \\ \noalign{\smallskip}
distance [pc]                                   & $10.6194 \pm 0.0032$ & & BJ18 \\
$M_*$ [M$_{\sun}$]                                  & $0.273 \pm 0.030$  &  & \\ \noalign{\smallskip}
$R_*$ [R$_{\sun}$]                                   & $0.303^{+0.026}_{-0.023}$  &  & \\
\noalign{\smallskip}
age [Myr]                                           & $> 800$ & & \\
${\rho_*}^{\bullet}$ [$\rho_\sun$]             & $9.15_{-1.4}^{+1.8}$  &   $\mathrm{JP}_{\textrm{transiting}}(\rho_*: \mathcal{N}(11.2, 1.9))$ & \\ \noalign{\smallskip}
${L_*}$ [L$_\sun$]             & $0.01128 \pm 0.00042$ &  &  \\ \noalign{\smallskip}
\teff\ [K]                                                   & $3415 \pm 135$ &  & \\ \noalign{\smallskip}
\logg\ [from cm.s$^{-2}$]                      & $4.86 \pm 0.13$ &   & \\ \noalign{\smallskip}
{[Fe/H]} [dex]                                       & $-0.46 \pm 0.26$ &  & \\ \noalign{\smallskip}
[Mg/H] [dex] \footnotemark[6]             & $-0.38 \pm 0.11$ &  & \\ \noalign{\smallskip}
[Si/H] [dex] \footnotemark[6]               & $-0.42 \pm 0.13$ &  & \\ \noalign{\smallskip}
${v0}^{\bullet}$ [\kms]                                              & $-5.57851_{-0.00069}^{+0.00072}$  & $\mathcal{N}(-5.5791, 0.0035)$  & \\ \noalign{\smallskip}
${A_{\rv}}^{\bullet}$ [\ms]                                              & $2.44_{-0.36}^{+0.43}$ & $\mathcal{U}(0, 17)$  & \\ \noalign{\smallskip}
${A_{\fwhm}}^{\bullet}$ [\ms]                                              & $8.6_{-1.1}^{+1.2}$  & $\mathcal{U}(0, 43)$  & \\ \noalign{\smallskip}
${P_{\mathrm{rot}}}^{\bullet}$ [\ms]                                    & $33_{-19}^{+43}$  & $\mathcal{J}(5, 520)$  & \\ \noalign{\smallskip}
${\tau_{\mathrm{decay}}}^{\bullet}$ [\ms]                                    & $49_{-10}^{+14}$ & $\mathcal{J}(2.5, 2600)$ + $\tau_{\mathrm{decay}} > P_{\mathrm{rot}} / 2$\footnotemark[2]  & \\ \noalign{\smallskip}
${\gamma}^{\bullet}$ [\ms]                                    & $3.2_{-1.6}^{+1.2}$  & $\mathcal{U}(0.05, 5)$ & \\ \noalign{\smallskip}
$u_{1,\tess}^{\bullet}$                                         & $0.156_{-0.042}^{+0.041}$ & $\mathcal{N}(0.147, 0.044)$  & \\ \noalign{\smallskip}
$u_{2,\tess}^{\bullet}$                                         & $1.593_{-0.038}^{+0.040}$ & $\mathcal{N}(1.583, 0.045)$ & \\ \noalign{\smallskip}
$u_{3,\tess}^{\bullet}$                                         & $-1.617_{-0.035}^{+0.033}$ & $\mathcal{N}(-1.627, 0.036)$ & \\ \noalign{\smallskip}
$u_{4,\tess}^{\bullet}$                                         & $0.542_{-0.016}^{+0.015}$ & $\mathcal{N}(0.539, 0.015)$ & \\ \noalign{\smallskip}

\\[-5pt]
\multicolumn{3}{l}{\textit{Parameters of instruments}} \\
\noalign{\smallskip}\hline\hline\noalign{\smallskip}
$\Delta\mathrm{RV}_{\mathrm{post/pre}}^{\bullet}$  [\ms]       & $1.2_{-1.1}^{+1.0}$  &  $\mathcal{N}(2.88, 4.8)$  &\\ \noalign{\smallskip}
${\Delta\mathrm{RV}_{\harps/\textrm{pre}}}^{\bullet}$ [\ms]    & $-99.13_{-0.34}^{+0.33}$  &  $\mathcal{N}(-99.5, 5.0)$  &\\ \noalign{\smallskip}
$\sigma_{\rv, \mathrm{pre}}^{\bullet}$ [\ms]                             & $0.88_{-0.31}^{+0.35}$ &  $\mathcal{U}(0, 4.5)$  & \\ \noalign{\smallskip}
$\sigma_{\rv, \mathrm{post}}^{\bullet}$ [\ms]                           & $0.91_{-0.55}^{+0.73}$ &  $\mathcal{U}(0, 3.6)$  & \\ \noalign{\smallskip}
$\sigma_{\rv, \mathrm{harp}}^{\bullet}$ [\ms]                           & $< 0.32$ &  $\mathcal{U}(0, 11)$  & \\ \noalign{\smallskip}
$C_{\mathrm{pre}}^{\bullet}$ [\kms]                                         & $ 4.5136_{-0.0028}^{+0.0030}$ & $\mathcal{N}(4.5057, 0.0089)$  & \\ \noalign{\smallskip}
$C_{\mathrm{post}}^{\bullet}$ [\kms]                                        & $4.5135_{-0.0028}^{+0.0029}$ & $\mathcal{N}(4.5171, 0.0099)$  & \\ \noalign{\smallskip}
$C_{\harps}^{\bullet}$ [\kms]                                                   & $3.0573_{-0.0022}^{+0.0022}$ &  $\mathcal{N}(3.0552, 0.0075)$  & \\ \noalign{\smallskip}
$\sigma_{\fwhm, \mathrm{pre}}^{\bullet}$ [\ms]                       & $5.42_{-0.95}^{+1.04}$ &  $\mathcal{U}(0, 9.0)$  & \\ \noalign{\smallskip}
$\sigma_{\fwhm, \mathrm{post}}^{\bullet}$ [\ms]                      & $< 1.0$ &  $\mathcal{U}(0, 7.2)$  & \\ \noalign{\smallskip}
$\sigma_{\fwhm, \harps}^{\bullet}$ [\ms]                                  & $4.16_{-0.68}^{+0.77}$ &  $\mathcal{U}(0, 21)$  & \\ \noalign{\smallskip}
$\sigma_{\tess}^{\bullet}$ [ppm]                                              &  $< 25$ & $\mathcal{U}(0, 4200)$  & \\ \noalign{\smallskip}
\noalign{\smallskip} 
\end{longtable}
\tablefoot{\\
- The values provided in the column "Posterior" were derived in this work, except when specified otherwise in the  column "Source". The references for these external sources are APASS DR9 \citep{henden2016}, 2MASS \citep{skrutskie2006}, GAIA-DR2 \citep{gaiacollaboration2018a}, \kostov\ \citep{kostov2019}, and BJ18 \citep{bailer-jones2018}.\\
- The justifications of the choices of \prior s can be found in \app{priors}. These priors were used for all the analyses performed in Sects. \sectref{lconly}, \sectref{rvonly} , and \sectref{finalana} with only one exception (see \footnotemark[3] below).\\
- $\mathcal{U}(vmin, vmax)$ and $\mathcal{J}(vmin, vmax)$ stand for uniform and Jeffreys probability distributions, respectively, with $vmin$ and $vmax$ as the minimum and maximum values.
$\mathrm{JP}$ stands for joint prior (see \app{priors} for more details).\\
$^{\bullet}$ indicates that the parameter is a main or jumping parameter for the \textsc{mcmc} explorations performed in Sects. \sectref{firstana} to \sectref{stabilityandfinalparam}.\\
\footnotemark[1] For the nontransiting planets, $a$ is computed from $a / R_*$.\\
\footnotemark[2] For the prior of ${\tau_{\mathrm{decay}}}$, "$+ {\tau_{\mathrm{decay}}}$, $\tau_{\mathrm{decay}} > P_{\mathrm{rot}} / 2$" indicates an additional condition imposed on the prior of this parameter.\\
\footnotemark[3] The only exception to the fact that the \prior s used are those provided in the column "Prior" of this table is for the ephemerides parameters $P$ and $t_{\textrm{ic}}$ of the three transiting planets in \sect{rvonly}. In these cases, the \prior s used are the \post s obtained for these parameters during the analysis of the \tess\ \lc\ alone (see \sect{lconly}). The \prior s are $P_b = \mathcal{N}(2.2531135, 1.7e-6)$, $t_{\textrm{ic}, b} = \mathcal{N}(1366.17057, 3.3e-4)$, $P_c = \mathcal{N}(3.6906776, 3.0e-6)$, $t_{\textrm{ic}, c} = \mathcal{N}(1367.27357, 2.8e-4)$, $P_d = \mathcal{N}(7.4507272, 7.8e-6)$, and $t_{\textrm{ic}, d} = \mathcal{N}(1362.73972, 4.8e-4)$.\\
\footnotemark[4] The parameters reported for planetary candidate 5 are obtained from the analysis presented in \sect{rvonly}. In contrast to the parameters of the other planets, which where obtained through the analysis described in \sect{stabilityandfinalparam}, they do not include any condition related to dynamical stability.\\
\footnotemark[5] $M_{\textrm{ref}}$ is the mean anomaly computed at the reference time 1354 BTJD, the time of the first \tess\ measurement.\\
\footnotemark[6] As described in \sect{abundances}, the abundance ratios [Mg/H] and [Si/H] are not directly measured on the observed spectra. They are statistical estimates obtained from a population of stars to which we believe L\,98-59 belongs. \\
}
\end{raggedleft}
\end{symbolfootnotes}
\twocolumn


\end{document}